\documentclass{aa}

\newcommand{\Msun}{M$_{\sun}$}

\newcommand{\hMsun}{{\ifmmode{h^{-1}{\rm {M_{\odot}}}}\else{$h^{-1}{\rm{M_{\odot}}}$}\fi}}  
\newcommand{\hMpc}{{\ifmmode{h^{-1}{\rm Mpc}}\else{$h^{-1}$Mpc }\fi}}  
\newcommand{\hkpc}{{\ifmmode{h^{-1}{\rm kpc}}\else{$h^{-1}$kpc }\fi}}

\newcommand{\xmm }{{\em XMM}}

\defcitealias{Verdugo2011}{VMM11}

\usepackage{multirow}
\usepackage{graphicx, subfig}
\usepackage{rotating}
\usepackage{txfonts}
\usepackage{natbib}
\usepackage{lscape}
\usepackage{array}
\usepackage{color}
\usepackage{soul}
\begin{document}
\title{Combining Strong Lensing and Dynamics in Galaxy Clusters:\\
integrating MAMPOSSt within LENSTOOL\\
 I. Application on \object{SL2S\,J02140-0535}
\thanks{SL2S: Strong Lensing Legacy Survey}$^{,}$  
\thanks{Based on observations obtained with MegaPrime/MegaCam, a joint project of CFHT and CEA/IRFU, at the Canada-France-Hawaii Telescope (CFHT) which is operated by the National Research Council (NRC) of Canada, the Institut National des Science de l'Univers of the Centre National de la Recherche Scientifique (CNRS) of France, and the University of Hawaii. This work is based in part on data products produced at Terapix available at the Canadian Astronomy Data Centre as part of the Canada-France-Hawaii Telescope Legacy Survey, a collaborative project of NRC and CNRS.
Also based on \emph{Hubble Space Telescope} (HST) data, VLT (FORS\,2) data, and \xmm\, data.} }
   	\subtitle{}
   \author{T. Verdugo\inst{1,2},
                  M. Limousin\inst{3},
                V. Motta\inst{2},
                G. A. Mamon\inst{4},
                G. Fo{\"e}x\inst{5},
                F. Gastaldello\inst{6},
                 E. Jullo\inst{3},
                A. Biviano\inst{7}, 
                K. Rojas\inst{2},
               R. P. Mu\~{n}oz\inst{8}, 
                R. Cabanac\inst{9}, 
              J.  Maga\~{n}a\inst{2},
               J. G. Fern\'andez-Trincado\inst{10}, 
               L.  Adame\inst{11}, 
               M. A. De Leo\inst{12} 
                }
   \offprints{tomasv$@$astrosen.unam.mx}

   \institute{Instituto de Astronom\'ia, Universidad Nacional Aut\'onoma de M\'exico, Apdo.Postal 106, Ensenada, B.C. 22860, M\'exico
   \and
    Instituto de F\'{\i}sica y Astronom\'{\i}a, Universidad de Valpara\'{\i}so,  Avenida Gran Breta\~{n}a 1111, Valpara\'{\i}so, Chile
       \and
       Laboratoire d'Astrophysique de Marseille, Universit\'e de Provence,\\ CNRS, 38 rue Fr\'ed\'eric Joliot-Curie, F-13388 Marseille Cedex 13, France
             \and
             Institut d'Astrophysique de Paris (UMR 7095: CNRS \& UPMC), 98 bis Bd Arago, F-75014 Paris, France           
             \and
             Max Planck Institute for Extraterrestrial Physics, Giessenbachstrasse, 85748 Garching, Germany           
             \and
             INAF-IASF Milano, via E. Bassini 15, I-20133 Milano, Italy             
              \and
             INAF-Osservatorio Astronomico di Trieste, via G.B. Tiepolo 11, 34143 Trieste, Italy
        \and
        Instituto de Astrof\'isica, Facultad de F\'isica, Pontificia Universidad Cat\'olica de Chile, Av.~Vicu\~na Mackenna 4860, 7820436 Macul, Santiago, Chile
        \and
        Laboratoire d'Astrophysique de Toulouse-Tarbes, Universit\'e de Toulouse, CNRS,      57 Avenue d'Azereix, 65 000 Tarbes, France   
     \and
     Institut Utinam, CNRS UMR 6213, Universit\'e de Franche-Comt\'e, OSU THETA Franche-Comt\'e-Bourgogne, Observatoire de Besan\c{c}on, BP 1615, 25010 Besan\c{c}on Cedex, France 
        \and 
          Facultad de Ciencias F\'isico-Matem\'aticas, Universidad Aut\'onoma de Nuevo Le\'on, San Nicol\'as de los Garza, M\'exico
         \and
      Department of Physics and Astronomy, University of California, Riverside, CA 92521, USA
   }

 \date{Preprint online version:}

 \abstract
  {The mass distribution in both galaxy clusters and groups is an important cosmological probe. 
It has become clear in the last years that mass profiles are best recovered when combining 
complementary probes of the gravitational potential. Strong lensing (SL) is very accurate in the inner regions, but other probes are required to constrain the mass distribution in the outer regions, such as weak lensing or dynamics studies.}
  {We constrain the mass distribution of a cluster showing gravitational arcs by combining a strong lensing method with a dynamical method using the velocities of its 24 member galaxies.}
  {We present a new framework were we simultaneously fit SL and dynamical data.
The SL analysis is based on the {\sc LENSTOOL} software, and the dynamical
analysis uses the {\sc MAMPOSS}t code, which we have integrated into {\sc LENSTOOL}. After describing the implementation of this new tool, we apply it on the galaxy group \object{SL2S\,J02140-0535} ($z_{\rm spec}=0.44$), which we have already studied in the past. We use new VLT/FORS2 spectroscopy of multiple images and group members, as well as shallow X-ray data from \xmm.
}
 {We confirm that the observed lensing features in \object{SL2S\,J02140-0535} belong to different background sources. One of this sources is located at $z_{\rm spec}$ = 1.017 $\pm$ 0.001, whereas the other source is located at  $z_{\rm spec}$ = 1.628 $\pm$ 0.001. With the analysis of our new and our previously reported spectroscopic data, we find 24 secure members for \object{SL2S\,J02140-0535}. Both data sets are well reproduced by a single NFW mass profile:  the dark matter halo coincides with the peak of the light distribution, with scale radius, concentration, and mass equal to  $r_s$ =$82^{+44}_{-17}$ kpc , $c_{200}$ = $10.0^{+1.7}_{-2.5}$, and $M_{200}$ =  $1.0^{+0.5}_{-0.2}$ $\times$ 10$^{14}$M$_{\odot}$ respectively. These parameters are better constrained when we fit simultaneously SL and dynamical information.  The mass contours of our best model agrees with the direction defined by the luminosity contours and the X-ray emission of \object{SL2S\,J02140-0535}. The simultaneous fit lowers the error in the mass estimate by 0.34 dex, when compared to the SL model, and in 0.15 dex when compared to the dynamical method.}
{The combination of SL and dynamics tools yields a more accurate probe of the mass profile of  \object{SL2S\,J02140-0535} up to $r_{200}$. However, there is tension between the best elliptical SL model and the best spherical dynamical model. The similarities in shape and alignment of the centroids of the total mass, light, and intracluster gas distributions add to the picture of a non disturbed system.}


   \titlerunning{Integrating MAMPOSSt within LENSTOOL}
   \authorrunning{Verdugo et~al.}

  \maketitle
%

 
 \section{Introduction}\label{Intro}

\begin{figure}[h!]\begin{center}
\includegraphics[scale=0.755]{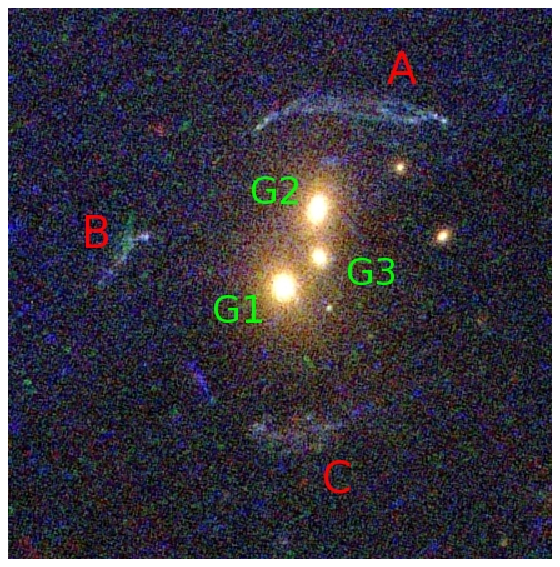}
\includegraphics[scale=0.411]{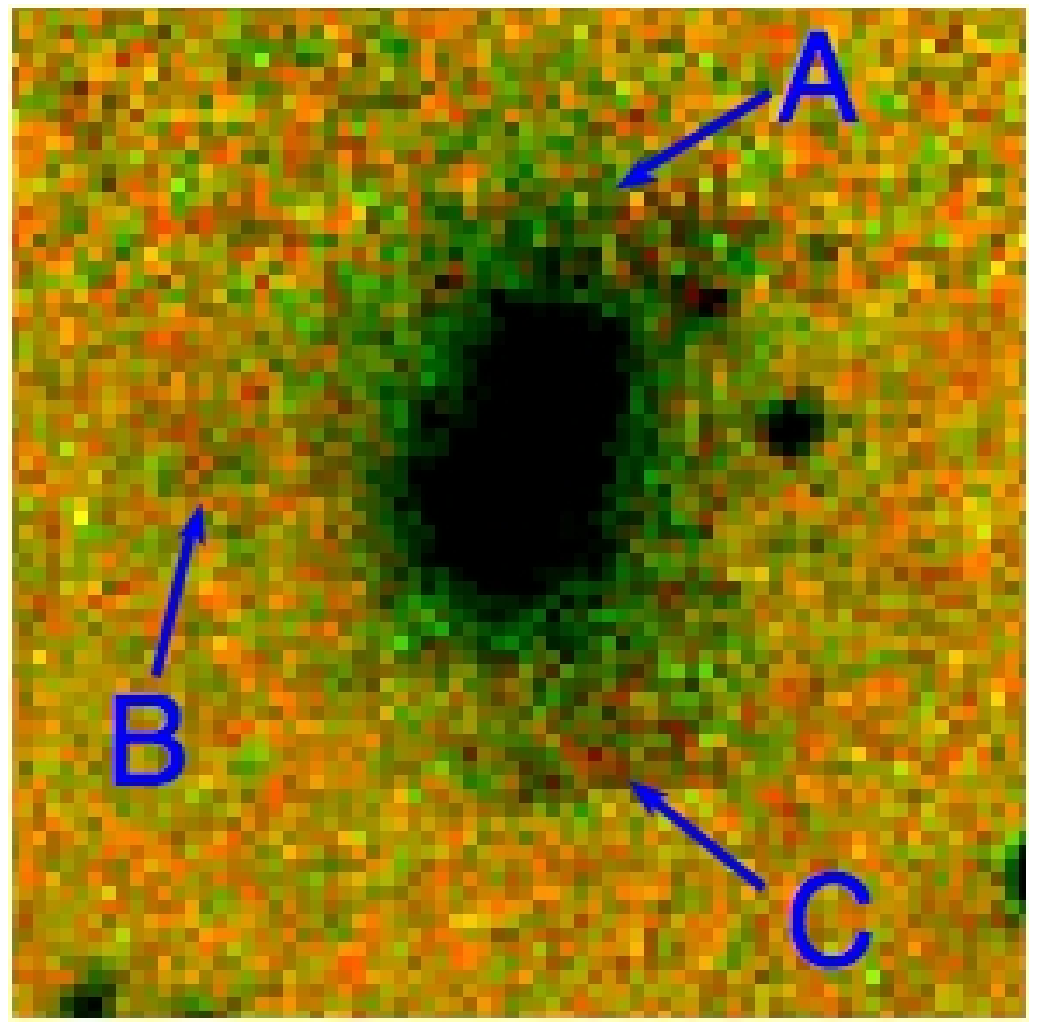}
\caption{ Images of the group SL2S\,J02140-0535 at $z_{\rm spec}=0.44$. \emph{Left:} composite HST/ACS F814, F606, F475 color image ($22\arcsec \times 22\arcsec$ = 125 $\times$ 125 kpc$^{2}$) showing the central region of the group \citepalias[from][]{Verdugo2011}. \emph{Right:} composite WIRCam $J, K_s$ color image ($22\arcsec \times 22\arcsec$).}
\label{presentlens} \end{center} 
\end{figure}

The Universe has evolved into the filamentary and clumpy structures \citep[dubbed the cosmic web,][]{Bond1996} that are observed in large redshift surveys  \citep[e.g.,][]{Colless2001}. Massive and rich galaxy clusters are located at the nodes of this cosmic web, being fed by accretion of individual galaxies and groups \citep[e.g.,][]{Frenk1996,Springel2005,Jauzac2012}. Galaxy clusters are the most massive gravitationally bound systems in the Universe, thus constituting one of the most important and crucial astrophysical objects to constrain cosmological parameters  \citep[see for example][]{Allen2011}. Furthermore, they provide  information about galaxy evolution \citep[e.g.,][]{Postman2005}. Also, galaxy groups are important cosmological probes \citep[e.g.,][]{Mulchaey2000,EkeVR2004} because they are tracers of  the large-scale structure of the Universe, but also because they are probes of the environmental dependence of the galaxy properties, the galactic content of dark matter haloes, and the clustering of galaxies \citep{Eke2004}. Although there is no clear boundary in mass between groups of galaxies and clusters of galaxies \citep[e.g.,][]{Tully2014}, it is commonly assumed that  groups of galaxies lie in the intermediate mass range between large elliptical galaxies and galaxy clusters (i.e., masses between $\sim 10^{13}$\Msun\, to  $\sim 10^{14}$\Msun).

The mass distribution in both galaxy groups and clusters has been studied extensively using different methods, such as the radial distribution of the gas through X-ray emission  \citep[e.g.,][and references therein]{Sun2012,Ettori2013}, the analysis of galaxy-based techniques that employs the positions, velocities and colors of the galaxies \citep[see][for a  comparison of the accuracies of dynamical methods in measuring $M_{200}$]{Old2014,Old2015}, or through gravitational lensing \citep[e.g.,][and references therein]{paperI,Limousin2010,Kneib2011}. Each probe has its own limitations and biases, which in turn impact the mass distribution measurements. Strong lensing (hereafter SL)  analysis provides the total amount of mass and its distribution with no assumptions on neither the dynamical state nor the nature of the matter producing the lensing effect. Nevertheless, the analysis has its own weakness; for example, it can solely constrain the two dimensional projected mass density, and it is limited to small projected radii. On the other hand, the analysis of galaxy kinematics do not has such limitations, but assumes local dynamical equilibrium (i.e., negligible rotation and streamings motions) and spherical symmetry. 

In this paper, we put forward a new method to overcome one of the limitations of the SL analysis, namely, the impossibility to constrain large-scale properties of the mass profile. Our method combines SL (in a parametric fashion, see Sect.\,\ref{Metho}) with the dynamics of the group or the cluster galaxy members, fitting simultaneously both data sets. Strong lensing and dynamics are both well recognized probes, the former providing an estimate of the projected two dimensional mass distribution within the core (typically a few dozens of arcsecs at most), whereas the latter is able to study the density profile at larger radii, using galaxies as test particles to probe the host potential.

There are several methods to constrain the mass profiles of galaxy clusters from galaxy kinematics. One can fit the 2\textit{nd} and 4\textit{th} moments of the line-of-sight velocities, in bins of projected radii \citep{Lokas2003}. One can assume a profile for the velocity anisotropy, and apply mass inversion techniques \citep[e.g.,][]{Mamon2010,Wolf2010,Sarli2014}. Both methods require binning of the data.  An alternative is to fit the observed distribution of galaxies in projected phase space (projected radii and line-of-sight velocities), which does not involve binning the data. This can be performed by assuming six-dimensional distribution functions (DFs), expressed as function of energy and angular momentum \citep[e.g.,][who used DFs derived by  \citealp{Wojtak2008} for  $\Lambda$CDM halos]{Wojtak2009}, but the method is very slow, as it involves triple integrals for every galaxy and every point in parameter space.  An accurate and efficient alternative is to assume a shape of the velocity DF as in the  {\sc MAMPOSS}t  method of  \citet[][]{Mamon2013}, which has been used to study the radial profiles of mass and velocity anisotropy of clusters \citep{Biviano2013,Munari2014,Guennou2014,Biviano2016}. {\sc MAMPOSS}t is ideal for the aims of the present study as: 1.- it is accurate for a dynamical model and very rapid\footnote{A valuable asset, since lensing modeling has become time demanding. See for example the discussion in \citet{Jauzac2014} about the  computing resources when modeling a SL cluster with many constraints.}; 2.-  it produces a likelihood as does the {\sc LENSTOOL} code used here for SL (see Sect.\,\ref{Metho}); 3.- can run with the same parametric form of the mass profile as used in {\sc LENSTOOL}.

 \begin{figure*}[!htp]
\centering
\includegraphics[width=0.75\textwidth]{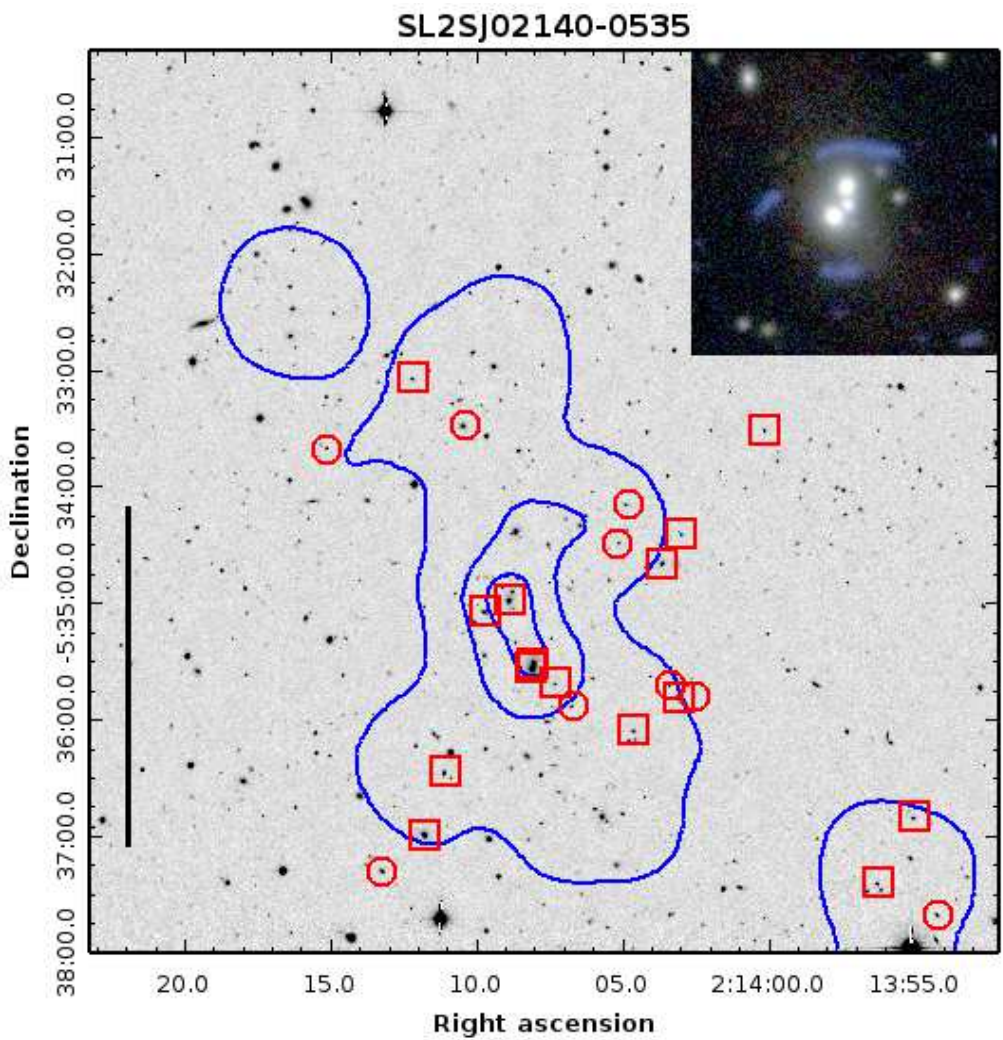}
\caption{CFHTLS $i$-band image with the luminosity density contours for \object{SL2S\,J02140-0535}. They represent $1\times10^6$, $5\times10^6$,  and $1.0\times10^7\;$L$_\odot\,$kpc$^{-2}$ from outermost to innermost contour, respectively. The red squares and circles show the location of the 24 confirmed members of the group, the squares represent the galaxies previously reported by \citet{Roberto2013}, and the circles the new observations.  The black vertical line on the left represents 1 Mpc  at the group rest-frame. The inset in the top-right corner shows a 30$\arcsec$$\times$30$\arcsec$  CFHTLS false color image of the system.\label{fig1}}
\end{figure*}

The idea of combining lensing and dynamics is not new. So far, dynamics have been used to probe the very centre
of the gravitational potential, through the measurement of the velocity dispersion profile of the central brightest cluster galaxy \citep{sand02,sand04,new09,new13}. 
However, the use of dynamical information at large scale (velocities of the galaxy members in a cluster or a group), together with SL analysis has not been fully explored. Through this approach, \citet{Thanjavur2010} showed that it is possible to characterize the mass distribution and the mass-to-light ratio of galaxy groups. \citet{Biviano2013} analyzed the cluster MACS\,J1206.2-0847, constraining its mass, velocity-anisotropy, and pseudo-phase-space density profiles,  finding a good agreement between the results obtained from cluster kinematics and those derived from lensing.  Similarly, \citet{Guennou2014} compared the mass profile inferred from lensing with different profiles obtained from three methods based on kinematics, showing that they are consistent  among themselves.

This work follows the analysis of  \citet{Verdugo2011}, hereafter  \citetalias{Verdugo2011}, where we combined SL and dynamics in the galaxy group \object{SL2S\,J02140-0535}. In \citetalias{Verdugo2011}, dynamics were used to constrain the scale radius of a NFW mass  profile, a quantity that is not accessible to SL constraints alone. These constraints were used as a prior in the SL analysis, allowing to probe the mass distribution from the centre to the virial radius of the galaxy group. However, the fit was not simultaneous. In this work we propose a framework aimed at fitting simultaneously SL and dynamics, combining the likelihoods obtained from both techniques in a consistent way. Our paper is arranged as follows: In Sect.\,\ref{Metho}  the methodology  is explained. In Sect.\,\ref{DATA} and Sect.\,\ref{MassM} we present the observational data images, spectroscopy, and the application of the method to the galaxy group \object{SL2S\,J02140-0535}. We summarize and discuss our results in Sect.\,\ref{Discus}. Finally in Sect.\,\ref{Conclusions}, we present the conclusions.  All our results are scaled to a flat, $\Lambda$CDM cosmology with $\Omega_{\rm{M}} = 0.3, \Omega_\Lambda = 0.7$ and a Hubble constant $H_0 = 70$ km\,s$^{-1}$ Mpc$^{-1}$. All images are aligned with WCS coordinates, i.e., North is up, East is left. Magnitudes are given in the AB system.

 \section{Methodology}\label{Metho}

In this section we explain how the SL and dynamical likelihoods are computed in our models.

 \subsection{Strong lensing}

The figure-of-merit-function, $\chi^{2}$, that quantifies the goodness of the fit for each trial of the lens model, has been introduced in several  works \citep[e.g.,][]{ver07,Limousin2007,jullo07},  therefore, we summarize the method here. Consider  a model whose parameters are $\vec {\theta}$, with $N$ sources, and $n_i$ the number of multiple images for source $i$. We compute, for every system $i$,  the position in the image plane $x^j(\theta)$ of image $j$, using the lens equation. Therefore, the contribution to the overall $\chi^{2}$ from multiple image system $i$ is

\begin{equation}\label{eq:Chi2Lens}
\chi_{i}^{2} =     \sum_{j=1}^{n_i}
\frac{\left[ x_{obs}^j - x^j(\theta)    \right]^2}{\sigma_{ij}^{2}},
\end{equation}

\noindent were $\sigma_{ij}$ is the error on the position of image $j$, and $x_{obs}^j$ is the observed position. Thus, we can write the likelihood as

\begin{equation}\label{eq:LikeLens}
\mathcal{L}_{\textrm{Lens}} = \prod_{i}^{N}\frac{1}{\prod_{j}\sigma_{ij}\sqrt{2\pi}}e^{-\chi^2_i/2},
\end{equation}

\noindent where it is assumed that the noise associated to the measurement of each image position is Gaussian and uncorrelated \citep[][]{jullo07}. This is not true in the case of images that are very close to each other, but it is a reasonable approximation for  \object{SL2S\,J02140-0535}. In this work we assume that the error in the image position is  $\sigma_{ij}$ = 0.5$\arcsec$, which is slightly greater than the value adopted in \citetalias{Verdugo2011}, but is half the value that has been suggested by other authors in order to take into account systematic errors in lensing modeling \citep[e.g.,][]{Jullo2010,DAloisio2011,Host2012,Zitrin2015}.

 \subsection{Dynamics}

{\sc MAMPOSS}t \citep[][]{Mamon2013} is a method that performs a maximum likelihood fit of the distribution of observed tracers in projected phase space (projected radii and line-of-sight velocity, hereafter  PPS). We refer the interested reader to \citet[][]{Mamon2013} for a detailed description, here we present a summary. {\sc MAMPOSS}t assumes parameterized radial profiles of mass and velocity anisotropy, as well as a shape for the three-dimensional velocity distribution (a Gaussian 3D velocity distribution). {\sc MAMPOSS}t fits the distribution of observed tracers in PPS. The method has been tested in cosmological simulations, showing the possibility to recover the virial radius, the tracer scale radius, and the dark matter scale radius when using 100 to 500 tracers \citep[][]{Mamon2013}. Moreover, \citet{Old2015} found that the mass normalization $M_{200}$ is recovered with 0.3 dex accuracy for as few as $\sim$30 tracers.

The velocity anisotropy is defined through the expression

\begin{equation}\label{eq:anisotropy}
\beta(r)  =   1 - \frac{\sigma^2_{\theta}(r) + \sigma^2_{\phi}(r)}{2\sigma^2_{r}(r)},
\end{equation}

\noindent where, in spherical symmetry,  $\sigma_{\phi}$($r$) = $\sigma_{\theta}$($r$). In the present work we adopt a constant anisotropy model with $\sigma_r/\sigma_{\theta}$ = (1 $-$ $\beta$)$^{-1/2}$, assuming spherical symmetry (see  Sect.\,\ref{MassM}).

The 3D velocity distribution is  assumed to be Gaussian:

\begin{equation}\label{eq:3Ddis}
f_{\upsilon}  =  \frac{1}{(2\pi)^{3/2}\sigma_{r}\sigma^2_{\theta}}\exp\left[ - \frac{\upsilon^2_{r}}{2\sigma^2_{r}}  - \frac{\upsilon^2_{\theta} + \upsilon^2_{\phi} }{2\sigma^2_{\theta}}    \right ],
\end{equation}

\noindent where $\upsilon_{r}$, $\upsilon_{\theta}$, and $\upsilon_{\phi}$ are the velocities in a spherical coordinate system. This Gaussian distribution assumes no rotation or radial streaming, which is a good assumption inside the virial radius, as has been shown by numerical simulations \citep{Prada2006,Cuesta2008}. The Gaussian 3D velocity model is a first-order approximation, which can be improved \citep[see][]{Beraldo2015}.

Thereby, {\sc MAMPOSS}t  fits the parameters using maximum likelihood estimation, i.e. by minimizing

\begin{equation}\label{eq:MLE}
- \ln{\mathcal{L}_{\textrm{Dyn}}}  = -\sum_{i=1}^n   \frac{\ln q(R_i,\upsilon_{z,i} \mid \bar{\theta})}{C(R_i)},
\end{equation}

\begin{figure}[h!]\begin{center}
\includegraphics[scale=0.46]{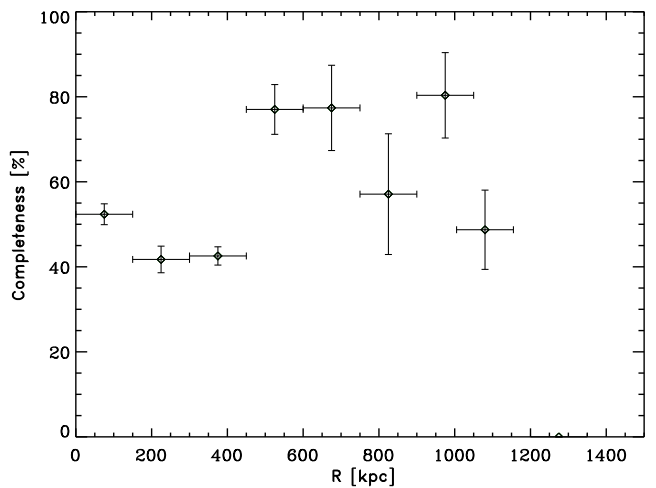}
\caption{Completeness as a function of the radius in \object{SL2S\,J02140-0535}. $C\sim60\%$ roughly constant up to 1 Mpc.}
\label{completeness} \end{center} 
\end{figure}

\noindent where $q$ is the probability density of observing an object at projected radius $R$, with line-of-sight (hereafter LOS) velocity $\upsilon_z$, for a N-parameter vector $\bar{\theta}$. $C(R_i)$  is the completeness of the data set (see section~\ref{NSD}).

\begin{figure*}[!htp]
\centering
\includegraphics[scale=0.46]{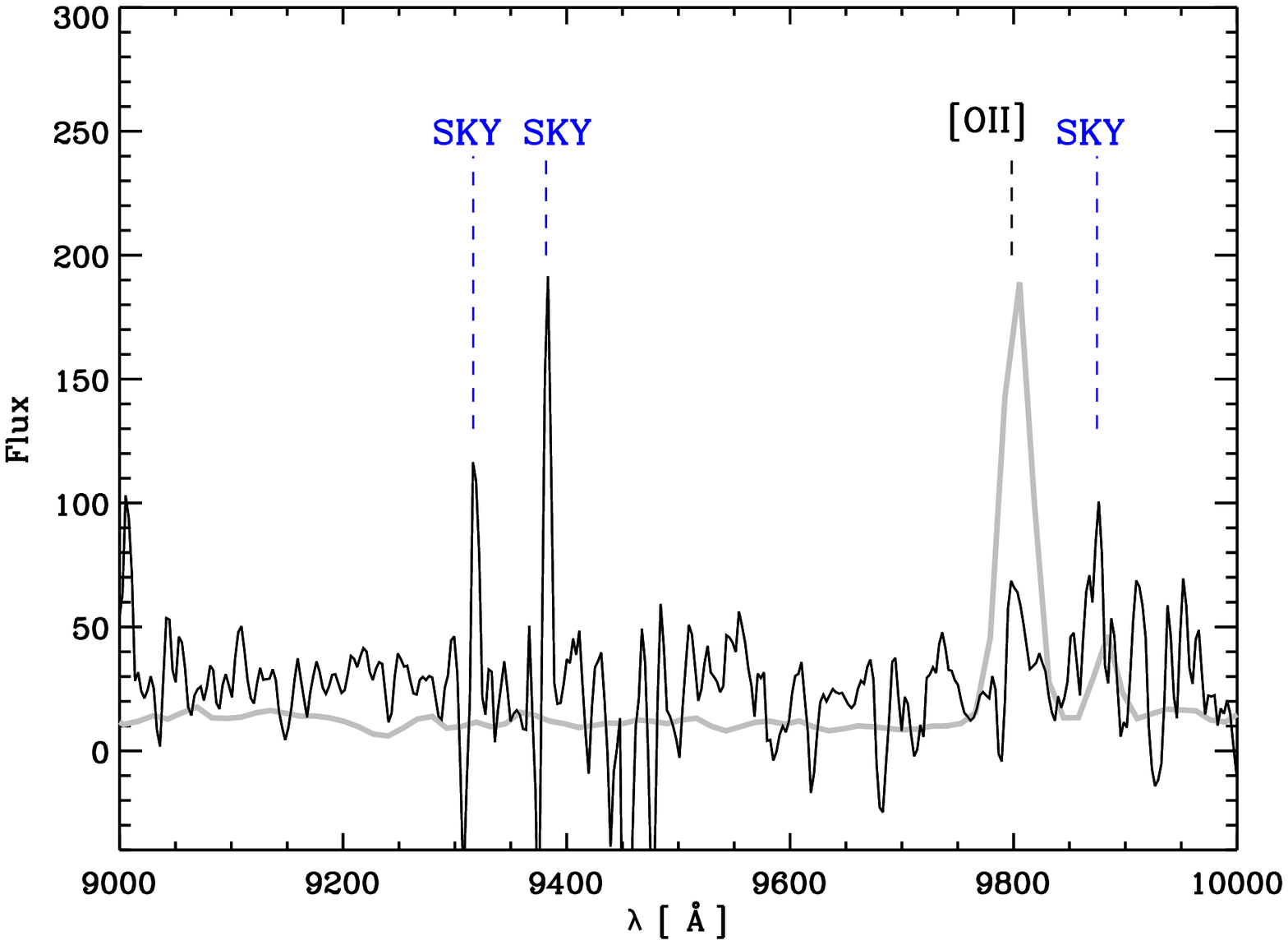} 
\includegraphics[scale=0.46]{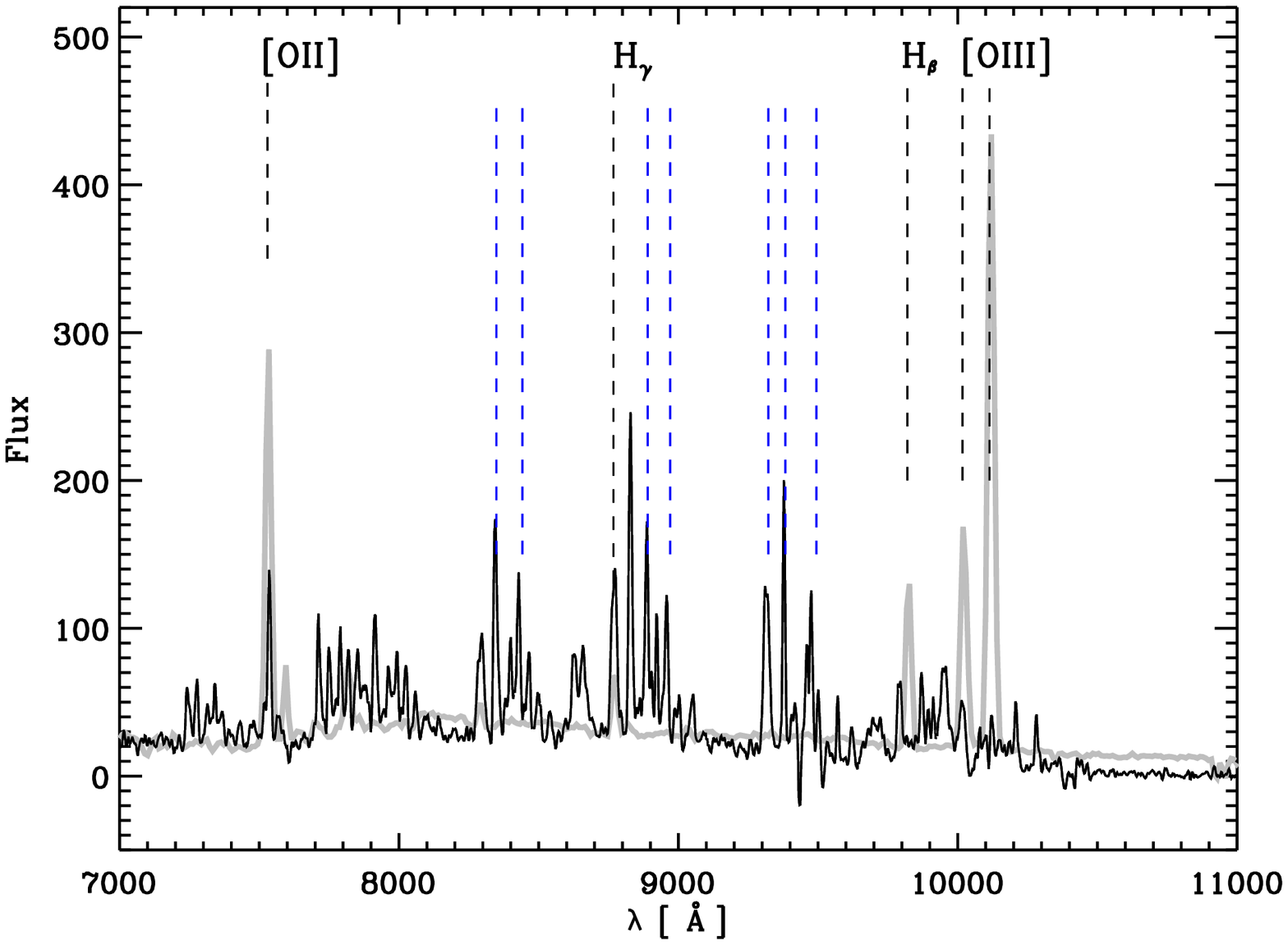}\\
\includegraphics[scale=0.37]{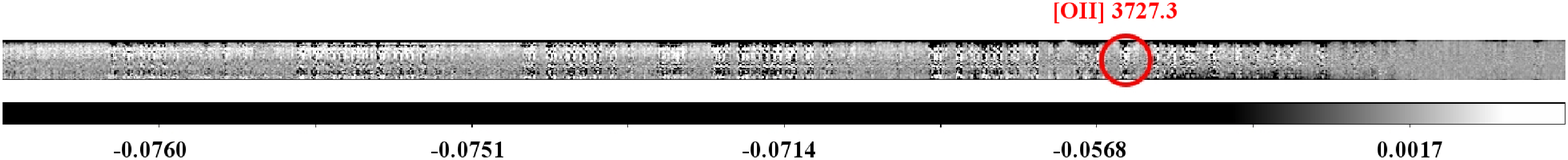}\\ 
\includegraphics[scale=0.37]{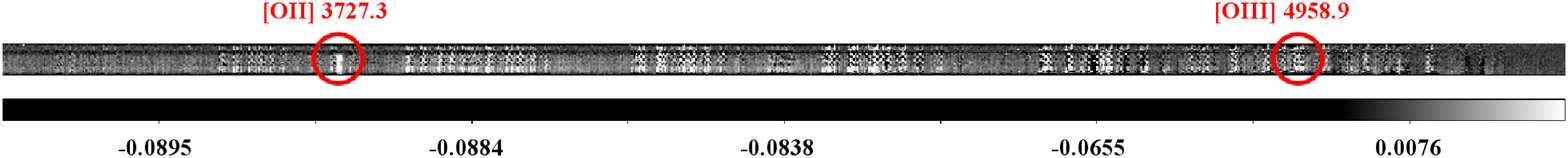} 
\caption{\textit{Top panel}.  As mentioned in section Sect.\,\ref{DATA3.3}, the slit length was limited by the position of the arcs, hence producing poor sky subtraction. \textit{Left}. Observed spectrum of arc A (black continuous line). In gray we depict a starburst template from \citet{Kinney1996} shifted at $z$ = 1.628. We marked a possible [OII]$\lambda$3727 emission line, and some sky lines in blue (see Sect.\,\ref{DATA3.3}). \textit{Right}.  Observed spectrum of arc C (black continuous line), as before we depicted in gray the starburst template from \citet{Kinney1996}, but shifted at $z$ = 1.02.  We marked some characteristic emission lines, along with the sky lines in blue, but we omit for clarity the labels of the last ones. \textit{Bottom panel.} Two-dimensional spectra of arc A (with a color bar in arbitrary units). Note the [OII]$\lambda$3727 emission line. Below, the two-dimensional spectra of arc C, with two emission lines clearly identified:  [OII]$\lambda$3727, and  [OIII]$\lambda$4958.9.
}
\label{spectraArcC}
\end{figure*}

 \subsection{Combining likelihoods}

In order to combine SL and dynamical constraints, we compute their respective
likelihoods. The SL likelihood is computed via {\sc LENSTOOL}\footnote{Publicly available at: http://www.oamp.fr/cosmology/ {\sc LENSTOOL}/} code. This software implements a Bayesian Monte Carlo Markov chain (MCMC) method to search for the most likely parameters in the modeling. It has been used in a large number of clusters studies, and characterized in 
\citet{jullo07}. The likelihood coming from dynamics of cluster members is computed using 
the {\sc MAMPOSS}t code \citep[][]{Mamon2013}, which has been tested and characterized on
simulations.
Technically, we have incorporated the {\sc MAMPOSS}t likelihood routine into {\sc LENSTOOL}.

Note that the SL likelihood (see  Eq.~\ref{eq:LikeLens}) depends on the image positions of the arcs and their respective errors. On the other hand, the {\sc MAMPOSS}t likelihood is calculated through the projected radii and line-of-sight velocity (Eq.~\ref{eq:MLE}). The errors on the inputs for the strong lensing on one hand and {\sc MAMPOSS}t on the other should not be correlated (in other words the joint lensing-dynamics covariance matrix should be diagonal). So, we can write:

\begin{equation}\label{eq:LikeTot}
\mathcal{L}_{T} = \mathcal{L}_{\textrm{Lens}}  \times  \mathcal{L}_{\textrm{Dyn}},
\end{equation}

\noindent where $\mathcal{L}_{\textrm{Dyn}}$ is given by Eq.~\ref{eq:MLE} and $\mathcal{L}_{\textrm{Lens}} $ is calculated through Eq.~\ref{eq:LikeLens}. This definition of a total likelihood, where the two techniques (lensing and dynamics) are considered independent, is not new, and has been used previously at different scale by other authors \citep[e.g.,][]{sand02,sand04}, here we are using the dynamics to obtain constraints in the outer regions\footnote{Note that here we are assuming the same weight of the SL and dynamics on the total likelihood. However this can not be the case, for example when combining SL and weak lensing \citep[see the discussion in][]{Umetsu2015}}. In this sense, the main difference with previous works, as for example \citet{Biviano2013}, \citet{Guennou2014} or \citetalias{Verdugo2011}, is that in this work we do a joint analysis, searching for a solution consistent with both methods, maximizing a total likelihood.

 \subsection{NFW mass profile}

We  adopt the NFW mass density profile that has been predicted in cosmological $N$-body simulations \citep{Navarro1996,nav97}, given by

\begin{equation}\label{eq:rho}
\rho(r) = \frac{\rho_{s}}{(r/r_s)(1+r/r_s)^{2}},
\end{equation}

\noindent where $r_s$ is the radius that corresponds to the region where the logarithmic slope of the density equals the isothermal value, and $\rho_s$ is a characteristic density. The scale radius is related to the virial radius $r_{200}$ through the expression $c_{200}$ =  $r_{200}/r_s$, which is the so-called concentration\footnote{$r_{200}$ is the radius of a spherical volume inside of which the mean density is $200$ times the critical density at the
given redshift $z$, $M_{200} = 200 \times (4\pi/3)r_{200}^{3} \rho_{crit}$ = $100H^2r^3_{200}/G$.}. The mass contained within a radius $r$ of  the NFW halo is given by

\begin{equation}\label{eq:mass}
 M(r)  =   4\pi r_s^{3}\rho_s \left[\ln{(1+r/r_s}) - \frac{r/r_s}{1+r/r_s}\right].
\end{equation}

\noindent Although other mass models (e.g., Hernquist or Burkert density profiles) have been studied within the {\sc MAMPOSS}t formalism \citep[see][]{Mamon2013}, and {\sc LENSTOOL} allows to probe different profiles, we adopt the NFW profile in order to compare our results with those obtained in \citetalias{Verdugo2011}.  Note that the  NFW profile is a spherical density profile, and {\sc MAMPOSS}t's formalism can only model spherical systems. However, with {\sc LENSTOOL}  the initial profile is spherical, but is transformed into a pseudo-elliptical NFW  \citep[see][]{Golse2002}, as is explained in \citet{jullo07}, in order to perform the lensing calculations. Although the simultaneous modeling share the same spherical parameters, the difference between the pseudo-elliptical and the spherical framework could influence our methodology (see Sect.\,\ref{MassM4.3}).

\begin{figure*}\begin{center}
\includegraphics[scale=0.45]{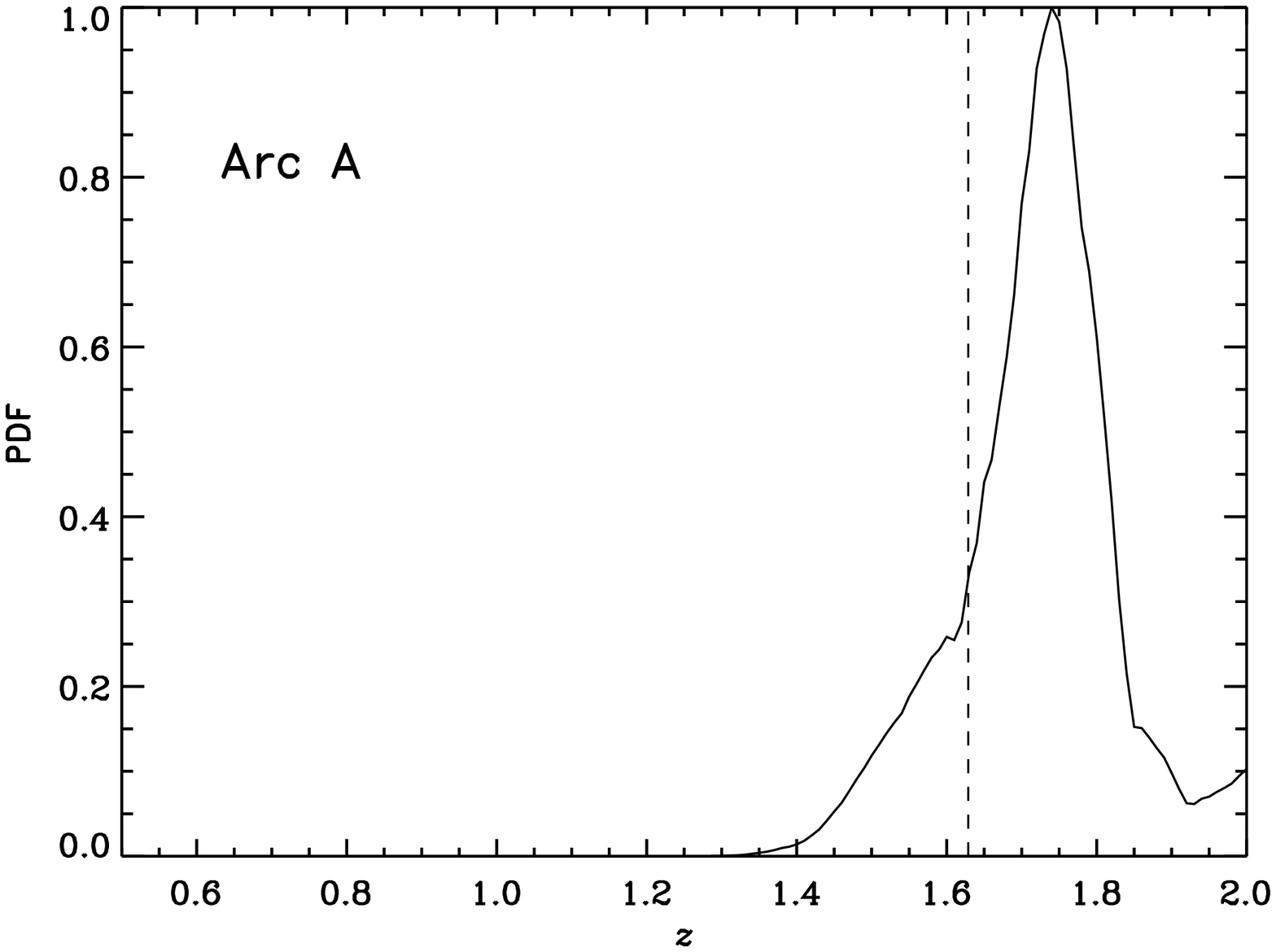} 
\includegraphics[scale=0.45]{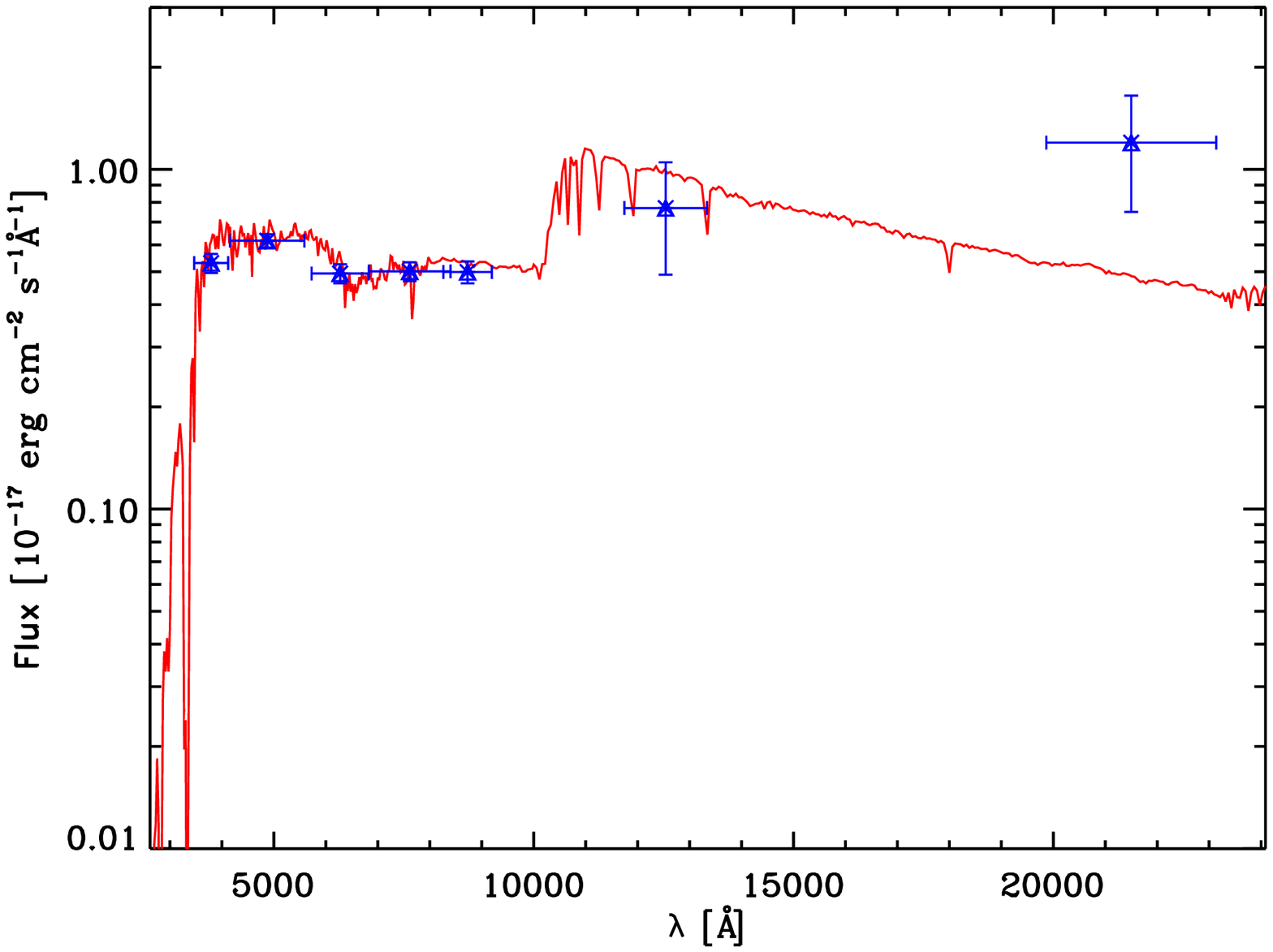}\\ 
\includegraphics[scale=0.45]{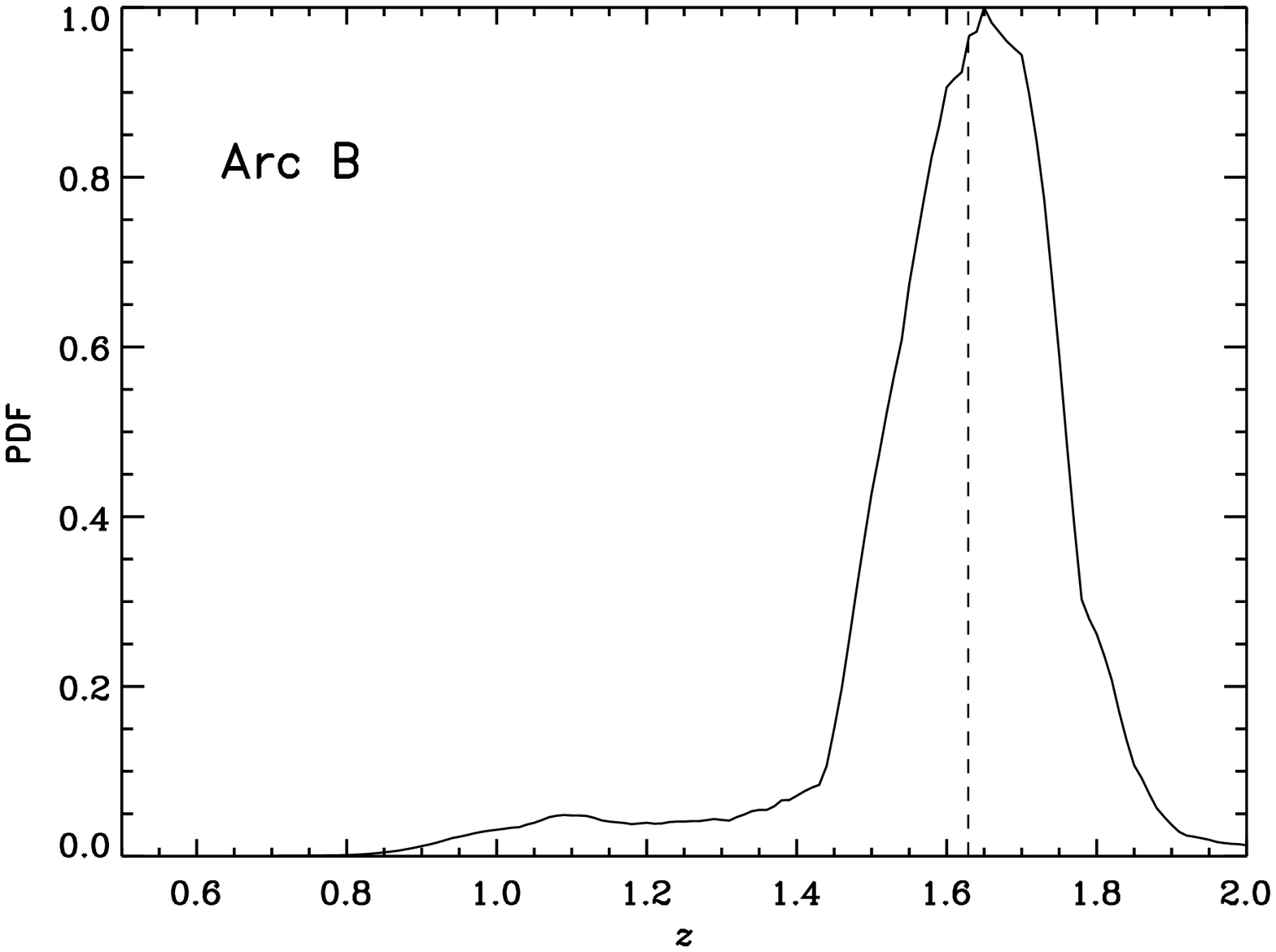} 
\includegraphics[scale=0.45]{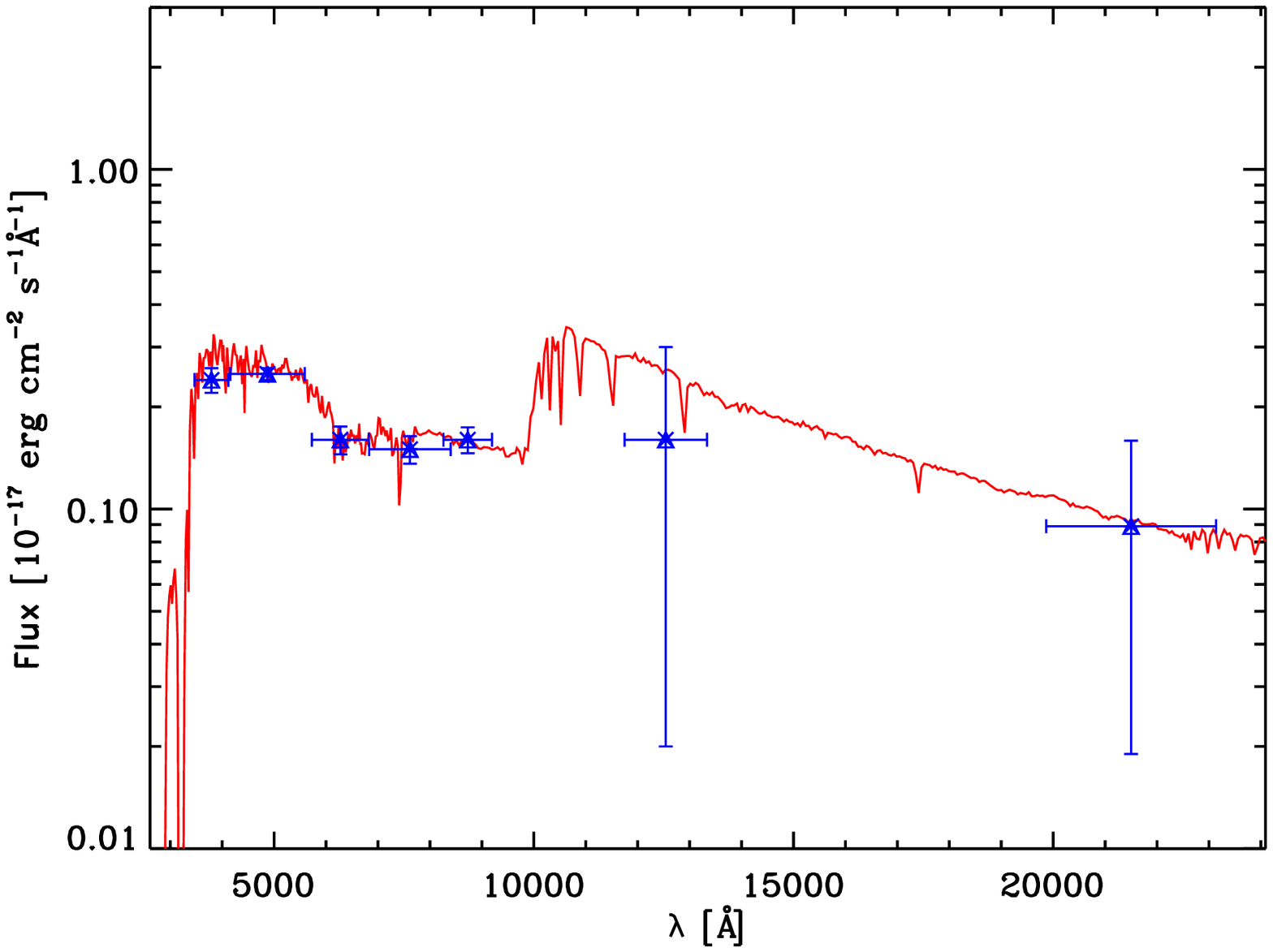}\\ 
\includegraphics[scale=0.45]{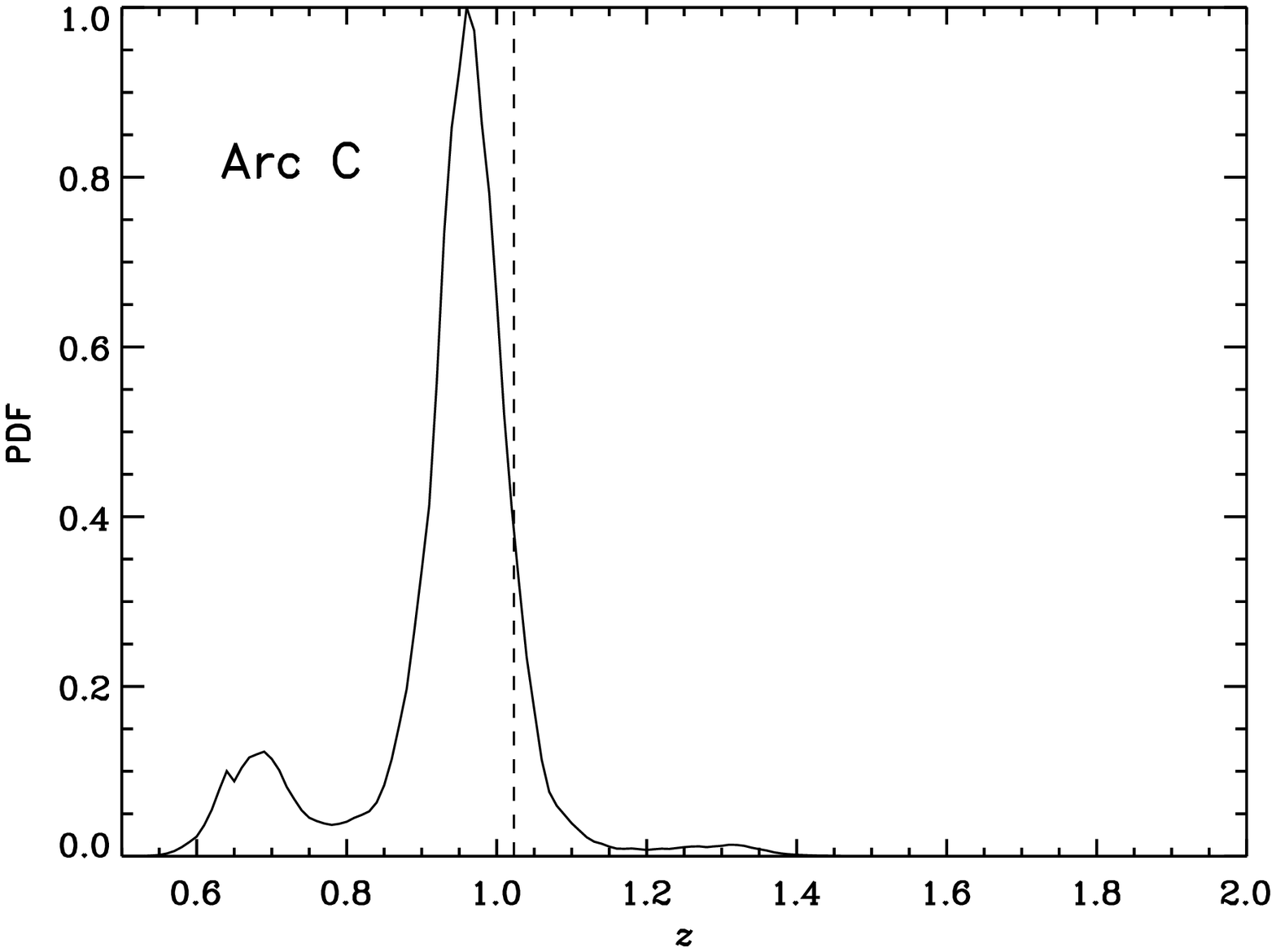} 
\includegraphics[scale=0.45]{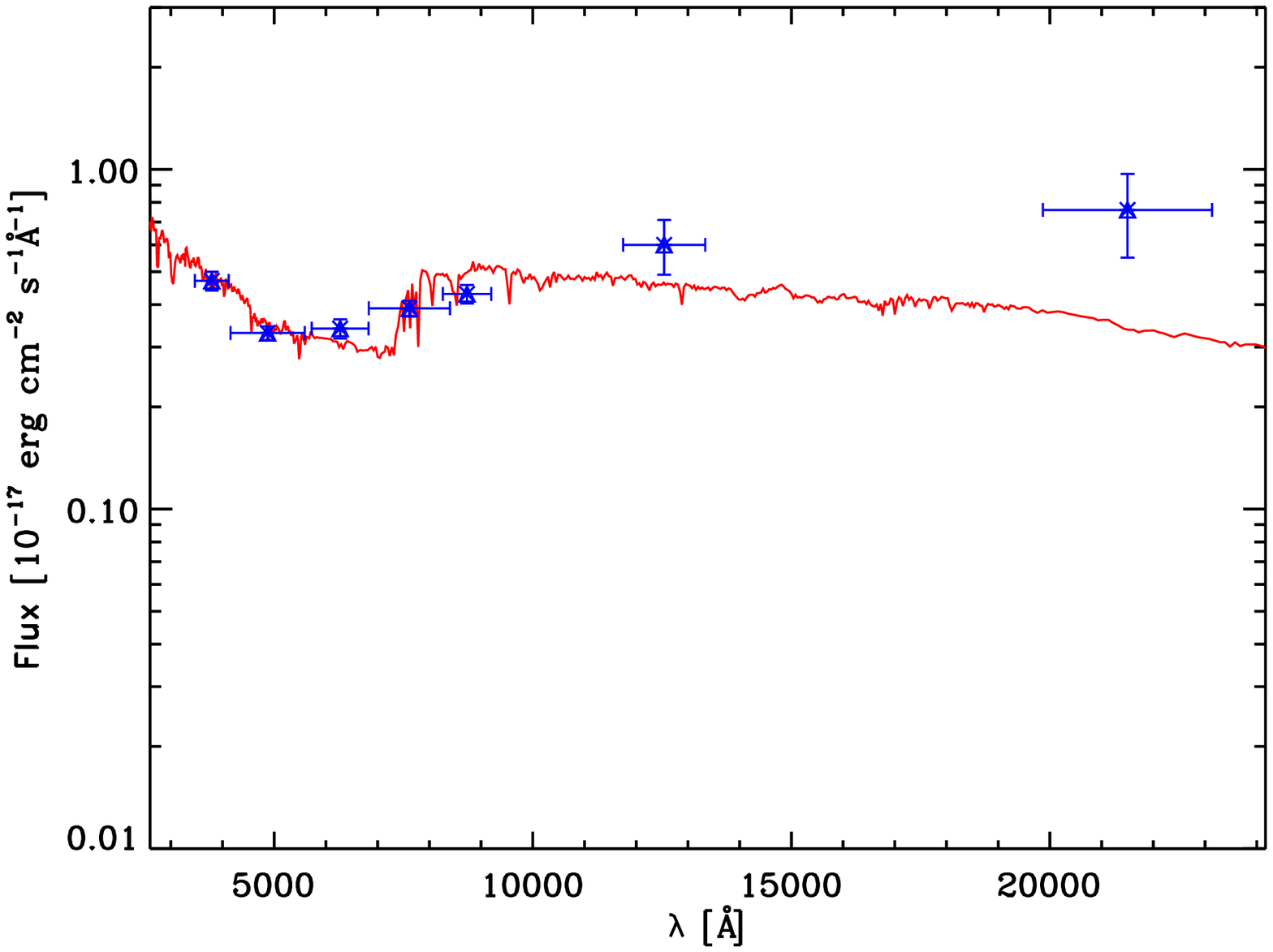}\\ 
\caption{\textit{Left column.} Photometric redshift PDF for the selected arcs (see text). The dashed vertical lines corresponds to the the spectroscopic value. \textit{Right column.} Best fit spectral energy distribution. Points represent the observed CFHTLS broad band magnitudes, and $J$ and $K_s$ from  WIRCam. Vertical and horizontal error bars correspond to photometric error  and wavelength range of each filter, respectively.} 
\label{SED_PDF}
\end{center}\end{figure*}

 \section{Data}\label{DATA}

In this section, we present the group \object{SL2S\,J02140-0535}, reviewing briefly our old data sets. Also, we present new data that we have obtained since  \citetalias{Verdugo2011}. From space, the lens was followed-up with \xmm\, Newton Space Telescope.  Additionally, a new spectroscopic follow-up of the arcs and group members have been carried out with the Very Large Telescope (VLT).
 
 \subsection{\object{SL2S\,J02140-0535} }
 
 This group, located at $z_{\rm spec}=0.44$, is populated by three central galaxies. We label them as G1 (the brightest group galaxy, BGG), G2, and G3 (see left panel of Fig.~\ref{presentlens}). The lensed images consist of three arcs surrounding these three galaxies:  arc  $A$, situated north of the deflector, composed by two merging images; a second arc in the east direction (arc $B$), which is associated to arc $A$, whereas a third arc, arc $C$, situated in the south, is singly imaged. \object{SL2S\,J02140-0535} \citep[first reported by][]{Cabanac2007} has been studied previously using strong lensing  by \citet{alardalone} and both strong and weak lensing by \citet{paperI}, and also kinematically by \citet{Roberto2013}.

\begin{figure}[h!]
\begin{center}
\includegraphics[scale=0.55]{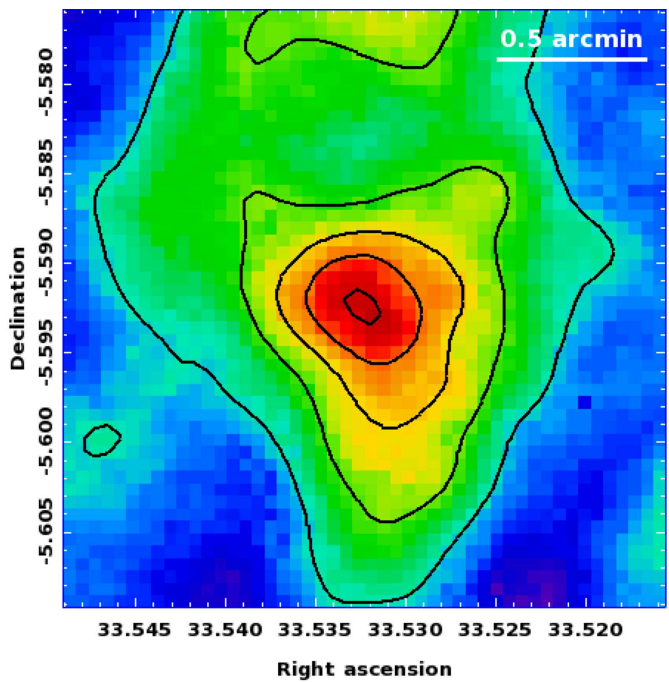}
\caption{The adaptively smoothed image of \object{SL2S\,J02140-0535} in the 0.5-2.0 keV band. The X-ray contours in black are linearly spaced from 5 to 20 cts/s/deg$^2$.}
\label{B1}
\end{center}\end{figure}

\object{SL2S\,J02140-0535} was observed in five filters ($u^*$, $g'$, $r'$, $i'$, $z'$) as part of the CFHTLS (Canada-France-Hawaii Telescope Legacy Survey)\footnote{http://www.cfht.hawaii.edu/Science/CFHLS/}  using the wide field imager \textsc{MegaPrime} \citep{Gwyn2011}, and in the infrared
using WIRCam (Wide-field InfraRed Camera, the near infrared mosaic imager at CFHT) as part of the proposal 07BF15 (P.I. G. Soucail), see  \citet{Verdugo2014} for more information. In the right  panel of Fig.~\ref{presentlens} we show a false-color image of \object{SL2S\,J02140-0535}, combining the two bands $J$ and  $K_s$. Note that arcs A and C appear mixed with the diffuse light of the central galaxies, and arc B is barely visible in the image. \object{SL2S\,J02140-0535} was also followed up spectroscopically using FORS\,2 at VLT \citepalias{Verdugo2011}.

From space, the lens was observed with the \emph{Hubble Space Telescope} (HST) in snapshot mode (C\,15, P.I. Kneib) using three bands with the ACS camera (F814, F606, and F475).

\subsection{New spectroscopic data}\label{NSD}

\textit{Selecting members.-} We used FORS\,2 on VLT with a medium resolution grism (GRIS\,600RI; 080.A-0610; PI V. Motta) to target the group members (see Mu\~noz et al. 2013) and a low resolution grism (GIRS\,300I; 086.A-0412; P.I. V. Motta) to observe the strongly lensed features. In the later observation, we use one mask with $2\times1300$~s on-target exposure time. 
Targets (other than strongly lensed features) were selected by a two-step process. First, we use the T0005 release of the CFHTLS survey (November, 2008) to obtain a photometric redshift--selected catalog which include galaxies within
$\pm0.01$ of the redshift of the main lens galaxy. The selected galaxies in this catalog have colors within $(g-i)_{lens}-0.15<g-i<(g-i)_{lens}+0.15$, where $(g-i)_{lens}$ is the color of the brightest galaxy within the Einstein radius. From this sample, we selected those candidates that were not observed previously. More details  will be presented in a forthcoming publication (Motta et~al., in prep.).

The spectroscopic redshifts of the galaxies were determined using the Radial Velocity SAO package \citep{Kurtz1998} within the IRAF software\footnote{IRAF is distributed by the National Optical Astronomy Observatory, which is operated by the Association of Universities for Research in Astronomy (AURA) under cooperative agreement with the National Science Foundation.}. By  visual inspection of the spectra, we identify several emission and absorption lines. Then, we determine the redshifts \citep[typical errors are discussed in][]{Roberto2013} by doing a cross-correlation between a spectrum and template spectra of known velocities. To determine the group membership of  \object{SL2S\,J02140-0535}, we follow the method presented in \citet{Roberto2013}, which in turn adopt the formalism of \citet{Wil05}. The group members are identified as follows: we assume initially that the group is located at  the redshift of the main bright lens galaxy, $z_{\textrm{lens}}$, with an initial observed-frame velocity dispersion of $\sigma_{\textrm{obs}}$ = 500(1+$z_{\textrm{lens}}$) km\,s$^{-1}$. After computing the required redshift range for group membership  \citep[see][]{Roberto2013}, and applying a biweight estimator \citep{Bee90}, the iterative process reached a stable membership solution with 24 secure members and a velocity dispersion of $\sigma$ = 562 $\pm$ 60 km\,s$^{-1}$. These galaxies are shown with red squares and circles in  Fig.~\ref{fig1}, and their respective redshifts are presented in  Table~\ref{tbl-A1}. The squares in Fig.~\ref{fig1} represent the galaxies previously reported by \citet{Roberto2013}.  Fig.~\ref{fig1} also shows the luminosity contours calculated according to  \citet[][]{Gael2013}. Fitting ellipses to the luminosity map, using the task \textit{ellipse} in IRAF, we find that the luminosity contours have position angles  equal to 99$^{\circ}$ $\pm$ 9$^{\circ}$, 102$^{\circ}$ $\pm$ 2$^{\circ}$, and 109$^{\circ}$ $\pm$ 2$^{\circ}$, from outermost to innermost contour respectively.

\begin{figure}[h!]\begin{center}
\includegraphics[scale=0.525]{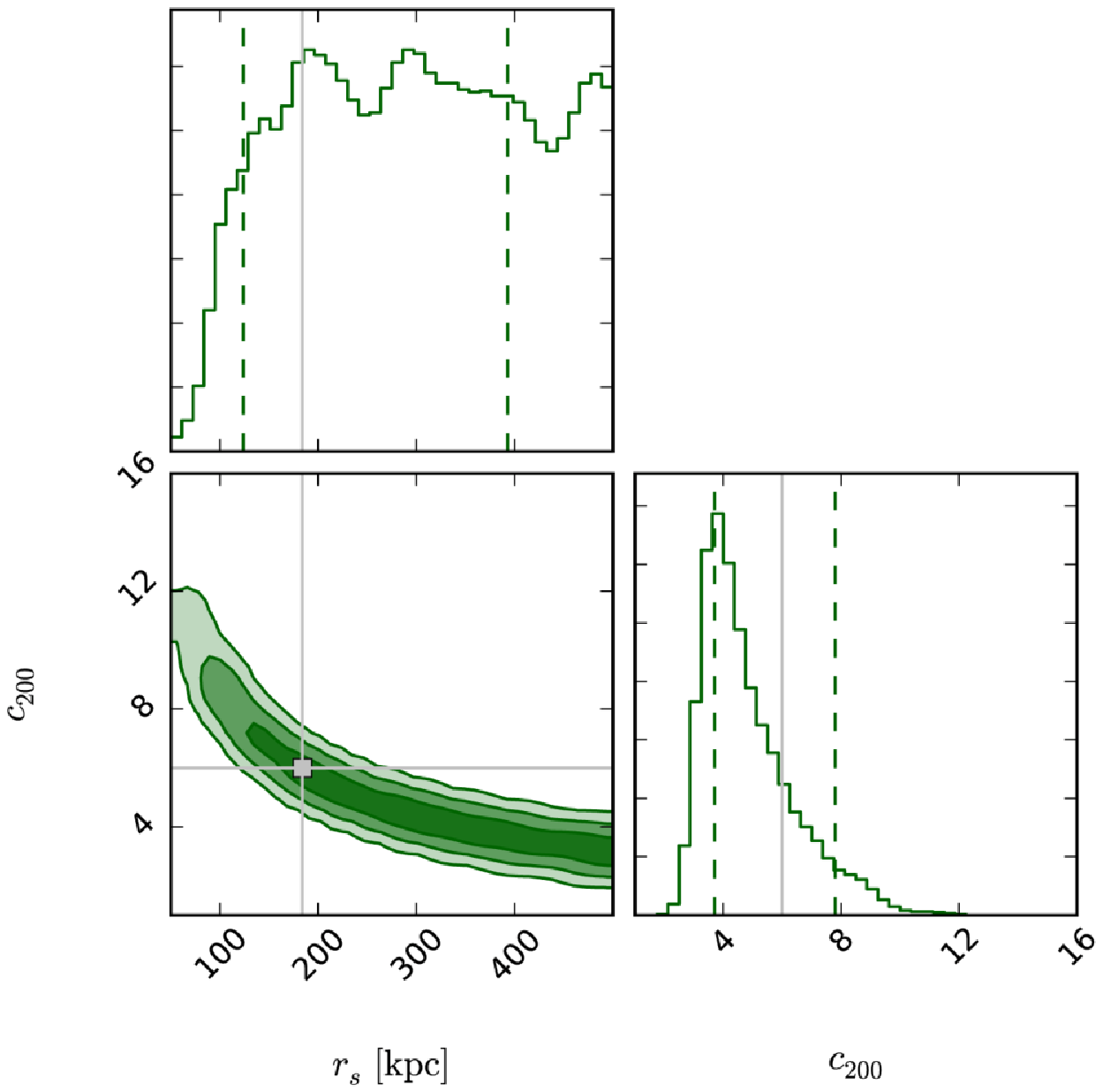}\\
 \vspace{0.1cm}
\includegraphics[scale=0.525]{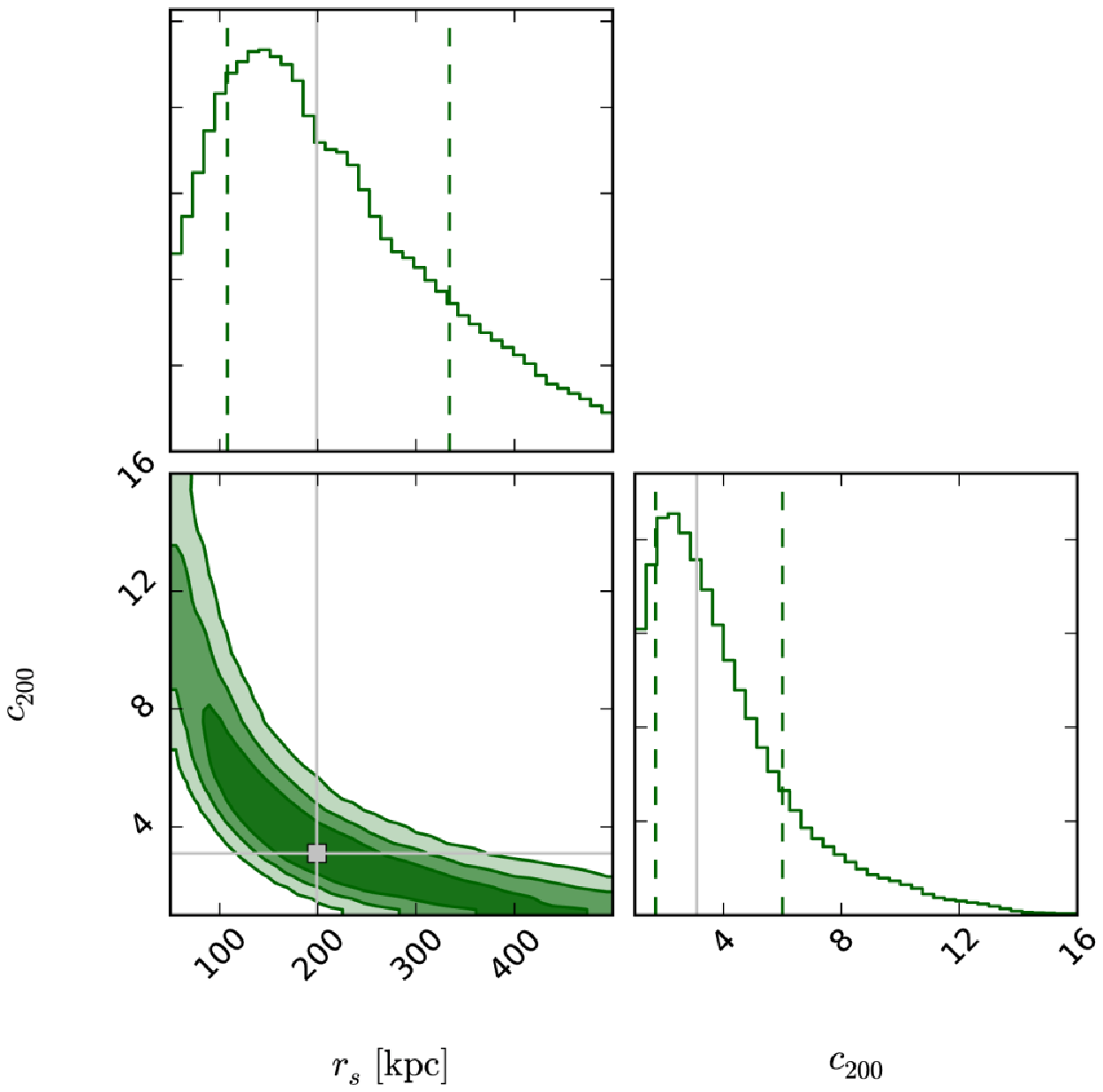}\\
\vspace{0.1cm}
\includegraphics[scale=0.525]{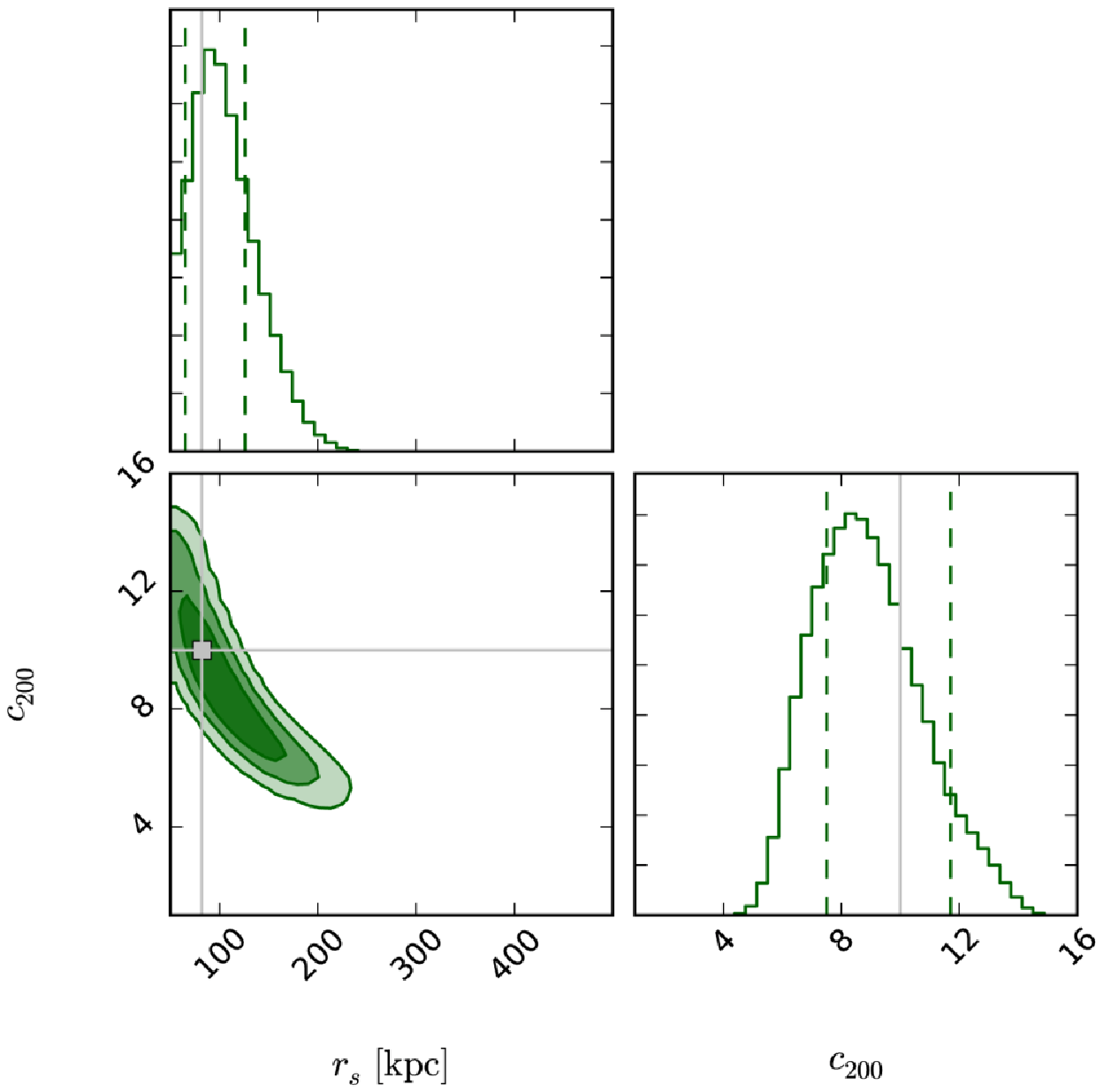}
\caption{PDFs and contours of the parameters $c_{200}$ and $r_{s}$. The three contours stand for the 68\%, 95\%, and 99\% confidence levels. The values obtained for our best-fit model are marked by a gray square, and with vertical lines in the 1D histograms (the asymmetric errors are presented in Table~\ref{tbl-1}).\textit{ Top panel.-} Results from the \textrm{SL\,Model}. \textit{Middle panel.-} Results from the \textrm{Dyn\,Model}. \textit{ Bottom panel.-} Results from the \textrm{SL+Dyn\,Model}.}
\label{PDFModels1} \end{center} 
\end{figure}

\textit{Completeness.-} Mu\~noz et al. (2013) presented the dynamical analysis of seven SL2S galaxy groups, including \object{SL2S\,J02140-0535}. They estimate the completeness within 1 Mpc of radius from the centre of the group to be  30\%. In the present work, as we increased the number of observed galaxies in the field of \object{SL2S\,J02140-0535}, and thus increasing the number of confirmed members (hereafter $gal_{\textrm{spec}}$), a new calculation is carried out to estimate the completeness as a function of the radius. We first define the color-magnitude cuts to be applied to the photometric catalog of the group, i.e. $0.7<(r-i)<0.92$ and $21.44<m_{i}<21.47$. These values correspond to the photometric properties of the $gal_{\textrm{spec}}$. Note that we exclude one galaxy because of its color $(r-i)=0.45$. Then, we select all the galaxies falling within the photometric ranges (hereafter $gal_{\textrm{phot}}$), and we estimate  the density of field galaxies within the $15'\times15'$ square arcminutes after excluding a central region of radius 1.3 Mpc (which is the largest distance from the center of the group of $gal_{\textrm{spec}}$). This density is then converted into an estimated total number of galaxies $N_{\textrm{field}}$ over the full field of view.

Given $gal_{\textrm{spec}}$, we bin the data and define $N_{\textrm{spec}}(r_{i})$ as the number of confirmed members in the $i$th radial bin. Thus, the radial profile of the completeness is given by

\begin{equation}\label{eq:CompG}
 C(r_{i}) \equiv \frac{N_{\textrm{spec}}(r_{i})} {N_{t}(r_{i})-N_{\textrm{field},r}(r_{i})},
\end{equation}

\noindent   where $N_{\textrm{field},r}(r_{i})$ is the number of field galaxies in the  $i$th bin, and  $N_{t}(r_{i})$ is the total number of $gal_{\textrm{phot}}$ present in the $i$th bin, i.e., its value is the sum of the number of group members and field galaxies.  To estimate $N_{\textrm{field},r}(r_{i})$, a Monte Carlo approach is adopted: we randomly draw the positions of the $N_{\textrm{field}}$ galaxies over the whole field of view, and then count the corresponding number of galaxies $N_{\textrm{field},r}(r_{i})$ falling in each bin. Thus, each Monte Carlo realization leads to an estimate of the completeness. Finally, we average the $C(r_{i})$, after excluding the realizations for which we obtain $N_{t}(r_{i})<N_{r}(r_{i})$ or $C(r_{i})>1$. In Fig.~\ref{completeness} we present the resulting profile and its estimated $1\sigma$ deviation, showing a completeness $C\sim60\%$ consistent with a constant profile up to 1 Mpc.

\subsection{Multiple images: confirming two different sources}\label{DATA3.3}

\textit{Spectroscopic redshifts.-} In \citetalias{Verdugo2011} we reported a strong emission line at 7538.4\,\AA\,  in the spectra of arc C, that we associated to [OII]$\lambda$3727 at   $z_{\rm spec}$ = 1.023 $\pm$ 0.001. We obtained new 2D spectra for the arcs consisting in two exposures of 1300s each. Due to the closeness of the components (inside a radius of $\sim$8\arcsec\ ), the slit length is limited by the relative position of the arcs, making sky-subtraction difficult (see bottom panel of Fig.~\ref{spectraArcC}). Most of the 2D spectra show a poor sky subtraction compared to which we would have obtained using longer slits. However, our new 2D spectra also shows the presence of [OII]$\lambda$3727 spectral feature, and additionally another emission line appears in the spectrum, [OIII]$\lambda$4958.9. In the same Fig.~\ref{spectraArcC} (top-right panel) we show the spectrum and marked some characteristic emission lines, along with a few sky lines. We compare it with a template of a starburst galaxy from  \citet{Kinney1996},  shifted at $z$ = 1.02. After performing a template fitting using RVSAO  we obtain $z_{\rm spec}$ = 1.017 $\pm$ 0.001, confirming our previously reported value.

On the other hand, in our previous work, we did not found any spectroscopic features in the arcs A and B due to the poor signal-to-noise.  In the top-left panel of  Fig.~\ref{spectraArcC} we show the new obtained spectrum of arc A. It reveals a weak  (but still visible) emission line at 9795.3\,\AA. This line probably corresponds to [OII]$\lambda$3727 at   $z\sim1.6$. However, we do not claim a clear detection (this region of the spectrum is affected by sky emission lines),  as we discuss below, but is worth to note that the photometric redshift estimate supports this detection (see Fig.~\ref{SED_PDF}). Assuming emission from [OII]$\lambda$3727 and  applying a Gaussian fitting, we obtain  $z_{\rm spec}$ = 1.628 $\pm$ 0.001. This feature is not present in the spectrum of arc B, since arc B is almost one magnitude fainter than arc A. Furthermore,  this line is not present in the spectrum of arc C, which confirms the previous finding of \citetalias{Verdugo2011}, i.e. system AB and arc C do come from two different sources.

\textit{Photometric redshifts.-} As a complementary test, and to extend the analysis presented in \citetalias{Verdugo2011}, we calculate the photometric redshifts of arcs A, B, and C  using the HyperZ  software \citep{hyperz}, adding the $J$ and $K_s$ bands to the original ones ($u^*$, $g'$, $r'$, $i'$, $z'$). The photometry in $J$ and $K_s$ bands was performed with the IRAF package \textit{apphot}. We employed polygonal apertures to obtain a more accurate flux measurement of the  arcs. For each arc, the vertices of the polygons were determined using the IRAF task \textit{polymark}, and the magnitudes inside these apertures were computed by the IRAF task \textit{polyphot}. The results are presented in Table~\ref{tbl-A2}.

\begin{figure}[h!]\begin{center}
\includegraphics[scale=0.525]{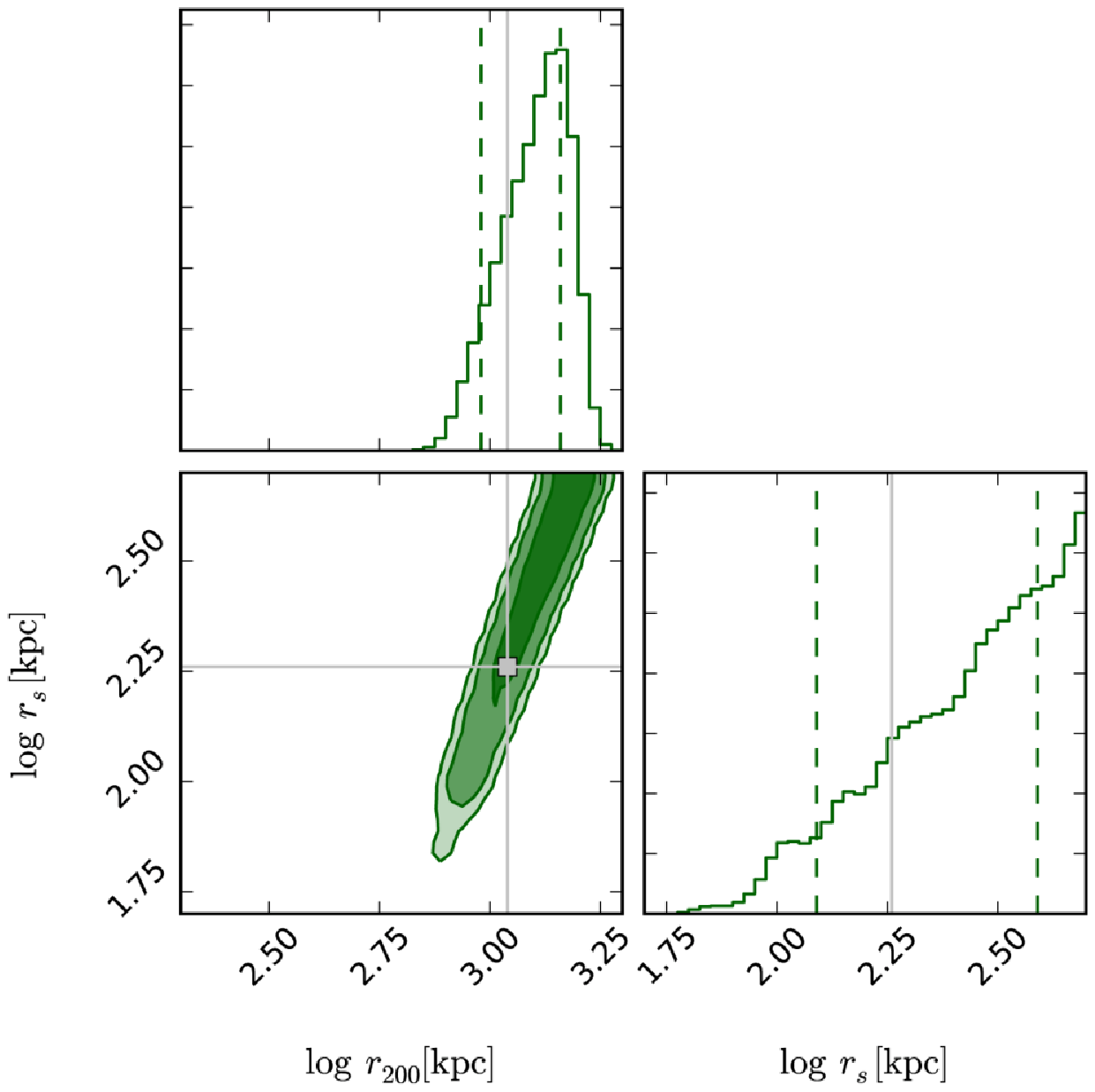}\\
 \vspace{0.1cm}
\includegraphics[scale=0.525]{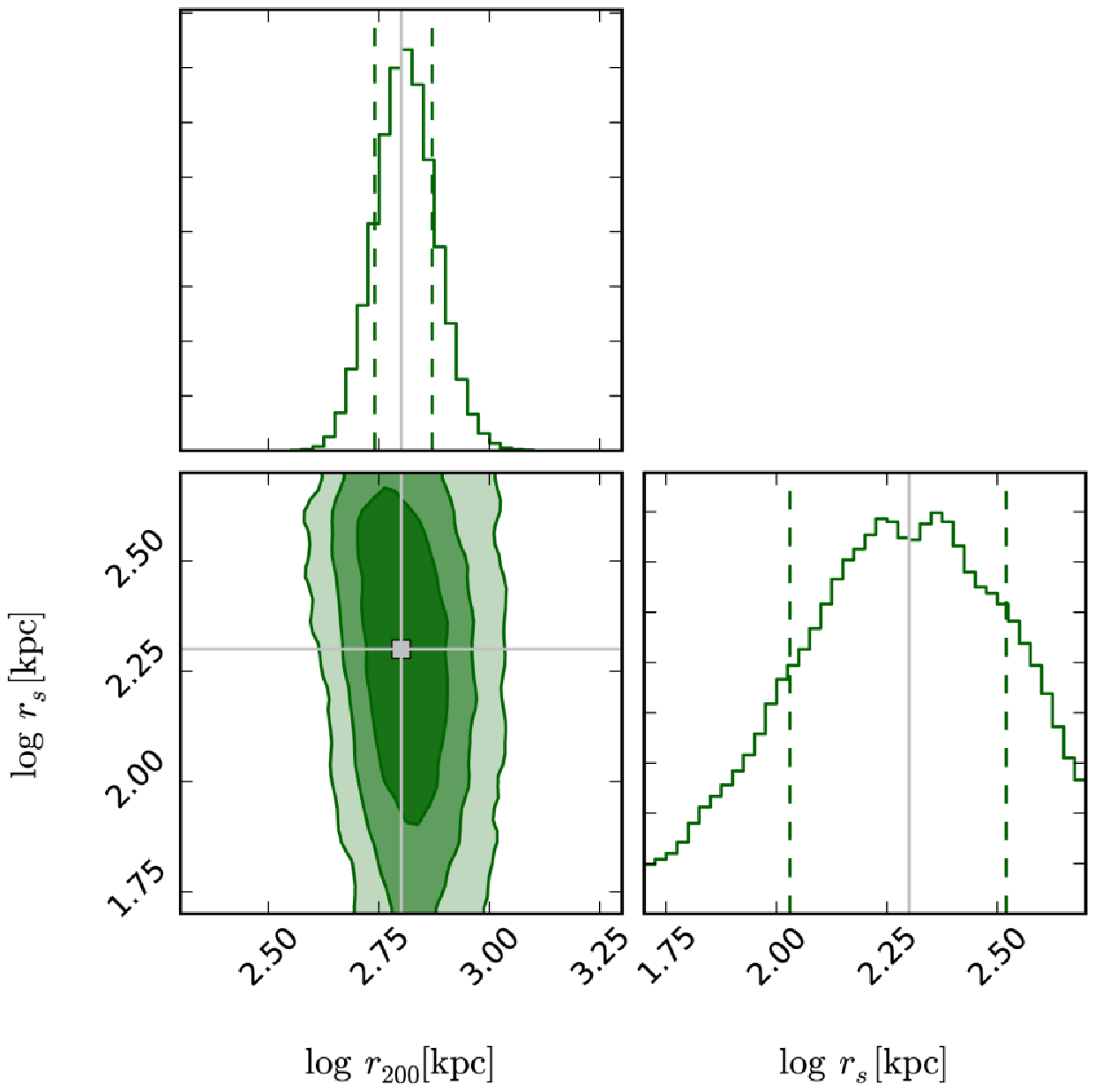}\\
\vspace{0.1cm}
\includegraphics[scale=0.525]{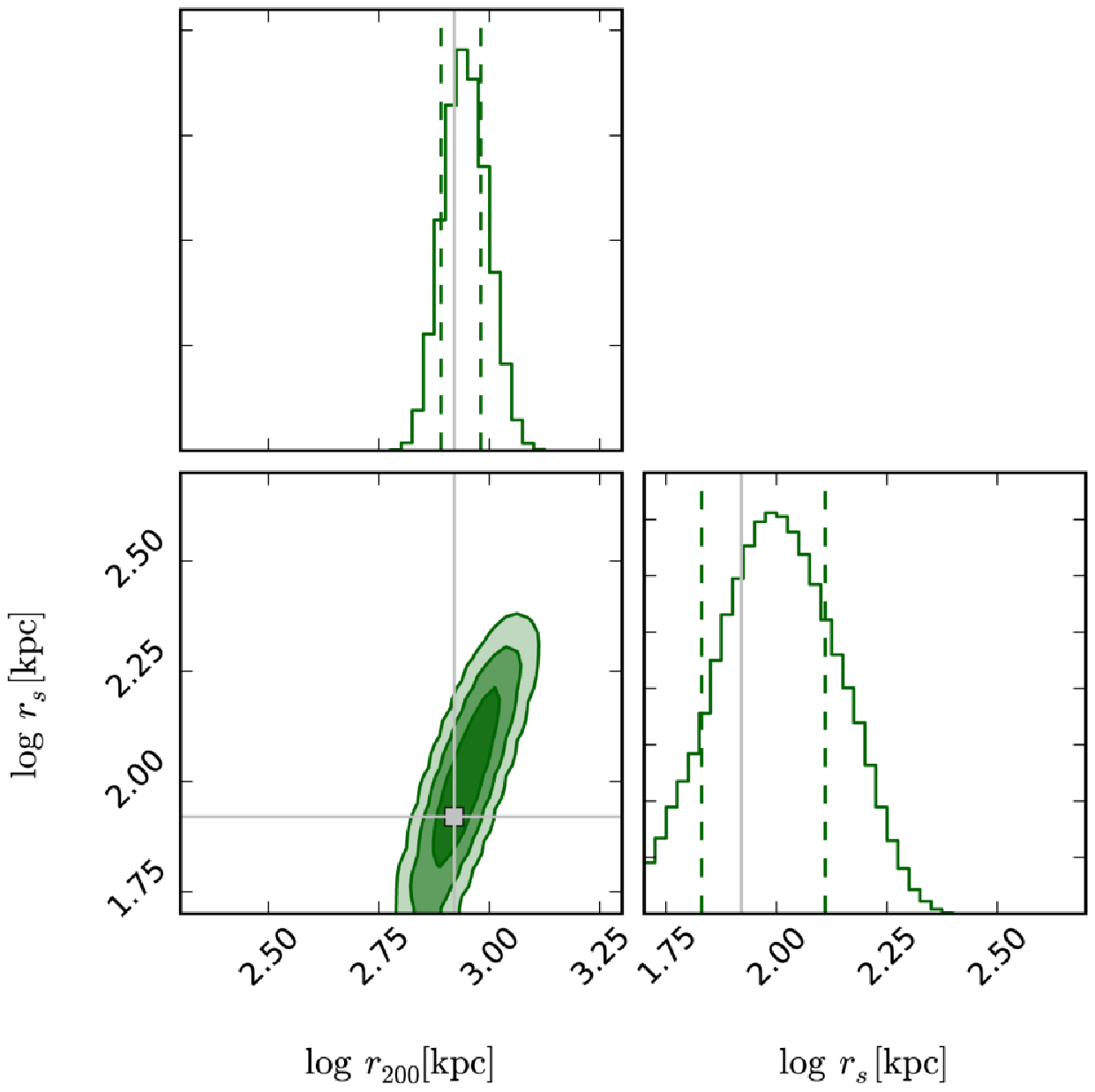}
\caption{PDFs and contours of the parameters log$r_s$ and log$r_{200}$. The three contours stand for the 68\%, 95\%, and 99\% confidence levels. The values obtained for our best-fit model are marked by a gray square, and with vertical lines in the 1D histograms (the asymmetric errors are presented in Table~\ref{tbl-1}).\textit{ Top panel.-} Results from  the \textrm{SL\,Model}. \textit{Middle panel.-} Results from the \textrm{Dyn\,Model}. \textit{ Bottom panel.-} Results from the \textrm{SL+Dyn\,Model}.}
\label{PDFModels2} \end{center} 
\end{figure}

It is evident in the right  panel of Fig.~\ref{presentlens} that the gravitational arcs are contaminated by the light of the central galaxies. In order to quantify the error in our photometric measurements in both $J$ and $K_s$ bands we proceed as follows: we subtract  the central galaxies of the group, and follow the procedure described in McLeod et al. (1998),that is, we fit a galaxy profile model convolved with a PSF (de Vaucouleurs profiles were fitted to the galaxies with synthetic PSFs). After the subtraction, we run again the IRAF task \textit{polyphot}.  The errors associated to the fluxes are defined as the quadratic sum of the errors on both measurements. 

The photometric redshifts for the arcs  were estimated from the magnitudes reported in Table~\ref{tbl-A2}, as well as those reported in \citetalias{Verdugo2011}. We present the output probability distribution function (PDF) from HyperZ in Fig.~\ref{SED_PDF}. We note in the same figure that the $K_s$ band data do not match with the best-fit spectral energy distribution, this is probably related to the fact that the photometry of the arcs is contaminated by the light of the central galaxies. Arc C is constrained to be at $z_{\rm phot} = 0.96 \pm 0.07$, which is in  good agreement with the $z_{\rm spec}$ = 1.017 $\pm$ 0.001 reported above. The multiple imaged system constituted by arcs A and B have $z_{\rm phot}$= 1.7 $\pm$ 0.1 and $z_{\rm phot}$= 1.6 $\pm$ 0.2,  respectively. The photometric redshift of arc A is in agreement with the identification of the emission line as [OII]$\lambda$3727 at $z_{\rm spec}$ = 1.628 $\pm$ 0.001. 

To summarize, both the spectroscopic and photometric data confirm the results of \citetalias{Verdugo2011}, namely, the system formed by  arcs A and B,  and the single arc C,  originate from two different sources, the former at  $z_{\rm spec}$ = 1.628, and the latter at $z_{\rm spec}$ = 1.017.

\subsection{X-ray data}

We observed \object{SL2S\,J02140-0535} with \xmm\ as part of an X-ray follow-up program 
of the SL2S groups to obtain an X-ray detection of these strong-lensing 
selected systems and to measure the X-ray luminosity and temperature (Gastaldello et~al., in prep.). 
SL2S J02140-0535 was observed by \xmm\ for 19 ks with the MOS detector and for
13 ks with the pn detector.
The data were reduced with SAS v14.0.0, using the tasks
{\em emchain} and {\em epchain}. We considered only event patterns
0-12 for MOS and 0-4 for pn, and the data were cleaned using the standard
procedures for bright pixels, hot columns removal, and pn out-of-time 
correction.
Periods of high backgrounds due to soft protons were filtered out 
leaving an effective exposure time of 11 ks for MOS and 8 ks for pn.

For each detector, we create images with point sources in the 0.5-2 keV band. The point sources were 
detected with the task {\em edetect\_chain}, and masked using circular
regions of 25\arcsec\ radius centered at the source position. 
The images were exposure-corrected and background-subtracted using the 
XMM-Extended Source Analysis Software (ESAS).
The \xmm\ image in the 0.5-2 keV band of the field of \object{SL2S\,J02140-0535} is 
shown in Fig.~\ref{B1}.

The X-ray peak is spatially coincident with the bright galaxies inside the arcs, and the X-ray isophotes are elongated in the same direction as the optical contours (see discussion in Sect.\,\ref{Discus}). The quality of the X-ray snapshot data is not sufficient for a detailed mass analysis assuming hydrostatic equilibrium \citep[e.g.,][]{Gastaldello2007}. In this case, the mass can only be obtained adopting a scaling relation, such as a mass-temperature relation  \citep[e.g.,][]{Gastaldello2014}. Therefore this mass determination is not of the same quality as the obtained with our lensing and dynamical information.  And, as we will discuss in Sect.\,\ref{Discus}, we need to be very cautious when assuming scaling relations for strong lensing clusters. We will only make use of the morphological information provided by the X-ray data hereinafter.

\begin{figure}[h!]\begin{center}
\includegraphics[scale=0.525]{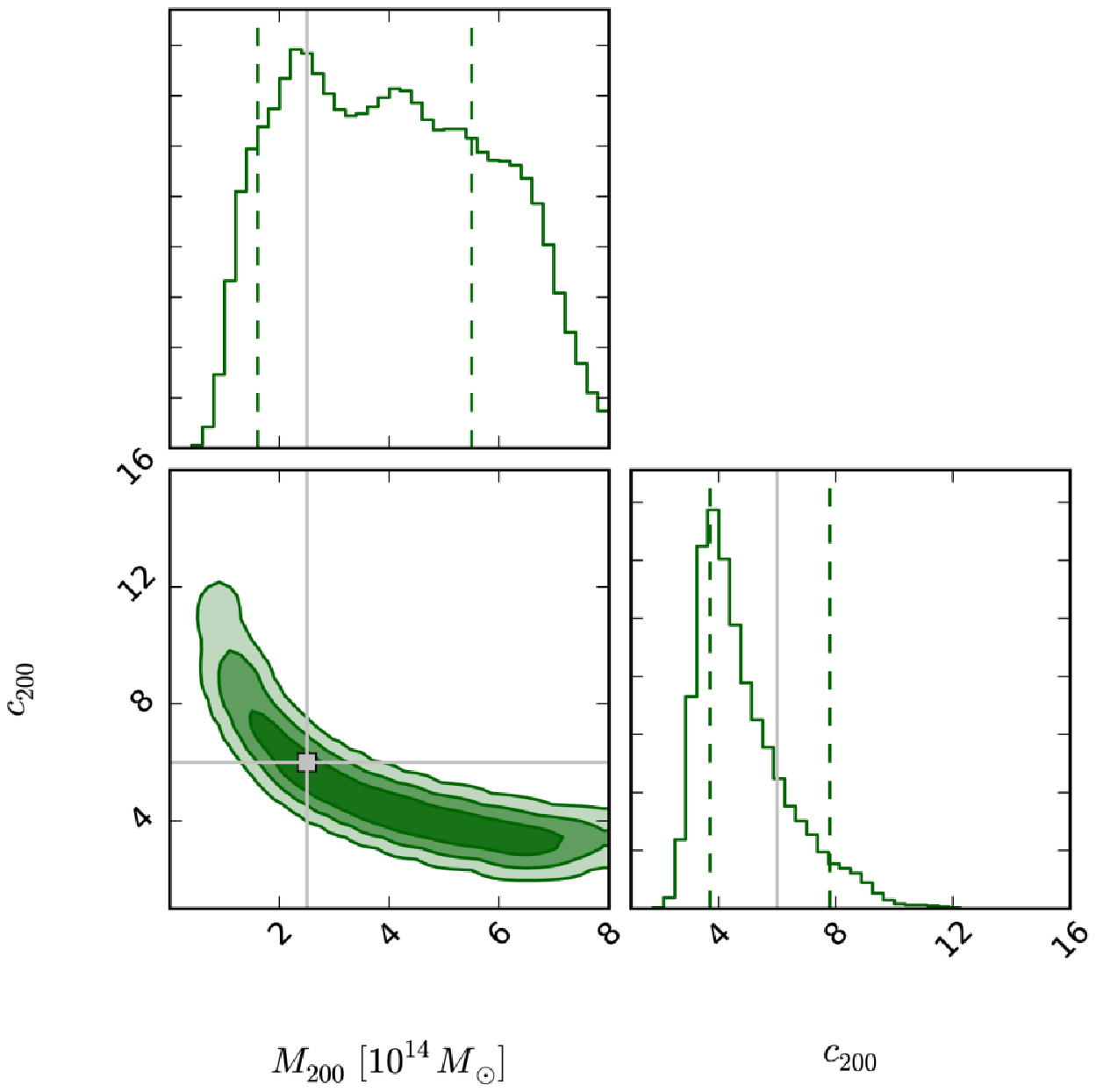}\\
 \vspace{0.1cm}
\includegraphics[scale=0.525]{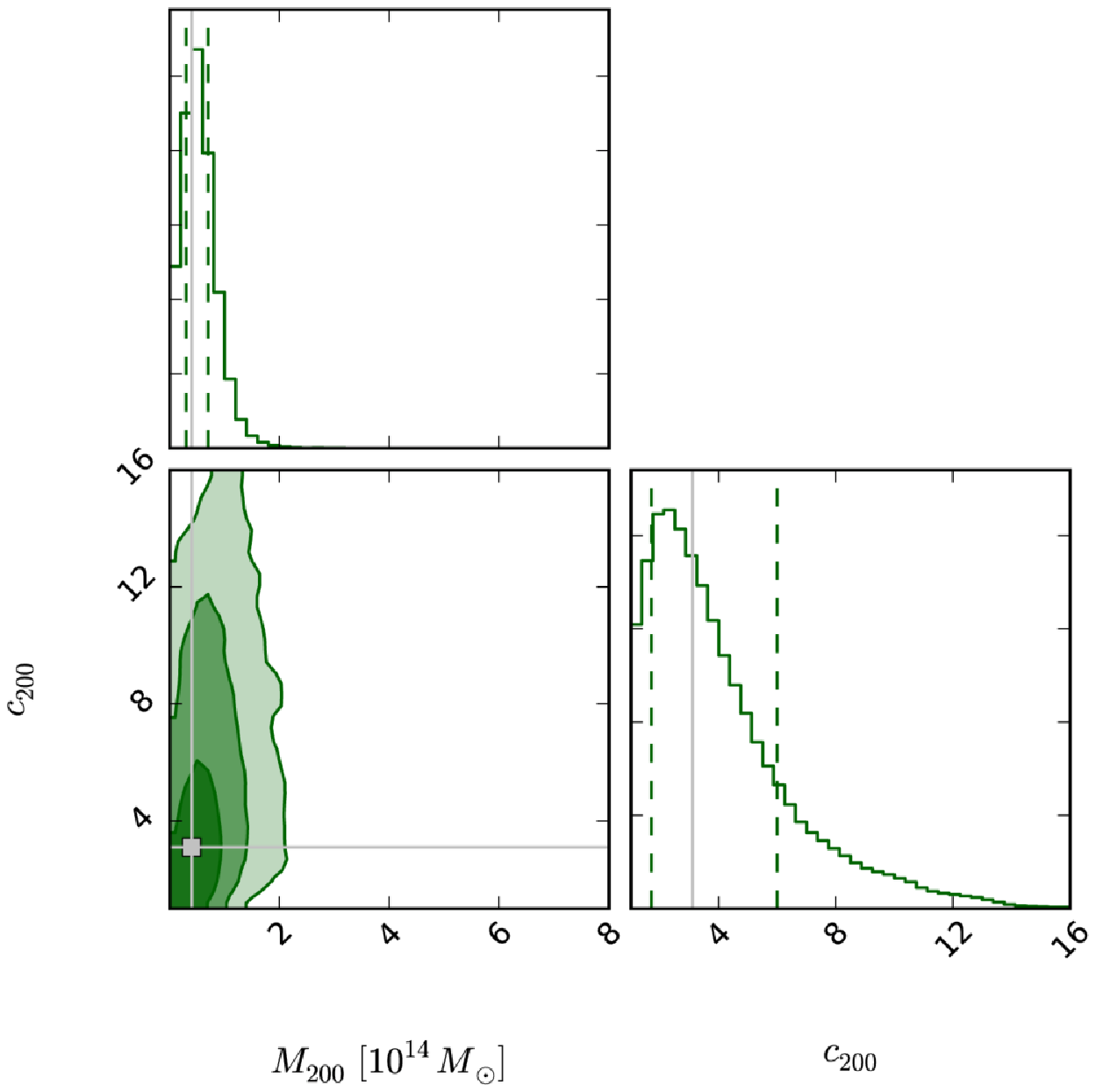}\\
\vspace{0.1cm}
\includegraphics[scale=0.525]{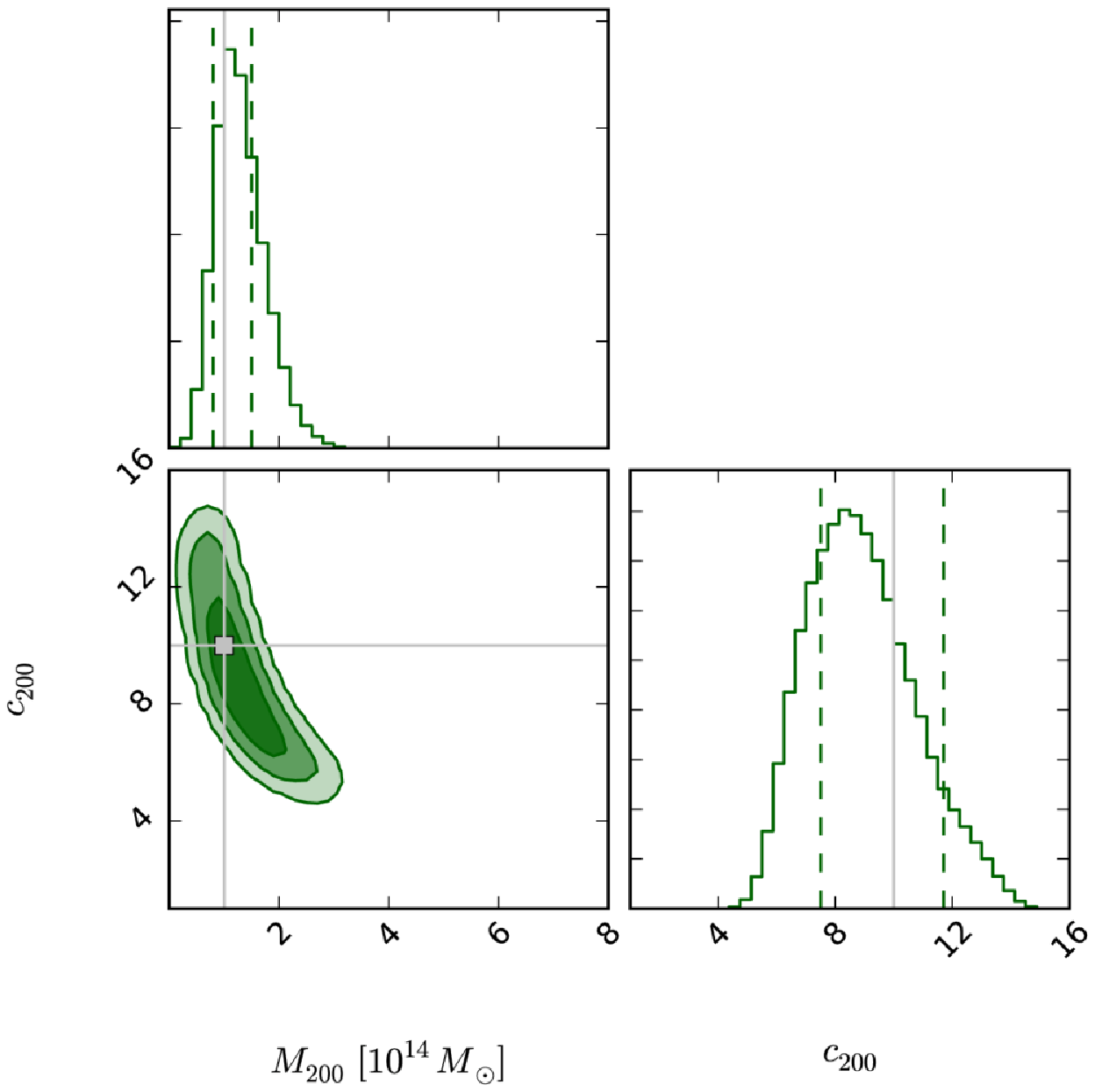}
\caption{PDFs and contours of the parameters $c_{200}$ and $M_{200}$. The three contours stand for the 68\%, 95\%, and 99\% confidence levels. The values obtained for our best-fit model are marked by a gray square, and with vertical lines in the 1D histograms (the asymmetric errors are presented in Table~\ref{tbl-1}).\textit{ Top panel.-} Results from the \textrm{SL\,Model}. \textit{Middle panel.-} Results from the \textrm{Dyn\,Model}. \textit{ Bottom panel.-} Results from the \textrm{SL+Dyn\,Model}.}
\label{Fig:9} \end{center} 
\end{figure}

\section{Results}\label{MassM}

In this Section, we apply the formalism outlined in Sect.\,\ref{Metho} on \object{SL2S\,J02140-0535}, using the data
presented in Sect.\,\ref{DATA}. In the subsequent analysis, \textrm{SL\,Model} refers to the SL modeling, \textrm{Dyn\,Model} to the dynamical analysis, and \textrm{SL+Dyn Model} to the combination of both methods.

\subsection{\textrm{SL\,Model}}
As we discussed in \citetalias{Verdugo2011}, the system AB show multiple subcomponents (surface brightness peaks) that can be conjugated as different multiple image systems, increasing the number of constraints as well as the degrees of freedom (for a fixed number of free parameters). Thus, AB system is transformed in four different systems, conserving C as a single-image arc \citepalias[see Fig. 4 in][]{Verdugo2011}. In this way, our model have five different arc systems in the optimization procedure, leading to 16 observational constraints. Based on the geometry of the multiple images, the absence of structure in velocity space, and the X-ray data, we model \object{SL2S\,J02140-0535} using a single large-scale mass clump
accounting for the dark matter component. This smooth component is modeled with a NFW mass density profile, characterized by its position, projected
ellipticity, position angle, scale radius, and concentration parameter. The position, $(X,Y)$  ranges from -5$\arcsec$ and 5$\arcsec$, the ellipticity from 0 <  $\epsilon$ < 0.7, and the position angle from 0 to 180 degrees. The parameters $r_s$ and $c_{200}$ are free to range between  50 kpc $\leq$ $r_s$ $\leq$ 500 kpc, and 1 $\leq$ $c_{200}$ $\leq$ 30, respectively.

Additionally, we add three smaller-scale clumps that are associated with the galaxies at the center of \object{SL2S\,J02140-0535}. We model them as follows: as 
in \citetalias{Verdugo2011}, we assume that the stellar mass distribution in these galaxies follows a pseudo isothermal elliptical mass distribution (PIEMD). A clump modeled with this profile is characterized by the seven following parameters: the center position, ($X,Y$), the ellipticity $\epsilon$, the position angle $\theta$, and the parameters, $\sigma_0$, $r_{\rm core}$, and $r_{\rm cut}$ \citep[see][for a detailed discussion of the properties of this mass profile]{Limousin2005, ardis2218}. The  center of the profiles, ellipticity, and position angle are assumed to be the same as for the luminous components. The remaining parameters in  the small-scale clumps, namely, $\sigma_0$, $r_{\rm core}$, and $r_{\rm cut}$, are scaled as a function of their galaxy luminosities \citep{Limousin2007}, using as a scaling factor the luminosity $L^{*}$ associated with the $g$-band magnitude of galaxy G1 (see Fig.~\ref{presentlens}), 

\begin{equation}\label{eq:scale}
\begin{array}{l}
\displaystyle r_{\rm core}=r_{\rm core}^{*}\left(\frac{L}{L^{*}}\right)^{1/2},\\ \\
\displaystyle r_{\rm cut}=r_{\rm cut}^{*}\left(\frac{L}{L^{*}}\right)^{1/2},\\ \\
\displaystyle  \sigma_0 = \sigma_0^{*}\left(\frac{L}{L^{*}}\right)^{1/4},
\end{array}
\end{equation}

\noindent setting $r_{\rm core}^{*}$ and  $\sigma_0^{*}$ to be 0.15 kpc and  253 km\,s$^{-1}$, respectively. This velocity dispersion is obtained from the LOS velocity dispersion of galaxy G1, with the use of the relation reported by \citet{ardis2218}. This LOS velocity dispersion has a value of $\sigma_{\rm los}^{*}$ = 215 $\pm$34  km s$^{-1}$, computed from the G-band absorption line profile \citepalias[see][]{Verdugo2011}. The last parameter, $r_{\rm cut}^{*}$, is constrained from the possible stellar masses for  galaxy G1 \citepalias[][]{Verdugo2011}, which in turn produce an interval  of 1 - 6 kpc.

\begin{figure}[h!]\begin{center}
\includegraphics[scale=0.49]{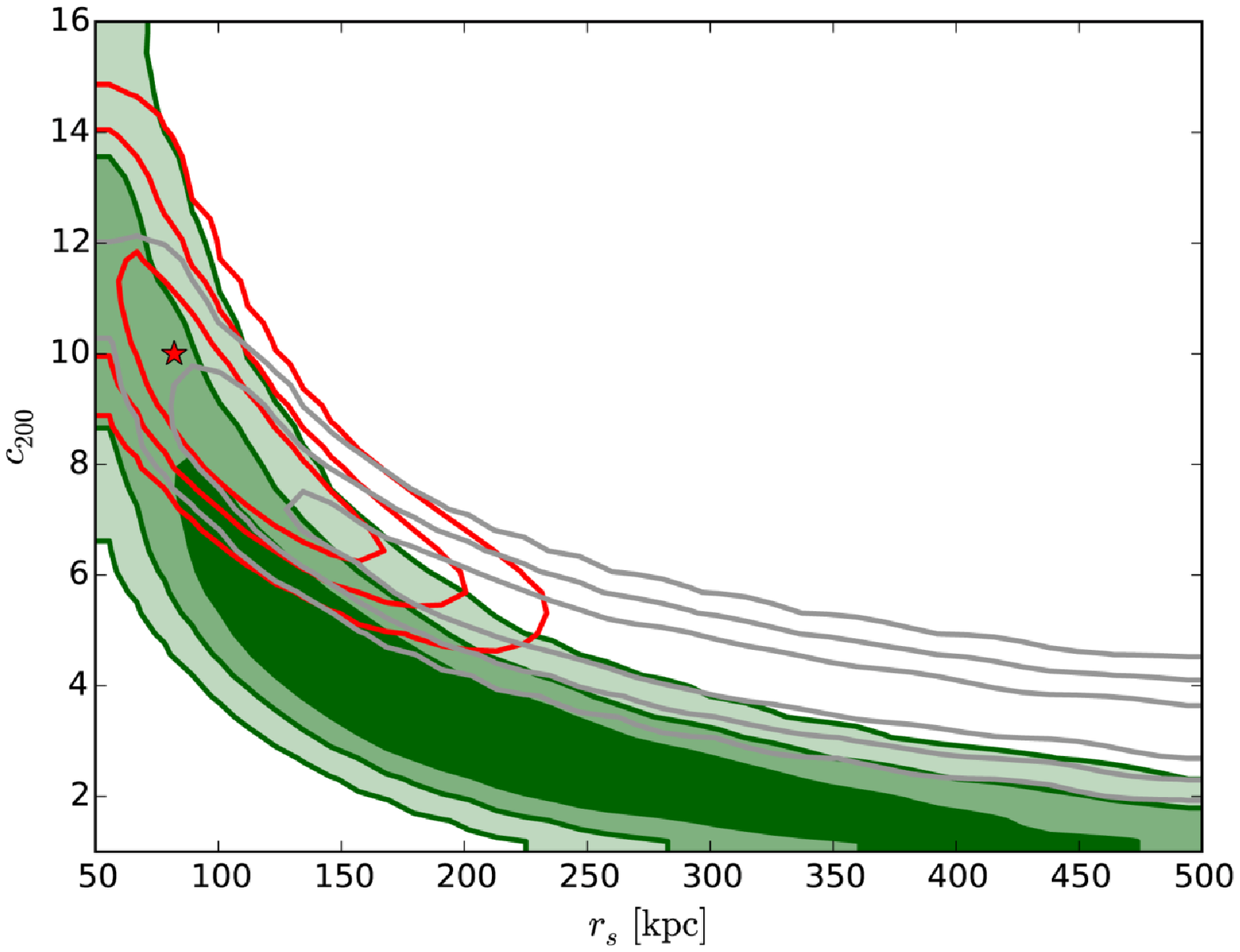}\\
 \vspace{0.77cm}
\includegraphics[scale=0.49]{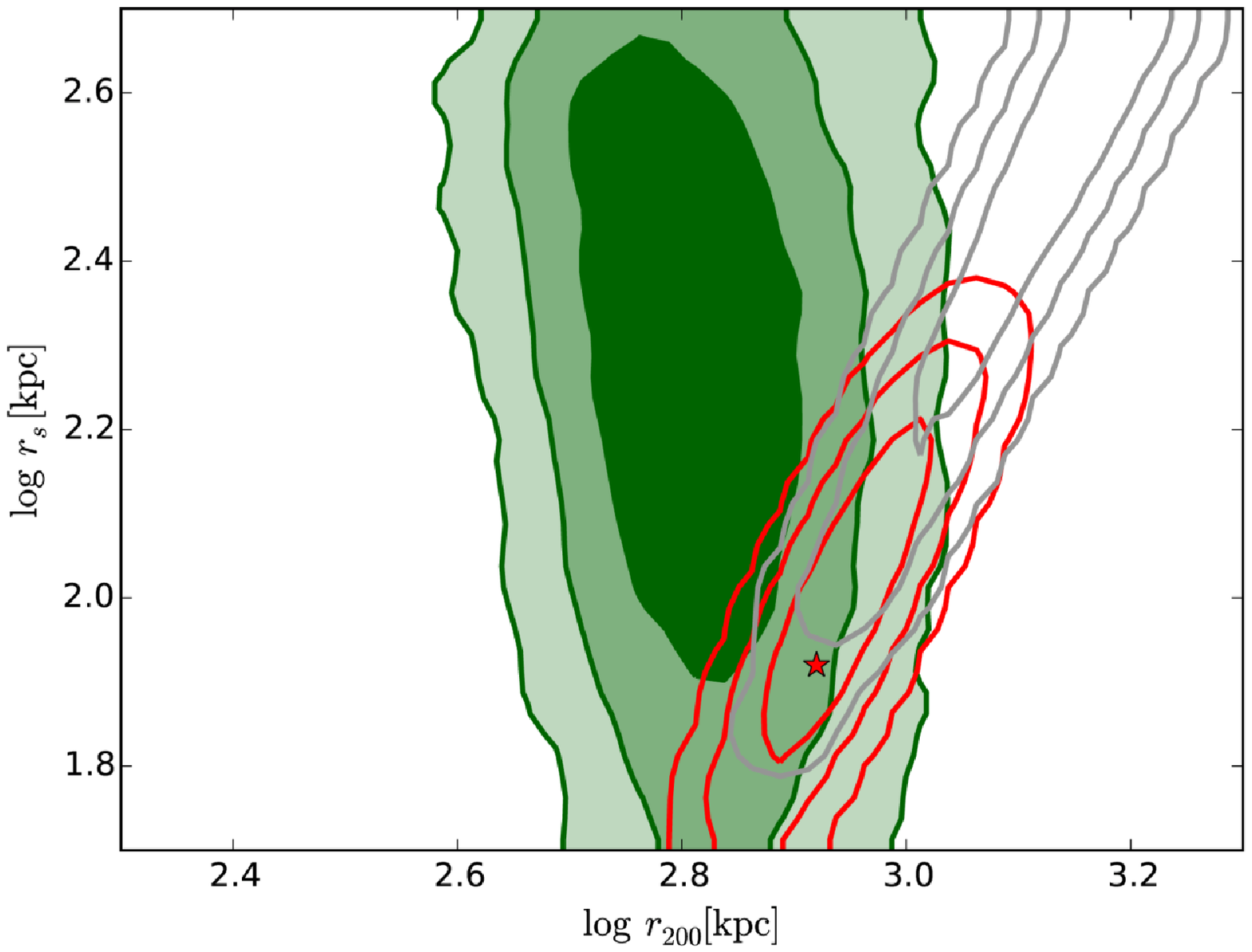}\\
 \vspace{0.77cm}
\includegraphics[scale=0.49]{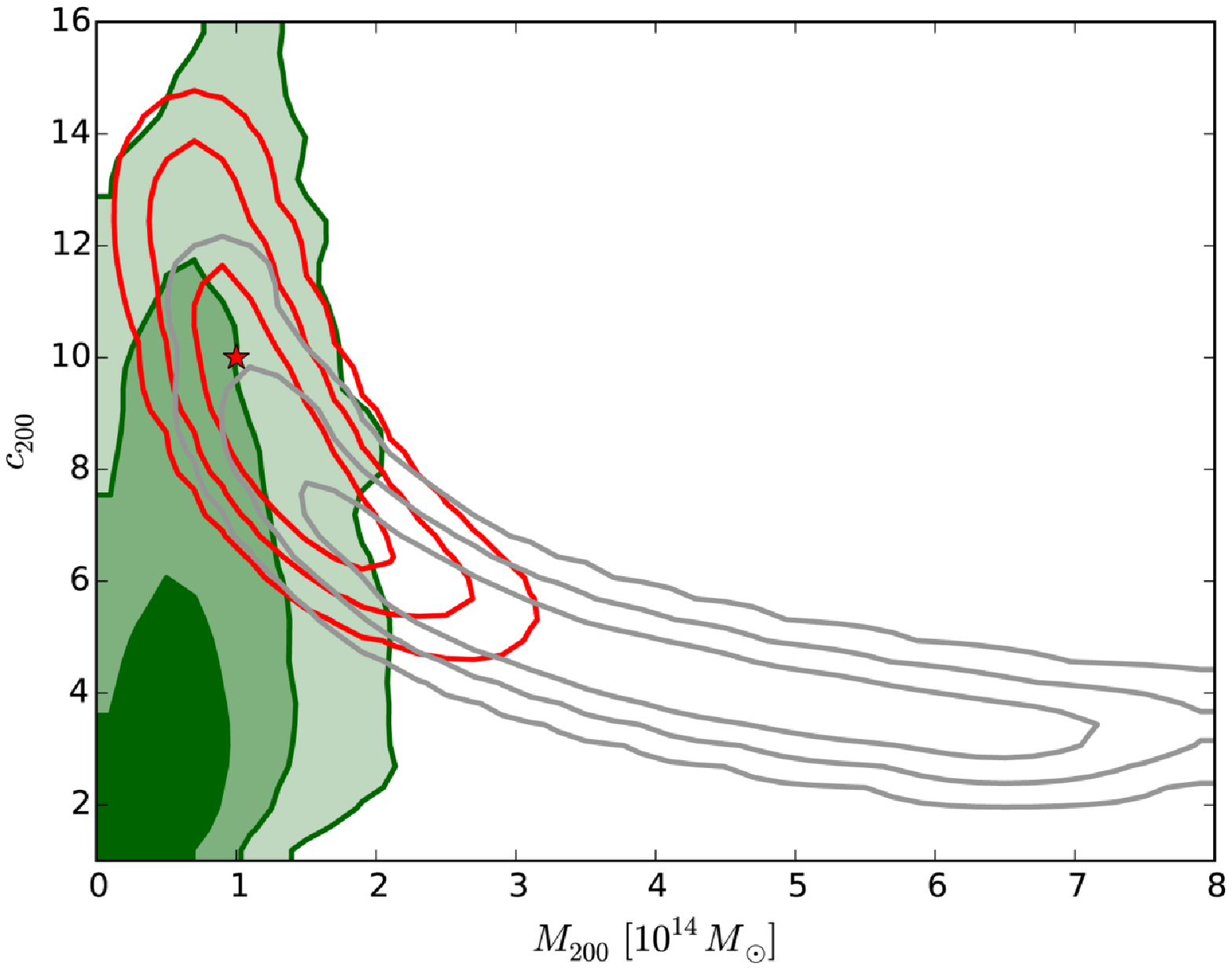}\\
\caption{Joint distributions. \textit{ Top panel.-} Scale radius and concentration \textit{Middle panel.-} log$r_s$ and log$r_{200}$. \textit{ Bottom panel.-} concentration and $M_{200}$. Green-filled contours are   1, 2 and 3-$\sigma$ regions from the \textrm{Dyn\,Model}. Grey contours  stand for the 68, 95, and 99\% confidence levels for the \textrm{SL\,Model}.  Red contours is the result of the \textrm{SL+Dyn\,Model},  with the best solution depicted with a  red star. \label{Fig:10}}
 \end{center} 
\end{figure}

Our model is computed and optimized in the image plane with the seven free parameters discussed above, namely \{$X$,  $Y$, $\epsilon$,  $\theta$, $r_s$, $c_{200}$, $r_{\rm cut}^{*}$\}. The first six parameters characterize the NFW profile, and the last parameter is related to the profile of the central galaxies. All the parameters are allowed to vary with uniform priors. We show the results (the PDF) of the SL analysis \emph{only} in top-panel of Fig.~\ref{PDFModels1}, and the best fit parameters are given in Table~\ref{tbl-1}. Additionally, in Fig.~\ref{PDFModels2} we show the plots of $r_s$ vs  $r_{200}$,  since they provide with a better understanding of how unconstrained is the lens model at large scale (see next section), and also to be consistent with the form in which the plots are presented in \citet{Mamon2013}. Figure~\ref{Fig:9} shows the results for the concentration and $M_{200}$, from which we can gain insight for the mass constraint.

\subsection{\textrm{Dyn\,Model}}

 Since we only have 24 group members, we assume that the group has an isotropic velocity dispersion (i.e., $\beta$ = 0 in  Eq.~\ref{eq:anisotropy}). This parameter $\beta$ might influence the parameters of the density profile ($r_s$ and $c_{200}$), however, it is not possible to constrain $\beta$ with only 24 galaxies, besides it is beyond the scope of this work to analyze its effect over the parameters. We defer this analysis to a forthcoming paper, in which we apply the method to a galaxy cluster with a greater number of members.

 The Jeans equation of dynamical equilibrium, as implemented in {\sc MAMPOSS}t, is only valid for values of $r \lesssim 2r_{vir} \simeq 2.7r_{200}$ \citep{Falco2013}.  Thus,  before  running MAMPOSt, we estimate the viral radius, $r_{200}$, of \object{SL2S\,J02140-0535}. From the scale radius and the concentration values reported in  \citetalias{Verdugo2011} we find the virial radius to be $r_{200}$ = 1 $\pm$ 0.2 Mpc. This value is considerably smaller than the previously reported value of 1.42 Mpc by  \citet{Lieu2015}\footnote{\object{SL2S\,J02140-0535} is identified as \object{{\sc XLSSC\,110 }} in the {\sc XXL} Survey.}. Table~\ref{tbl-A1} shows that there are 3 galaxies with 1 Mpc $<$ R $<$ 1.4 Mpc, i.e. within 1.4$r_{200}$, which seems sufficiently small to keep in our analysis. The galaxy members lie between  7.9 kpc to  1392.3 kpc (with a mean distance of 650 kpc from the center). Given the scarce number of members in \object{SL2S\,J02140-0535} we keep this galaxy in our calculations.

To further simplify our analysis, we assume that the completeness, as a function of the radius, is a constant (see Sect.\,\ref{NSD}). Also, we assume that both the tracer scale radius  \citep[$r_{\nu}$ in][]{Mamon2013} and the dark matter scale radius $r_s$ are the same, that is, the total mass density profile is forced to be proportional to the galaxy number density profile: \textit{we assume that mass follows light}. As we will see in the next section, this is not a bad assumption. Therefore our model has only two free parameters, namely, the scale radius $r_s$, and the concentration $c_{200}$. These parameters have broad priors, with  50 kpc $\leq$ $r_s$ $\leq$ 500 kpc, and 1 $\leq$ $c_{200}$ $\leq$ 30.  The middle panels of Fig.~\ref{PDFModels1},  Fig.~\ref{PDFModels2}, and Fig.~\ref{Fig:9}  show the PDF for this model; the best values of the fit are presented in Table~\ref{tbl-1}.

\begin{table*}
\caption{Best-fit model parameters.}
\begin{center}
\label{tbl-1} 
\begin{tabular}{lccccccccc}
\hline\hline 
\\
\multirow{2}{*}{Parameter} &
      \multicolumn{2}{c}{\textrm{SL\,Model}} &
      \multicolumn{2}{c}{\textrm{Dyn\,Model}} &
      \multicolumn{2}{c}{\textrm{SL+Dyn\,Model}} \\
     & Group & $L^{*}$ & Group & $L^{*}$ & Group & $L^{*}$ \\  \\
    \hline
    \\
X$^{\dagger}$ [\arcsec] & $1.2^{+0.2}_{-0.4}$ & -- &   --  &  -- &   $0.9\pm0.2$  &  -- \\ \\
Y$^{\dagger}$ [\arcsec] & $0.7^{+0.9}_{-0.4}$ & -- &   --  &  -- &   $1.5^{+0.4}_{-0.3}$  &  -- \\ \\
$\epsilon^{\dagger\dagger}$  & $0.23^{+0.04}_{-0.05}$ & -- &   --  &  -- &   $0.298^{+0.002}_{-0.045}$  &  -- \\ \\
$\theta\,[^{\circ}]$  & $111.2^{+1.6}_{-1.3}$ & -- &   --  &  -- &   $111.1^{+1.4}_{-1.3}$  &  -- \\ \\
$r_s$ [kpc] & $184^{+209}_{-60}$ & -- &   $199^{+135}_{-\phantom{1}91}$  &  -- &   $82^{+44}_{-17}$  &  -- \\ \\
log$r_{200}$ [kpc] & $3.04^{+0.12}_{-0.06}$ & -- &   $2.80^{+0.07}_{-0.06}$  &  -- &   $2.92^{+0.06}_{-0.03}$  &  -- \\ \\
$c_{200}$  & $6.0^{+1.8}_{-2.3}$ & -- &   $3.1^{+2.9}_{-1.4}$  &  -- &   $10.0^{+1.7}_{-2.5}$  &  -- \\ \\
$M_{200}$ [10$^{14}$M$_\odot$]  & $2.5^{+3.0}_{-0.9}$ & -- &   $0.4^{+0.3}_{-0.1}$  &  -- &   $1.0^{+0.5}_{-0.2}$  &  -- \\ \\
$r_{\rm core}$ [kpc]   & -- & $[0.15]$ &   --  &  -- &   --  &  $[0.15]$ \\ \\
$r_{\rm cut}$ [kpc]   & -- & $2.6^{+2.1}_{-1.1}$ &   --  &  -- &   --  &  $2.4^{+2.1}_{-1.0}$ \\ \\
$\sigma_0$ [km s$^{-1}$]  & -- & $[253]$ &   --  &  -- &   --  &  $[253]$ \\ \\
$\chi^{2}_{DOF}$ & \phantom{000000}-- \phantom{000}0.1 & -- &   \phantom{000000}-- \phantom{0000.9}  &  -- &  \phantom{000000}-- \phantom{000}0.9  &  -- \\ \\
\\
\hline 
\end{tabular}
\end{center}
\tablefoot{($\dagger$): The position in arc seconds relative to the BGG.\\
($\dagger$$\dagger$):  The ellipticity is defined as $\epsilon$ = ($a^2$ $-$ $b^2$)/($a^2$ $+$ $b^2$), where a and b are the semimajor and semiminor axis, respectively, of the elliptical shape.\\
The first column identifies the model parameters. In columns 2-10 we provide the results for each model, using square brackets for those values which are not optimized.  Columns $L^{*}$ indicate the parameters associated to the small-scale clumps. Asymmetric errors are calculated following \citet{Andrae2010} and \citet{Barlow1989}.\\}
\end{table*}

\subsection{\textrm{SL+Dyn Model}}\label{MassM4.3}

The main difference of our work, when compared to previous works \citep[e.g.,][]{Biviano2013,Guennou2014}  is that we apply a joint analysis, seeking a solution consistent with both the SL and the dynamical methods, maximizing the total likelihood. In the bottom-panels of Fig.~\ref{PDFModels1}, Fig.~\ref{PDFModels2}, and Fig.~\ref{Fig:9} we show the PDF of this combined model. The best-fit values are presented in Table~\ref{tbl-1}. 

From the figures it is clear that exists tension between the results from the \textrm{SL\,Model} and the \textrm{Dyn\,Model}, the models are in disagreement at 1-$\sigma$ level. The discrepancy is related to the oversimplified assumption of the spherical \textrm{Dyn\,Model}.  Although in some cases it is expected to recover a spherical mass distribution at large scale \citep[e.g.,][]{Gavazzi2005}, at smaller scale, i.e. at strong lensing scales, the mass distribution tends to be aspherical. In order to investigate such tension between the results, we construct a strong lensing spherical model, with the same constrains as before. In the left panel of Fig.~\ref{figA2} we show the results. It is clear that in this case the model is not well constrained, a natural result given the lensing images in \object{SL2S\,J02140-0535}. However, the comparison between the joint distributions in the top panel of Fig.~\ref{Fig:10} and the one in the right panel of Fig.~\ref{figA2} shows that the change in the contours is small, which indicates that the assumption of spherical symmetry has little impact in the final result\footnote{Note that the agreement between contours is related to the the shallow distribution from {\sc LENSTOOL}, which produce a joint distribution that follows the top edge of the narrower {\sc MAMPOSS}t distribution.}.  Note also that the combined model has a bimodal distribution (see right panel of Fig.~\ref{figA2}), with higher values of concentration inherited from the lensing constraints. 

A possible way to shed more light on the systematics errors of our method is to test it with simulations. For example, comparing between spherical and non-spherical halos, or quantifying  the bias when a given mass distribution is assumed and the underlaying one is different. Such kind of analysis is out of the scope of the present work, however it could be performed in the near future since the state-of-the-art simulations on lensing galaxy clusters has reached an incredible quality \citep[e.g.,][]{Meneghetti2016}.

\section{Discusion}\label{Discus}

\subsection{Lensing and dynamics as complementary probes}

From Fig.~\ref{PDFModels1} and Fig.~\ref{PDFModels2}, it is clear that \textrm{SL\,Model} is not able to constrain the NFW mass profile. This result is expected since SL constraints are available in the very central part of \object{SL2S\,J02140-0535}, whereas the scale radius is generally several times the SL region. The degeneracy between $c_{200}$ and $r_s$ (or $r_s$ and $r_{200}$), which is related  to the mathematical definition of the gravitational potential,  was previously discussed in  \citetalias{Verdugo2011}. This degeneracy occurs commonly in lensing modeling \citep[e.g.,][]{jullo07}. From \textrm{SL\,Model} we obtain the following values:  $r_s$  = $184^{+209}_{-60}$ kpc, and $c_{200}$  = $6.0^{+1.8}_{-2.3}$. Thus, the model is not well constrained (similarly, we obtain for the virial radius a value of log$r_{200}$ = 3.04$^{+0.12}_{-0.06}$ kpc).  Moreover, the mass $M_{200}$ is not constrained in the \textrm{SL\,Model} (see Fig.~\ref{Fig:9}).

The same conclusion holds when considering dynamics only, i.e., \textrm{Dyn\,Model}: the constraints are so broad that both parameters ($r_s$ and $c$) can be considered as unconstrained (see medium-panel of Fig.~\ref{PDFModels1} and medium-panel of Fig.~\ref{PDFModels2}). In this case we obtained a scale radius of $r_s$  = $199^{+135}_{-\phantom{1}91}$ kpc, and a concentration of $c_{200}$ = $3.1^{+2.9}_{-1.4}$. However, in this case the scale radius is slightly more constrained when compared to the value obtained with the  \textrm{SL\,Model}. This is due to employing  the distribution of  the galaxies to estimate the value when using {\sc MAMPOSS}t. Furthermore, the viral radius $r_{200}$ is even more constrained, with log$r_{200}$ = 2.80$^{+0.07}_{-0.06}$  kpc. Note that these weak constraints are related to the small number of galaxy members (24) in the group. Nonetheless, even with the low  number of galaxies the error in our mass $M_{200}$ (see Table~\ref{tbl-1} and Fig.~\ref{Fig:9}) is approximately a factor of two, i.e.,  $\sim$ 0.3 dex, consistent with the analysis of  \citet{Old2015}.

\begin{figure*}[!htp]
\centering
\includegraphics[scale=0.87]{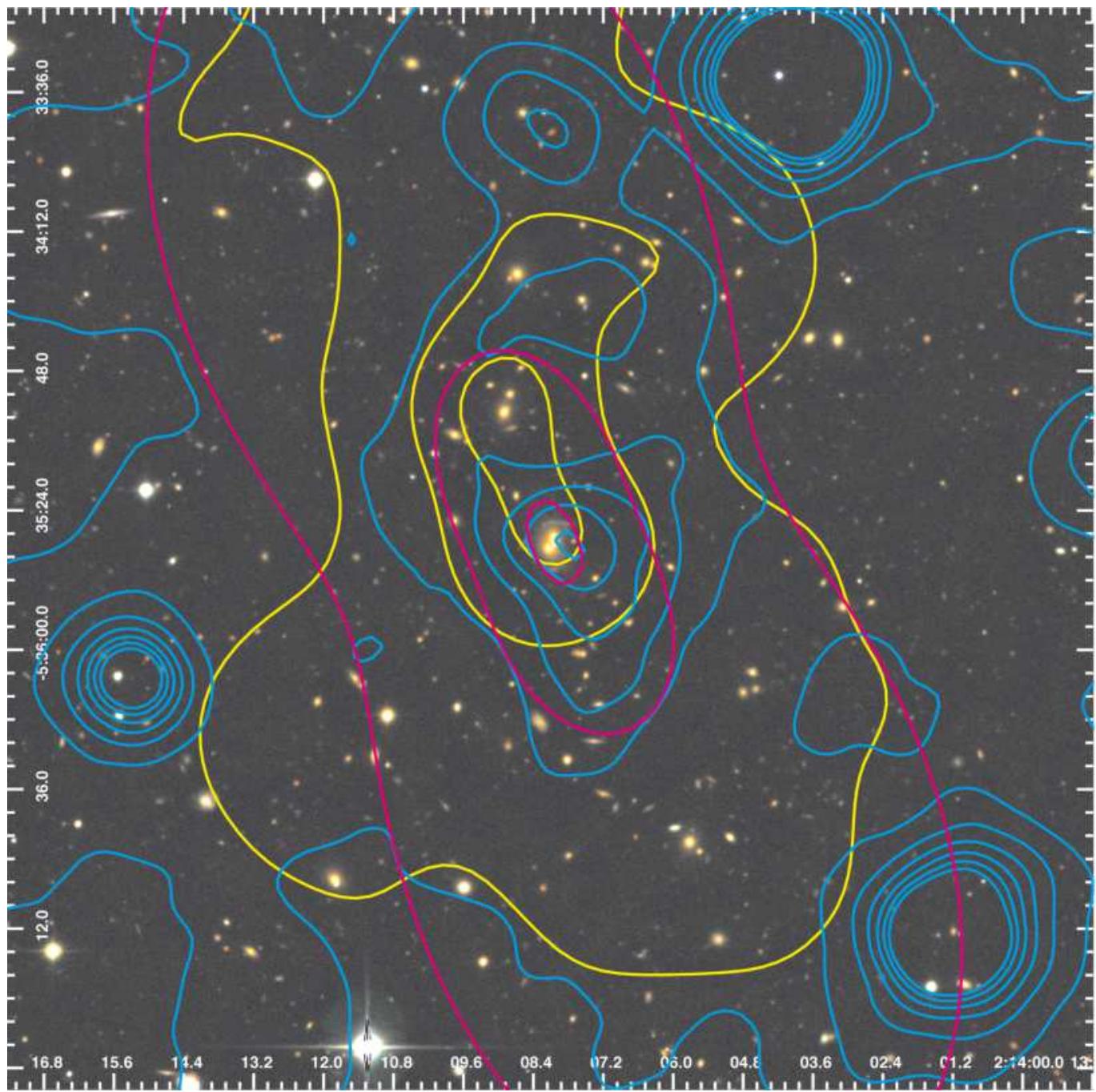}
\caption{The distribution of mass, gas and galaxies in \object{SL2S\,J02140-0535}, as reflected
by the total mass derived from the combined lensing and dynamics analysis (magenta),
the adaptively smoothed $i$-band luminosity of group galaxies (yellow) and the
surface brightness from \xmm\, observations (blue).  The lensing mass contours (magenta lines) correspond to projected surface densities of 0.2$\times$10$^{9}$, 
1.2$\times$10$^{9}$, and 7.4$\times$10$^{9}$ \Msun\,arcsec$^{-2}$. The size of the image is 1.5$\times$1.5 Mpc.}
\label{Fig:11}
\end{figure*}

Interestingly, when combining both probes, \textrm{SL+Dyn\,Model}, it is possible to constrain both the scale radius and the concentration parameter.  SL is sensitive to the mass distribution at inner radii (within 10$\arcsec$), whereas the dynamics provide constraints at larger radius  (see bottom-panels of Fig.~\ref{PDFModels1} and Fig.~\ref{PDFModels2}). For this model, we find the values $r_s$ =$82^{+44}_{-17}$ kpc, $c_{200}$ = $10.0^{+1.7}_{-2.5}$, and $M_{200}$ = $1.0^{+0.5}_{-0.2}$ $\times$ 10$^{14}$ M$_\odot$. The errors in the mass, although big, are smaller when compared to the  two previous models, by a factor of 2.2 (0.34 dex) and by a factor of 1.4 (0.15 dex), for 
\textrm{SL\,Model} and \textrm{Dyn\,Model}, respectively.

To highlight how the combined \textrm{SL+Dyn\,Model} is better constrained than both the \textrm{SL\,Model} and the \textrm{Dyn\,Model}, we show in Fig.~\ref{Fig:10} the 2D contours for $c_{200} - r_s$, log$r_s$ $-$ log$r_{200}$, and $c_{200} - M_{200}$, for the three models discussed in this work. We note the overlap of the solutions of the \textrm{SL\,Model} and the \textrm{Dyn\,Model}, as well as the stronger constraints of the \textrm{SL+Dyn\,Model}. The shift in the solutions for \textrm{SL+Dyn\,Model} that it is seen in Fig.~\ref{PDFModels1}, i.e., $r_s$ is much lower (greater $c_{200}$) than the values for the \textrm{SL\,Model} and the \textrm{Dyn\,Model}, can be understood in the light of the discussion presented in Sect.\,\ref{MassM4.3}, and additionally explained with the analysis of Fig.~\ref{Fig:10}. On one hand, the tension between both results (lack of agreement between solutions at 1$\sigma$) is the result of assuming a spherical mass distribution in the \textrm{Dyn\,Model}. On the other hand, the joint solution of \textrm{SL+Dyn\,Model} (red contours) is consistent with the region where both the \textrm{SL\,Model} and the \textrm{Dyn\,Model} overlap.

\subsection{Mass, light \& gas}

We find that the centre of the mass distribution coincides with that of the light (see Fig.~\ref{Fig:11}). In  \citetalias{Verdugo2011} we showed that the position angle of the halo was consistent with the orientation of the luminosity contours and the spatial distribution of the group-galaxy members. In the present work we confirm these results. The measured position angles of the luminosity contours presented in Fig.~\ref{fig1} and Fig.~\ref{Fig:11} (the values are equal to 109$^{\circ}$ $\pm$ 2$^{\circ}$, 102$^{\circ}$ $\pm$ 2$^{\circ}$, and 99$^{\circ}$ $\pm$ 9$^{\circ}$, from innermost to outermost contour), agree with the orientation of the position angle of $111.1^{+1.4}_{-1.3}$ degrees of the halo.

In addition to the distribution of mass and light, Fig.~\ref{Fig:11} shows the distribution of the gas component of \object{SL2S\,J02140-0535}, which was obtained from our X-ray analysis. The agreement between these independent observational tracers of the three group constituents (dark matter, gas, and galaxies) is remarkable. This supports a scenario where the mass is traced by light, and argues in favor of  a non disturbed structure, i.e., the opposite to a disturbed one, where the different tracers are separated, such as in the Bullet Group \citep[][]{Gastaldello2014} or as in the more extreme cluster mergers \citep[e.g.,][]{Bradac2008,Randall2008,Menanteau2012}.

\subsection{Comparison with our previous work}

In  \citetalias{Verdugo2011} we analyzed \object{SL2S\,J02140-0535} using the dynamical information to constrain and build a reliable SL model for this galaxy group. However, it is not expected to have a perfect agreement between the best value of the parameters computed in the former work and the values reported in the present paper, mainly because the difference in methodologies, and also due to the new spectroscopic number of members reported in this work. For example,  in \citetalias{Verdugo2011} we found the values $c_{200}$ = 6.0 $\pm$ 0.6, and $r_s$ = 170 $\pm$ 18 kpc, whereas for our \textrm{SL+Dyn\,Model}, we find the values  $c_{200}$ = $10.0^{+1.7}_{-2.5}$, and $r_s$ = $82^{+44}_{-17}$ kpc. However, it is important to note  that the latter values lie within the range predicted by \citetalias{Verdugo2011}  with  dynamics (cf. 2 < $c_{200}$ <  8 , and 50 kpc < $r_s$ <200  kpc). Furthermore, those ranges need to be corrected by using the new velocity dispersion. This correction will shift the confidence interval to larger values in $c$, and smaller values in $r_s$, thus improving the agreement between both works.

Additionally, as presented in Sect.\,\ref{DATA} , we found the velocity dispersion to be $\sigma$ = 562 $\pm$ 60 km\,s$^{-1}$ with the 24 confirmed members. This velocity dispersion is in  good agreement with the velocity reported in \citetalias{Verdugo2011},  $\sigma$ = 630 $\pm$ 107 km\,s$^{-1}$, which was computed with only 16 members \footnote{The projected viral radius, $\widetilde{R_{v}}$  = 0.9 $\pm$ 0.3 Mpc  \citep[the projected harmonic mean radius, e.g.,][]{Irg02}, is also consistent with the value reported in our previous work, $\widetilde{R_{v}}$  = 0.8 $\pm$ 0.3 Mpc, which is worth to note as it was used in  \citetalias{Verdugo2011} to estimate the priors in the SL modeling.}.  It is also in agreement with the value obtained from weak lensing analysis \citep[$\sigma$ = 638 $^{+101}_{-152}$    km\,s$^{-1}$,][]{Gael2013}.

\begin{figure}[h!]\begin{center}
\includegraphics[scale=0.5]{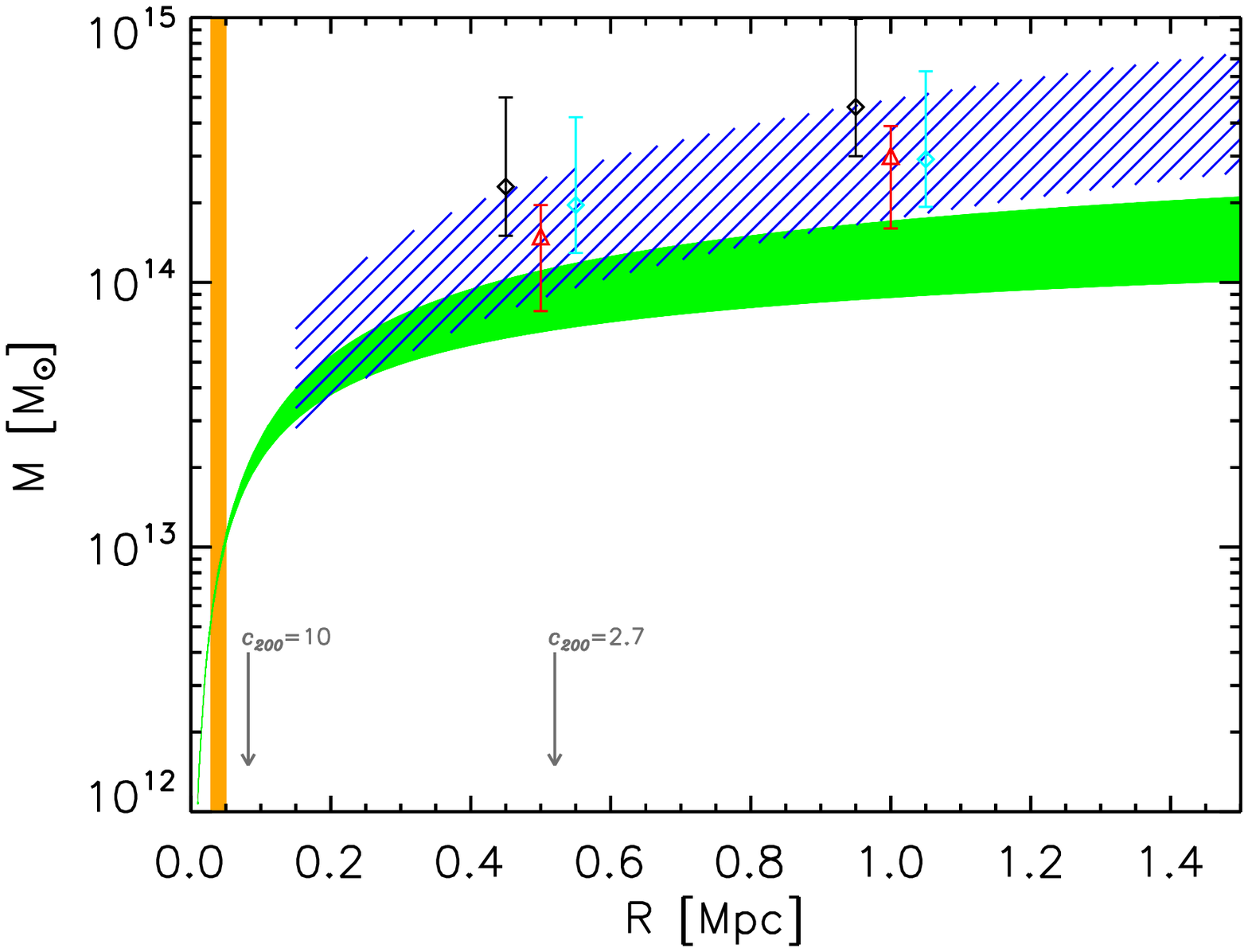}
\caption{2D projected mass as a function of the radius measured from the BGG. The green and blue shaded areas corresponds to
the mass profile within 1-$\sigma$ errors for the SL model (\textrm{SL+Dyn Model}) and the weak
lensing model reported in \citetalias{Verdugo2011}, respectively. The orange-shaded region shows the area where the arc systems lie. The two red triangles with error bars, show two estimates (at 0.5 and 1 Mpc) of the weak lensing mass from \citet{Gael2013}. Black diamonds (shifted in $r$-0.05 for clarity) show the predicted mass calculated from the work of \citet{Lieu2015}. Cyan diamonds (shifted in $r$+0.05 for clarity) are the masses calculated from \citet{Lieu2015} data but assuming $c_{200}$ = 10 (see text).  We also depict with two arrows, our best $r_s$ (for $c_{200}$ = 10)  and the $r_s$ (for $c_{200}$ = 2.7) reported in \citet{Lieu2015}.\label{Fig:12}}
 \end{center} 
\end{figure}

\subsection{An over-concentrated galaxy group?}

The concentration value of \object{SL2S\,J02140-0535} is clearly higher than the expected from $\Lambda$CDM numerical simulations. Assuming a dark matter halo at $z$ = 0.44 with $M_{200}$ $\approx$ 1 $\times$ 10$^{14}$ M$_{\sun}$ the concentration is  $c_{200}$  $\approx$ 4.0 \citep[computed with the procedures of][]{Duff08}. 
\object{SL2S\,J02140-0535} has also been studied by \citet{Gael2014}, who were able to constrain the 
scale radius and the concentration parameters of galaxy groups using stacking techniques.
\object{SL2S\,J02140-0535}, with an Einstein radius of $\sim$7$\arcsec$, belong to their stack
"R3", which was characterized to have $c_{200} \sim 10$
and M$_{200} \sim 10^{14}$ M$_{\sun}$. Those values are in agreement with our computed values. 
As discussed thoroughly in \citet{Gael2014}, this over-concentration seems to be due to an alignment of the major axis with the line of sight. Even in the case  a cluster displays  mass contours elongated in the plane of the sky, the major axis could be near to the line of sight \citep[see for example][]{Limousin2007,Limousin2013}.

Finally, figure~\ref{Fig:12} shows the comparison between the mass obtained from our \textrm{SL+Dyn\,Model}  with the weak lensing mass  previously obtained by \citetalias{Verdugo2011}. Both models overlap up to $\sim$1 Mpc; this is consistent with the scarce number of galaxies located at radii larger than 1 Mpc. Therefore, the dynamic  constraints are not strong, also it is worth to note that the weak lensing mass estimate can be slightly overestimated at large radii, since the mass is calculated assuming a singular isothermal sphere. The red  triangles in  Fig.~\ref{Fig:12} show two estimates (at 0.5 and 1 Mpc) of the weak lensing mass, calculated using  the values reported in \citet{Gael2013}. Those values are also consistent with the above mentioned measurements. For comparison, we show in black diamonds the predicted mass (at 0.5 and 1 Mpc) derived  from \citet{Lieu2015}. The discrepancy between these values and our values calculated from lensing arises from the fact that \citet{Lieu2015} set the  cluster concentration from  a mass-concentration relation derived from N-body simulations, thus obtaining the values $c_{200}$ = 2.7 and $r_s$ = 0.52 Mpc. To prove this assertion, we perform  a simple test. We use the values of $M_{200}$ and $c_{200}$  from \citet{Lieu2015}, and then we generate a shear profile with their same radial range and number of bins. We fitted their  data  seeking  the best $M_{200}$  value,  assuming a concentration value of  $c_{200}$ = 10. The projected mass from this estimate is shown  as cyan symbols in Fig.~\ref{Fig:12}. This change in concentration not only solves the difference in mass estimates, but also explains  why   \object{SL2S\,J02140-0535} ({\sc XLSSC\,110})  is an  outlier  in the sample of \citet{Lieu2015}. This highlights the risk of assuming a $c$-$M$ relation for  some particular objects, such as  strong lensing clusters. The discussion of the bias and the effect  on the $M$-$T$ scaling relation will be discussed in a forthcoming publication (Fo{\"e}x et al. in prep.).

 \section{Conclusions}\label{Conclusions}

 We have presented a framework that allows to fit simultaneously strong lensing and dynamics. We apply our method  to probe the gravitational potential of the lens galaxy group \object{SL2S\,J02140-0535} on a large radial range, by combining two well known codes, namely {\sc MAMPOSS}t \citep[][]{Mamon2013} and {\sc LENSTOOL} \citep{jullo07}. We performed a fit adopting  a NFW profile and three galaxy-scale mass components as perturbations to the group potential, as previously done by  \citetalias{Verdugo2011}, but now including  the dynamical information in a new  consistent way.  The number of galaxies increased to 24, when  new VLT (FORS2) spectra  were analyzed.  This new  information was included to perform the combined  strong lensing and dynamics analysis. Moreover, we studied the gas distribution within the group from X-ray data obtained with \xmm.
 
 We list below our results:
 
 \begin{enumerate}
\item   Our new observational  data set confirms the results presented previously  in \citetalias{Verdugo2011}. We also present supporting X-ray analysis.

\begin{itemize}

\item  Spectroscopic analysis confirms  that the  arcs AB and the arc  C of \object{SL2S\,J02140-0535}  belong to different sources, the former  at $z_{\rm spec}$ = 1.017 $\pm$ 0.001, and the latter  at  $z_{\rm spec}$ = 1.628 $\pm$ 0.001. These redshift  values are consistent with the  photometric redshift  estimation. 

\item We find 24 secure members of \object{SL2S\,J02140-0535}, from the analysis of our new and previously reported spectroscopic data.  The completeness $C\sim60\%$ is roughly constant up to 1 Mpc. We also computed the velocity dispersion, obtaining  $\sigma$ = 562 $\pm$ 60 km\,s$^{-1}$,  a value comparable to the previous estimate of  \citetalias{Verdugo2011}.

\item The X-ray contours show an elongated shape  consistent with the spatial distribution of  the confirmed members. This argues in favor of an  unimodal structure, since the X-ray emission is unimodal and centered on the lens.

\end{itemize}

\item   Our method fits simultaneously strong lensing and dynamics, allowing to probe the mass distribution from the centre of the group up to its  viral radius. However, there is a tension between the results of the \textrm{Dyn\,Model} and the \textrm{SL\,Model}, related to the assumed spherical symmetry of the former. While our result shows that deviation from spherical symmetry can in some cases induce a bias in the {\sc MAMPOSS}t solution for the cluster $M(r)$, this does not need to be the rule. In another massive cluster at $z$ = 0.44, \citet{Biviano2013} found good agreement between the spherical {\sc MAMPOSS}t solution and the non-spherical solution from strong lensing. In addition, MAMPOSSt has been shown to provide unbiased results for the mass profiles of cluster-sized halos extracted from cosmological simulations \citep{Mamon2013}.

\begin{itemize}
\item  Models relying solely on either lensing (\textrm{SL\,Model}) or dynamical information (\textrm{Dyn\,Model}) are not able to constrain the scale radius of the NFW profile. We obtain for the best \textrm{SL\,Model} a scale radius of  $r_s$  = $184^{+209}_{-60}$ kpc, whereas  for the best \textrm{Dyn\,Model} model we obtain a value of  $r_s$  = $199^{+135}_{-\phantom{1}91}$ kpc. We find  that the concentration parameter is unconstrained as well.

\item  However, it is possible to constrain both the scale radius and the concentration parameter when combining both lensing and dynamics analysis \citepalias[as previously discussed in][]{Verdugo2011}. We find a scale radius of $r_s$ =$82^{+44}_{-17}$ kpc,  and  a concentration value of $c_{200}$ = $10.0^{+1.7}_{-2.5}$. The  \textrm{SL+Dyn\,Model}  reduces  the error in the mass estimation in 0.34 dex (a factor of 2.2), when compared to  the \textrm{SL\,Model}, and in 0.15 dex (a factor of 1.4), compared to  the \textrm{Dyn\,Model}.

\item  Our joint \textrm{SL+Dyn\,Model} allows to probe,  in a reliable fashion,  the mass profile of  the group  \object{SL2S\,J02140-0535} at large scale. We find  a good agreement between the luminosity contours, the mass contours, and the X-ray emission. This result confirms that  the  mass is traced by light.

\end{itemize}

\end{enumerate}

 The joint lensing-dynamical analysis presented in this paper,  applied to the lens galaxy group \object{SL2S\,J02140-0535}, is aimed to show a consistent method to probe the mass density profile of groups and clusters of galaxies. This is the first paper in a series in which we extend our methodology to the galaxy clusters, for which  the number of constraints is  larger  both in lensing images and in galaxy members. Therefore,  we should be able to probe with our new method more parameters,  such as the anisotropy parameter and the tracer radius (Verdugo et al. in preparation).

 \begin{acknowledgements}
 
The authors thank the anonymous referee for invaluable remarks and suggestions. T. V. thanks the staff of the Instituto de F\'isica y Astronom\'ia of the Universidad de Valpara\'iso.   ML acknowledges the Centre National de la Recherche Scientifique (CNRS) for its support. V. Motta gratefully acknowledges support from FONDECYT through grant 1120741,  ECOS-CONICYT through grant C12U02, and Centro de Astrof\'{\i}sica de Valpara\'{\i}so. M.L. and E.J. also acknowledge support from ECOS-CONICYT C12U02. A.B. acknowledges partial financial support from the PRIN INAF 2014: "Glittering kaleidoscopes in the sky: the multifaced nature and role of galaxy clusters" P.I.: M. Nonino.  K. Rojas acknowledges support from Doctoral scholarship FIB-UV/2015 and ECOS-CONICYT C12 U02. J.M. acknowledges support from FONDECYT through grant 3160674. J.G.F-T is currently supported by Centre National d'Etudes Spatiales (CNES) through PhD grant 0101973 and the R\'egion de Franche-Comt\'e and by the French Programme National de Cosmologie et Galaxies (PNCG). M. A. De Leo would like to thank the NASA-funded FIELDS program, in partnership with JPL on a MUREP project, for their support.
\end{acknowledgements}

\bibliographystyle{aa} 
\bibliography{references}

\begin{thebibliography}{78}
\expandafter\ifx\csname natexlab\endcsname\relax\def\natexlab#1{#1}\fi

\bibitem[{{Alard}(2009)}]{alardalone}
{Alard}, C. 2009, \aap, 506, 609

\bibitem[{{Allen} {et~al.}(2011){Allen}, {Evrard}, \& {Mantz}}]{Allen2011}
{Allen}, S.~W., {Evrard}, A.~E., \& {Mantz}, A.~B. 2011, \araa, 49, 409

\bibitem[{{Andrae}(2010)}]{Andrae2010}
{Andrae}, R. 2010, ArXiv e-prints [\eprint[arXiv]{1009.2755}]

\bibitem[{{Barlow}(1989)}]{Barlow1989}
{Barlow}, R. 1989, {Statistics. A guide to the use of statistical methods in
  the physical sciences}

\bibitem[{{Beers} {et~al.}(1990){Beers}, {Flynn}, \& {Gebhardt}}]{Bee90}
{Beers}, T.~C., {Flynn}, K., \& {Gebhardt}, K. 1990, \aj, 100, 32

\bibitem[{{Beraldo e Silva} {et~al.}(2015){Beraldo e Silva}, {Mamon}, {Duarte},
  {Wojtak}, {Peirani}, \& {Bou{\'e}}}]{Beraldo2015}
{Beraldo e Silva}, L., {Mamon}, G.~A., {Duarte}, M., {et~al.} 2015, \mnras,
  452, 944

\bibitem[{{Biviano} {et~al.}(2013){Biviano}, {Rosati}, {Balestra}, {Mercurio},
  {Girardi}, {Nonino}, {Grillo}, {Scodeggio}, {Lemze}, {Kelson}, {Umetsu},
  {Postman}, {Zitrin}, {Czoske}, {Ettori}, {Fritz}, {Lombardi}, {Maier},
  {Medezinski}, {Mei}, {Presotto}, {Strazzullo}, {Tozzi}, {Ziegler},
  {Annunziatella}, {Bartelmann}, {Benitez}, {Bradley}, {Brescia}, {Broadhurst},
  {Coe}, {Demarco}, {Donahue}, {Ford}, {Gobat}, {Graves}, {Koekemoer},
  {Kuchner}, {Melchior}, {Meneghetti}, {Merten}, {Moustakas}, {Munari}, {Reg{\H
  o}s}, {Sartoris}, {Seitz}, \& {Zheng}}]{Biviano2013}
{Biviano}, A., {Rosati}, P., {Balestra}, I., {et~al.} 2013, \aap, 558, A1

\bibitem[{{Biviano} {et~al.}(2016){Biviano}, {van der Burg}, {Muzzin},
  {Sartoris}, {Wilson}, \& {Yee}}]{Biviano2016}
{Biviano}, A., {van der Burg}, R.~F.~J., {Muzzin}, A., {et~al.} 2016, ArXiv
  e-prints [\eprint[arXiv]{1605.06510}]

\bibitem[{{Bolzonella} {et~al.}(2000){Bolzonella}, {Miralles}, \& {Pell{\'
  o}}}]{hyperz}
{Bolzonella}, M., {Miralles}, J.-M., \& {Pell{\' o}}, R. 2000, \aap, 363, 476

\bibitem[{{Bond} {et~al.}(1996){Bond}, {Kofman}, \& {Pogosyan}}]{Bond1996}
{Bond}, J.~R., {Kofman}, L., \& {Pogosyan}, D. 1996, \nat, 380, 603

\bibitem[{{Brada{\v c}} {et~al.}(2008){Brada{\v c}}, {Allen}, {Treu},
  {Ebeling}, {Massey}, {Morris}, {von der Linden}, \& {Applegate}}]{Bradac2008}
{Brada{\v c}}, M., {Allen}, S.~W., {Treu}, T., {et~al.} 2008, \apj, 687, 959

\bibitem[{{Cabanac} {et~al.}(2007){Cabanac}, {Alard}, {Dantel-Fort}, {Fort},
  {Gavazzi}, {Gomez}, {Kneib}, {Le F{\`e}vre}, {Mellier}, {Pello}, {Soucail},
  {Sygnet}, \& {Valls-Gabaud}}]{Cabanac2007}
{Cabanac}, R.~A., {Alard}, C., {Dantel-Fort}, M., {et~al.} 2007, \aap, 461, 813

\bibitem[{{Colless} {et~al.}(2001){Colless}, {Dalton}, {Maddox}, {Sutherland},
  {Norberg}, {Cole}, {Bland-Hawthorn}, {Bridges}, {Cannon}, {Collins}, {Couch},
  {Cross}, {Deeley}, {De Propris}, {Driver}, {Efstathiou}, {Ellis}, {Frenk},
  {Glazebrook}, {Jackson}, {Lahav}, {Lewis}, {Lumsden}, {Madgwick}, {Peacock},
  {Peterson}, {Price}, {Seaborne}, \& {Taylor}}]{Colless2001}
{Colless}, M., {Dalton}, G., {Maddox}, S., {et~al.} 2001, \mnras, 328, 1039

\bibitem[{{Cuesta} {et~al.}(2008){Cuesta}, {Prada}, {Klypin}, \&
  {Moles}}]{Cuesta2008}
{Cuesta}, A.~J., {Prada}, F., {Klypin}, A., \& {Moles}, M. 2008, \mnras, 389,
  385

\bibitem[{{D'Aloisio} \& {Natarajan}(2011)}]{DAloisio2011}
{D'Aloisio}, A. \& {Natarajan}, P. 2011, \mnras, 411, 1628

\bibitem[{{Duffy} {et~al.}(2008){Duffy}, {Schaye}, {Kay}, \& {Dalla
  Vecchia}}]{Duff08}
{Duffy}, A.~R., {Schaye}, J., {Kay}, S.~T., \& {Dalla Vecchia}, C. 2008,
  \mnras, 390, L64

\bibitem[{{Eke} {et~al.}(2004{\natexlab{a}}){Eke}, {Baugh}, {Cole}, {Frenk},
  {Norberg}, {Peacock}, {Baldry}, {Bland-Hawthorn}, {Bridges}, {Cannon},
  {Colless}, {Collins}, {Couch}, {Dalton}, {de Propris}, {Driver},
  {Efstathiou}, {Ellis}, {Glazebrook}, {Jackson}, {Lahav}, {Lewis}, {Lumsden},
  {Maddox}, {Madgwick}, {Peterson}, {Sutherland}, \& {Taylor}}]{Eke2004}
{Eke}, V.~R., {Baugh}, C.~M., {Cole}, S., {et~al.} 2004{\natexlab{a}}, \mnras,
  348, 866

\bibitem[{{Eke} {et~al.}(2004{\natexlab{b}}){Eke}, {Frenk}, {Baugh}, {Cole},
  {Norberg}, {Peacock}, {Baldry}, {Bland-Hawthorn}, {Bridges}, {Cannon},
  {Colless}, {Collins}, {Couch}, {Dalton}, {de Propris}, {Driver},
  {Efstathiou}, {Ellis}, {Glazebrook}, {Jackson}, {Lahav}, {Lewis}, {Lumsden},
  {Maddox}, {Madgwick}, {Peterson}, {Sutherland}, \& {Taylor}}]{EkeVR2004}
{Eke}, V.~R., {Frenk}, C.~S., {Baugh}, C.~M., {et~al.} 2004{\natexlab{b}},
  \mnras, 355, 769

\bibitem[{{El{\'{\i}}asd{\'o}ttir} {et~al.}(2007){El{\'{\i}}asd{\'o}ttir},
  {Limousin}, {Richard}, {Hjorth}, {Kneib}, {Natarajan}, {Pedersen}, {Jullo},
  \& {Paraficz}}]{ardis2218}
{El{\'{\i}}asd{\'o}ttir}, {\'A}., {Limousin}, M., {Richard}, J., {et~al.} 2007,
  ArXiv e-prints, 710 [\eprint{0710.5636}]

\bibitem[{{Ettori} {et~al.}(2013){Ettori}, {Donnarumma}, {Pointecouteau},
  {Reiprich}, {Giodini}, {Lovisari}, \& {Schmidt}}]{Ettori2013}
{Ettori}, S., {Donnarumma}, A., {Pointecouteau}, E., {et~al.} 2013, \ssr, 177,
  119

\bibitem[{{Falco} {et~al.}(2013){Falco}, {Mamon}, {Wojtak}, {Hansen}, \&
  {Gottl{\"o}ber}}]{Falco2013}
{Falco}, M., {Mamon}, G.~A., {Wojtak}, R., {Hansen}, S.~H., \& {Gottl{\"o}ber},
  S. 2013, \mnras, 436, 2639

\bibitem[{{Fo{\"e}x} {et~al.}(2014){Fo{\"e}x}, {Motta}, {Jullo}, {Limousin}, \&
  {Verdugo}}]{Gael2014}
{Fo{\"e}x}, G., {Motta}, V., {Jullo}, E., {Limousin}, M., \& {Verdugo}, T.
  2014, \aap, 572, A19

\bibitem[{{Fo{\"e}x} {et~al.}(2013){Fo{\"e}x}, {Motta}, {Limousin}, {Verdugo},
  {More}, {Cabanac}, {Gavazzi}, \& {Mu{\~n}oz}}]{Gael2013}
{Fo{\"e}x}, G., {Motta}, V., {Limousin}, M., {et~al.} 2013, \aap, 559, A105

\bibitem[{{Frenk} {et~al.}(1996){Frenk}, {Evrard}, {White}, \&
  {Summers}}]{Frenk1996}
{Frenk}, C.~S., {Evrard}, A.~E., {White}, S.~D.~M., \& {Summers}, F.~J. 1996,
  \apj, 472, 460

\bibitem[{{Gastaldello} {et~al.}(2007){Gastaldello}, {Buote}, {Humphrey},
  {Zappacosta}, {Bullock}, {Brighenti}, \& {Mathews}}]{Gastaldello2007}
{Gastaldello}, F., {Buote}, D.~A., {Humphrey}, P.~J., {et~al.} 2007, \apj, 669,
  158

\bibitem[{{Gastaldello} {et~al.}(2014){Gastaldello}, {Limousin}, {Fo{\"e}x},
  {Mu{\~n}oz}, {Verdugo}, {Motta}, {More}, {Cabanac}, {Buote}, {Eckert},
  {Ettori}, {Fritz}, {Ghizzardi}, {Humphrey}, {Meneghetti}, \&
  {Rossetti}}]{Gastaldello2014}
{Gastaldello}, F., {Limousin}, M., {Fo{\"e}x}, G., {et~al.} 2014, \mnras, 442,
  L76

\bibitem[{{Gavazzi}(2005)}]{Gavazzi2005}
{Gavazzi}, R. 2005, \aap, 443, 793

\bibitem[{{Golse} \& {Kneib}(2002)}]{Golse2002}
{Golse}, G. \& {Kneib}, J.-P. 2002, \aap, 390, 821

\bibitem[{{Guennou} {et~al.}(2014){Guennou}, {Biviano}, {Adami}, {Limousin},
  {Lima Neto}, {Mamon}, {Ulmer}, {Gavazzi}, {Cypriano}, {Durret}, {Clowe},
  {LeBrun}, {Allam}, {Basa}, {Benoist}, {Cappi}, {Halliday}, {Ilbert},
  {Johnston}, {Jullo}, {Just}, {Kubo}, {M{\'a}rquez}, {Marshall}, {Martinet},
  {Maurogordato}, {Mazure}, {Murphy}, {Plana}, {Rostagni}, {Russeil},
  {Schirmer}, {Schrabback}, {Slezak}, {Tucker}, {Zaritsky}, \&
  {Ziegler}}]{Guennou2014}
{Guennou}, L., {Biviano}, A., {Adami}, C., {et~al.} 2014, \aap, 566, A149

\bibitem[{{Gwyn}(2011)}]{Gwyn2011}
{Gwyn}, S.~D.~J. 2011, ArXiv e-prints [\eprint[arXiv]{1101.1084}]

\bibitem[{{Host}(2012)}]{Host2012}
{Host}, O. 2012, \mnras, 420, L18

\bibitem[{{Irgens} {et~al.}(2002){Irgens}, {Lilje}, {Dahle}, \&
  {Maddox}}]{Irg02}
{Irgens}, R.~J., {Lilje}, P.~B., {Dahle}, H., \& {Maddox}, S.~J. 2002, \apj,
  579, 227

\bibitem[{{Jauzac} {et~al.}(2014){Jauzac}, {Cl{\'e}ment}, {Limousin},
  {Richard}, {Jullo}, {Ebeling}, {Atek}, {Kneib}, {Knowles}, {Natarajan},
  {Eckert}, {Egami}, {Massey}, \& {Rexroth}}]{Jauzac2014}
{Jauzac}, M., {Cl{\'e}ment}, B., {Limousin}, M., {et~al.} 2014, \mnras, 443,
  1549

\bibitem[{{Jauzac} {et~al.}(2012){Jauzac}, {Jullo}, {Kneib}, {Ebeling},
  {Leauthaud}, {Ma}, {Limousin}, {Massey}, \& {Richard}}]{Jauzac2012}
{Jauzac}, M., {Jullo}, E., {Kneib}, J.-P., {et~al.} 2012, \mnras, 426, 3369

\bibitem[{{Jullo} {et~al.}(2007){Jullo}, {Kneib}, {Limousin},
  {El{\'{\i}}asd{\'o}ttir}, {Marshall}, \& {Verdugo}}]{jullo07}
{Jullo}, E., {Kneib}, J.-P., {Limousin}, M., {et~al.} 2007, New Journal of
  Physics, 9, 447

\bibitem[{{Jullo} {et~al.}(2010){Jullo}, {Natarajan}, {Kneib}, {D'Aloisio},
  {Limousin}, {Richard}, \& {Schimd}}]{Jullo2010}
{Jullo}, E., {Natarajan}, P., {Kneib}, J.-P., {et~al.} 2010, Science, 329, 924

\bibitem[{{Kinney} {et~al.}(1996){Kinney}, {Calzetti}, {Bohlin}, {McQuade},
  {Storchi-Bergmann}, \& {Schmitt}}]{Kinney1996}
{Kinney}, A.~L., {Calzetti}, D., {Bohlin}, R.~C., {et~al.} 1996, \apj, 467, 38

\bibitem[{{Kneib} \& {Natarajan}(2011)}]{Kneib2011}
{Kneib}, J.-P. \& {Natarajan}, P. 2011, \aapr, 19, 47

\bibitem[{{Kurtz} \& {Mink}(1998)}]{Kurtz1998}
{Kurtz}, M.~J. \& {Mink}, D.~J. 1998, \pasp, 110, 934

\bibitem[{{Lieu} {et~al.}(2015){Lieu}, {Smith}, {Giles}, {Ziparo}, {Maughan},
  {D{\'e}mocl{\`e}s}, {Pacaud}, {Pierre}, {Adami}, {Bah{\'e}}, {Clerc},
  {Chiappetti}, {Eckert}, {Ettori}, {Lavoie}, {Le Fevre}, {McCarthy},
  {Kilbinger}, {Ponman}, {Sadibekova}, \& {Willis}}]{Lieu2015}
{Lieu}, M., {Smith}, G.~P., {Giles}, P.~A., {et~al.} 2015, ArXiv e-prints
  [\eprint[arXiv]{1512.03857}]

\bibitem[{{Limousin} {et~al.}(2009){Limousin}, {Cabanac}, {Gavazzi}, {Kneib},
  {Motta}, {Richard}, {Thanjavur}, {Foex}, {Pello}, {Crampton}, {Faure},
  {Fort}, {Jullo}, {Marshall}, {Mellier}, {More}, {Soucail}, {Suyu},
  {Swinbank}, {Sygnet}, {Tu}, {Valls-Gabaud}, {Verdugo}, \& {Willis}}]{paperI}
{Limousin}, M., {Cabanac}, R., {Gavazzi}, R., {et~al.} 2009, \aap, 502, 445

\bibitem[{{Limousin} {et~al.}(2010){Limousin}, {Jullo}, {Richard}, {Cabanac},
  {Suyu}, {Halkola}, {Kneib}, {Gavazzi}, \& {Soucail}}]{Limousin2010}
{Limousin}, M., {Jullo}, E., {Richard}, J., {et~al.} 2010, \aap, 524, A95

\bibitem[{{Limousin} {et~al.}(2005){Limousin}, {Kneib}, \&
  {Natarajan}}]{Limousin2005}
{Limousin}, M., {Kneib}, J.-P., \& {Natarajan}, P. 2005, \mnras, 356, 309

\bibitem[{{Limousin} {et~al.}(2013){Limousin}, {Morandi}, {Sereno},
  {Meneghetti}, {Ettori}, {Bartelmann}, \& {Verdugo}}]{Limousin2013}
{Limousin}, M., {Morandi}, A., {Sereno}, M., {et~al.} 2013, \ssr, 177, 155

\bibitem[{{Limousin} {et~al.}(2007){Limousin}, {Richard}, {Jullo}, {Kneib},
  {Fort}, {Soucail}, {El{\'{\i}}asd{\'o}ttir}, {Natarajan}, {Ellis}, {Smail},
  {Czoske}, {Smith}, {Hudelot}, {Bardeau}, {Ebeling}, {Egami}, \&
  {Knudsen}}]{Limousin2007}
{Limousin}, M., {Richard}, J., {Jullo}, E., {et~al.} 2007, \apj, 668, 643

\bibitem[{{{\L}okas} \& {Mamon}(2003)}]{Lokas2003}
{{\L}okas}, E.~L. \& {Mamon}, G.~A. 2003, \mnras, 343, 401

\bibitem[{{Mamon} {et~al.}(2013){Mamon}, {Biviano}, \& {Bou{\'e}}}]{Mamon2013}
{Mamon}, G.~A., {Biviano}, A., \& {Bou{\'e}}, G. 2013, \mnras, 429, 3079

\bibitem[{{Mamon} \& {Bou{\'e}}(2010)}]{Mamon2010}
{Mamon}, G.~A. \& {Bou{\'e}}, G. 2010, \mnras, 401, 2433

\bibitem[{{Menanteau} {et~al.}(2012){Menanteau}, {Hughes}, {Sif{\'o}n},
  {Hilton}, {Gonz{\'a}lez}, {Infante}, {Barrientos}, {Baker}, {Bond}, {Das},
  {Devlin}, {Dunkley}, {Hajian}, {Hincks}, {Kosowsky}, {Marsden}, {Marriage},
  {Moodley}, {Niemack}, {Nolta}, {Page}, {Reese}, {Sehgal}, {Sievers},
  {Spergel}, {Staggs}, \& {Wollack}}]{Menanteau2012}
{Menanteau}, F., {Hughes}, J.~P., {Sif{\'o}n}, C., {et~al.} 2012, \apj, 748, 7

\bibitem[{{Meneghetti} {et~al.}(2016){Meneghetti}, {Natarajan}, {Coe},
  {Contini}, {De Lucia}, {Giocoli}, {Acebron}, {Borgani}, {Bradac}, {Diego},
  {Hoag}, {Ishigaki}, {Johnson}, {Jullo}, {Kawamata}, {Lam}, {Limousin},
  {Liesenborgs}, {Oguri}, {Sebesta}, {Sharon}, {Williams}, \&
  {Zitrin}}]{Meneghetti2016}
{Meneghetti}, M., {Natarajan}, P., {Coe}, D., {et~al.} 2016, ArXiv e-prints
  [\eprint[arXiv]{1606.04548}]

\bibitem[{{Mu{\~n}oz} {et~al.}(2013){Mu{\~n}oz}, {Motta}, {Verdugo}, {Garrido},
  {Limousin}, {Padilla}, {Fo{\"e}x}, {Cabanac}, {Gavazzi}, {Barrientos}, \&
  {Richard}}]{Roberto2013}
{Mu{\~n}oz}, R.~P., {Motta}, V., {Verdugo}, T., {et~al.} 2013, \aap, 552, A80

\bibitem[{{Mulchaey}(2000)}]{Mulchaey2000}
{Mulchaey}, J.~S. 2000, \araa, 38, 289

\bibitem[{{Munari} {et~al.}(2014){Munari}, {Biviano}, \& {Mamon}}]{Munari2014}
{Munari}, E., {Biviano}, A., \& {Mamon}, G.~A. 2014, \aap, 566, A68

\bibitem[{{Navarro} {et~al.}(1996){Navarro}, {Frenk}, \& {White}}]{Navarro1996}
{Navarro}, J.~F., {Frenk}, C.~S., \& {White}, S.~D.~M. 1996, \apj, 462, 563

\bibitem[{{Navarro} {et~al.}(1997){Navarro}, {Frenk}, \& {White}}]{nav97}
{Navarro}, J.~F., {Frenk}, C.~S., \& {White}, S.~D.~M. 1997, \apj, 490, 493

\bibitem[{{Newman} {et~al.}(2013){Newman}, {Treu}, {Ellis}, {Sand}, {Nipoti},
  {Richard}, \& {Jullo}}]{new13}
{Newman}, A.~B., {Treu}, T., {Ellis}, R.~S., {et~al.} 2013, \apj, 765, 24

\bibitem[{{Newman} {et~al.}(2009){Newman}, {Treu}, {Ellis}, {Sand}, {Richard},
  {Marshall}, {Capak}, \& {Miyazaki}}]{new09}
{Newman}, A.~B., {Treu}, T., {Ellis}, R.~S., {et~al.} 2009, \apj, 706, 1078

\bibitem[{{Old} {et~al.}(2014){Old}, {Skibba}, {Pearce}, {Croton}, {Muldrew},
  {Mu{\~n}oz-Cuartas}, {Gifford}, {Gray}, {der Linden}, {Mamon}, {Merrifield},
  {M{\"u}ller}, {Pearson}, {Ponman}, {Saro}, {Sepp}, {Sif{\'o}n}, {Tempel},
  {Tundo}, {Wang}, \& {Wojtak}}]{Old2014}
{Old}, L., {Skibba}, R.~A., {Pearce}, F.~R., {et~al.} 2014, \mnras, 441, 1513

\bibitem[{{Old} {et~al.}(2015){Old}, {Wojtak}, {Mamon}, {Skibba}, {Pearce},
  {Croton}, {Bamford}, {Behroozi}, {de Carvalho}, {Mu{\~n}oz-Cuartas},
  {Gifford}, {Gray}, {der Linden}, {Merrifield}, {Muldrew}, {M{\"u}ller},
  {Pearson}, {Ponman}, {Rozo}, {Rykoff}, {Saro}, {Sepp}, {Sif{\'o}n}, \&
  {Tempel}}]{Old2015}
{Old}, L., {Wojtak}, R., {Mamon}, G.~A., {et~al.} 2015, \mnras, 449, 1897

\bibitem[{{Postman} {et~al.}(2005){Postman}, {Franx}, {Cross}, {Holden},
  {Ford}, {Illingworth}, {Goto}, {Demarco}, {Rosati}, {Blakeslee}, {Tran},
  {Ben{\'{\i}}tez}, {Clampin}, {Hartig}, {Homeier}, {Ardila}, {Bartko},
  {Bouwens}, {Bradley}, {Broadhurst}, {Brown}, {Burrows}, {Cheng}, {Feldman},
  {Golimowski}, {Gronwall}, {Infante}, {Kimble}, {Krist}, {Lesser}, {Martel},
  {Mei}, {Menanteau}, {Meurer}, {Miley}, {Motta}, {Sirianni}, {Sparks}, {Tran},
  {Tsvetanov}, {White}, \& {Zheng}}]{Postman2005}
{Postman}, M., {Franx}, M., {Cross}, N.~J.~G., {et~al.} 2005, \apj, 623, 721

\bibitem[{{Prada} {et~al.}(2006){Prada}, {Klypin}, {Simonneau},
  {Betancort-Rijo}, {Patiri}, {Gottl{\"o}ber}, \& {Sanchez-Conde}}]{Prada2006}
{Prada}, F., {Klypin}, A.~A., {Simonneau}, E., {et~al.} 2006, \apj, 645, 1001

\bibitem[{{Randall} {et~al.}(2008){Randall}, {Markevitch}, {Clowe}, {Gonzalez},
  \& {Brada{\v c}}}]{Randall2008}
{Randall}, S.~W., {Markevitch}, M., {Clowe}, D., {Gonzalez}, A.~H., \&
  {Brada{\v c}}, M. 2008, \apj, 679, 1173

\bibitem[{{Sand} {et~al.}(2002){Sand}, {Treu}, \& {Ellis}}]{sand02}
{Sand}, D.~J., {Treu}, T., \& {Ellis}, R.~S. 2002, \apjl, 574, L129

\bibitem[{{Sand} {et~al.}(2004){Sand}, {Treu}, {Smith}, \& {Ellis}}]{sand04}
{Sand}, D.~J., {Treu}, T., {Smith}, G.~P., \& {Ellis}, R.~S. 2004, \apj, 604,
  88

\bibitem[{{Sarli} {et~al.}(2014){Sarli}, {Meyer}, {Meneghetti}, {Konrad},
  {Majer}, \& {Bartelmann}}]{Sarli2014}
{Sarli}, E., {Meyer}, S., {Meneghetti}, M., {et~al.} 2014, \aap, 570, A9

\bibitem[{{Springel} {et~al.}(2005){Springel}, {White}, {Jenkins}, {Frenk},
  {Yoshida}, {Gao}, {Navarro}, {Thacker}, {Croton}, {Helly}, {Peacock}, {Cole},
  {Thomas}, {Couchman}, {Evrard}, {Colberg}, \& {Pearce}}]{Springel2005}
{Springel}, V., {White}, S.~D.~M., {Jenkins}, A., {et~al.} 2005, \nat, 435, 629

\bibitem[{{Sun}(2012)}]{Sun2012}
{Sun}, M. 2012, New Journal of Physics, 14, 045004

\bibitem[{{Thanjavur} {et~al.}(2010){Thanjavur}, {Crampton}, \&
  {Willis}}]{Thanjavur2010}
{Thanjavur}, K., {Crampton}, D., \& {Willis}, J. 2010, \apj, 714, 1355

\bibitem[{{Tully}(2014)}]{Tully2014}
{Tully}, R.~B. 2014, ArXiv e-prints [\eprint[arXiv]{1411.1511}]

\bibitem[{{Umetsu} {et~al.}(2015){Umetsu}, {Sereno}, {Medezinski}, {Nonino},
  {Mroczkowski}, {Diego}, {Ettori}, {Okabe}, {Broadhurst}, \&
  {Lemze}}]{Umetsu2015}
{Umetsu}, K., {Sereno}, M., {Medezinski}, E., {et~al.} 2015, \apj, 806, 207

\bibitem[{{Verdugo} {et~al.}(2007){Verdugo}, {de Diego}, \& {Limousin}}]{ver07}
{Verdugo}, T., {de Diego}, J.~A., \& {Limousin}, M. 2007, \apj, 664, 702

\bibitem[{{Verdugo} {et~al.}(2014){Verdugo}, {Motta}, {Fo{\"e}x},
  {Forero-Romero}, {Mu{\~n}oz}, {Pello}, {Limousin}, {More}, {Cabanac},
  {Soucail}, {Blakeslee}, {Mej{\'{\i}}a-Narv{\'a}ez}, {Magris}, \&
  {Fern{\'a}ndez-Trincado}}]{Verdugo2014}
{Verdugo}, T., {Motta}, V., {Fo{\"e}x}, G., {et~al.} 2014, \aap, 571, A65

\bibitem[{{Verdugo} {et~al.}(2011){Verdugo}, {Motta}, {Mu{\~n}oz}, {Limousin},
  {Cabanac}, \& {Richard}}]{Verdugo2011}
{Verdugo}, T., {Motta}, V., {Mu{\~n}oz}, R.~P., {et~al.} 2011, \aap, 527, A124

\bibitem[{{Wilman} {et~al.}(2005){Wilman}, {Balogh}, {Bower}, {Mulchaey},
  {Oemler}, {Carlberg}, {Morris}, \& {Whitaker}}]{Wil05}
{Wilman}, D.~J., {Balogh}, M.~L., {Bower}, R.~G., {et~al.} 2005, \mnras, 358,
  71

\bibitem[{{Wojtak} {et~al.}(2009){Wojtak}, {{\L}okas}, {Mamon}, \&
  {Gottl{\"o}ber}}]{Wojtak2009}
{Wojtak}, R., {{\L}okas}, E.~L., {Mamon}, G.~A., \& {Gottl{\"o}ber}, S. 2009,
  \mnras, 399, 812

\bibitem[{{Wojtak} {et~al.}(2008){Wojtak}, {{\L}okas}, {Mamon},
  {Gottl{\"o}ber}, {Klypin}, \& {Hoffman}}]{Wojtak2008}
{Wojtak}, R., {{\L}okas}, E.~L., {Mamon}, G.~A., {et~al.} 2008, \mnras, 388,
  815

\bibitem[{{Wolf} {et~al.}(2010){Wolf}, {Martinez}, {Bullock}, {Kaplinghat},
  {Geha}, {Mu{\~n}oz}, {Simon}, \& {Avedo}}]{Wolf2010}
{Wolf}, J., {Martinez}, G.~D., {Bullock}, J.~S., {et~al.} 2010, \mnras, 406,
  1220

\bibitem[{{Zitrin} {et~al.}(2015){Zitrin}, {Fabris}, {Merten}, {Melchior},
  {Meneghetti}, {Koekemoer}, {Coe}, {Maturi}, {Bartelmann}, {Postman},
  {Umetsu}, {Seidel}, {Sendra}, {Broadhurst}, {Balestra}, {Biviano}, {Grillo},
  {Mercurio}, {Nonino}, {Rosati}, {Bradley}, {Carrasco}, {Donahue}, {Ford},
  {Frye}, \& {Moustakas}}]{Zitrin2015}
{Zitrin}, A., {Fabris}, A., {Merten}, J., {et~al.} 2015, \apj, 801, 44

\end{thebibliography}

  
\clearpage

\onecolumn

\begin{appendix}
\section{Spectroscopic and photometric data} \label{sec:ap1}

\clearpage

 

\begin{table}
\caption{Spectroscopic data of the confirmed members of the group.}
\label{tbl-A1} 
\centering 
\begin{tabular}{lccc}
\hline\hline 
\\

 \multicolumn{1}{c}{RA} & \multicolumn{1}{c}{DEC} & \multicolumn{1}{c}{$z_{\rm spec}$}     & \multicolumn{1}{c}{R}  
\\
\multicolumn{1}{c}{ } & \multicolumn{1}{c}{ } & \multicolumn{1}{c}{ }    & \multicolumn{1}{c}{[kpc]} 
       
\\
\hline 
\\
33.533779 &  -5.592632 &   0.4446$^{\dagger}$       &  -- \\
 33.533501   &    -5.591930   &    0.4449$^{\dagger}$        &   7.9 \\
33.530430   &  -5.594814   &    0.4474$^{\dagger}$       &    79.8  \\
33.527908    &    -5.597961    &     0.4443           &     161.0 \\
33.536942   &    -5.582868   &  0.4430$^{\dagger}$     &     204.3   \\
33.540424    &   -5.584474    &    0.4427$^{\dagger}$         &    210.7 \\
33.519188    &    -5.601521   &    0.4459$^{\dagger}$       &    349.6 \\
33.546135   &   -5.607511   &  0.4436$^{\dagger}$     &      398.0 \\
33.514061   &   -5.595108   &    0.4455        &     405.4 \\
33.521729    &   -5.574636   &  0.4462         &       439.8 \\
33.512676    &   -5.596797    &  0.4473$^{\dagger}$      &   440.5 \\
33.510559    &  -5.596503   &   0.4442       &     480.4 \\
33.515137    &    -5.577593   &   0.4442$^{\dagger}$     &      486.3 \\
33.519920    &    -5.569031    &     0.4449            &    556.8 \\
33.548912    &   -5.616460   &    0.4426$^{\dagger}$       &      580.2 \\
33.512367    &   -5.573329   &    0.4440$^{\dagger}$     &  586.5 \\
33.543442    &    -5.557844     &     0.4424         &   735.2 \\
33.555248    &    -5.621617    &    0.4471       &       745.7 \\
33.563099    &    -5.561144   &    0.4465          &   883.5 \\
33.550777    &    -5.551144    &     0.4438$^{\dagger}$        &      916.6 \\
33.500538    &    -5.558484   &     0.4459$^{\dagger}$      &       976.6 \\
33.484375    &   -5.623324   &     0.4436$^{\dagger}$       &     1194.2 \\
33.479259    &    -5.613928    &     0.4435$^{\dagger}$      &      1201.0\\
33.475819    &    -5.627750    &    0.4438       &   1392.3 \\
\\
\hline 
\end{tabular}
\tablefoot{($\dagger$): Previously reported in \citet{Roberto2013} \\
Column (1) and (2): Right ascension and declination. Column (3): Redshift. Column (4): The projected radius measured with respect to the BGG.}
\end{table}

\begin{table}
\caption{Results from photometry in arc A.}
\label{tbl-A2} 
\centering 
\begin{tabular}{lcccc}
\hline\hline 
\\
\multicolumn{1}{c}{ID} & \multicolumn{1}{c}{$J$} & \multicolumn{1}{c}{$K_s$} & \multicolumn{1}{c}{$z_{\rm phot}$}  & \multicolumn{1}{c}{$z_{\rm spec}$}
\\
\hline 
\\
$A$  &  19.9 $\pm$  0.4  & 18.2 $\pm$ 0.4  &   1.7 $\pm$ 0.1  &   1.628 $\pm$ 0.001 \\
$B$  &  21.6 $\pm$  1.0  & 21.1 $\pm$ 0.9  &   1.6 $\pm$ 0.2  &   -- \\
$C$  &  20.2 $\pm$  0.2  & 18.7 $\pm$ 0.3  &   0.96 $\pm$ 0.07  &   1.017 $\pm$ 0.001 \\
\\
\hline 
\end{tabular}
\tablefoot{Column 1 lists the identification for each object as in Fig. 1. Cols. 2 and 3 are the WIRCam magnitudes. Columns 4 and 5 list the photometric and spectroscopic redshifts, respectively.}
\end{table}

\clearpage
 
 \section{Spherical lensing model} \label{sec:ap2}
 
\begin{figure}[h!]\begin{center}
 \vspace{0.5cm}
\includegraphics[scale=0.5]{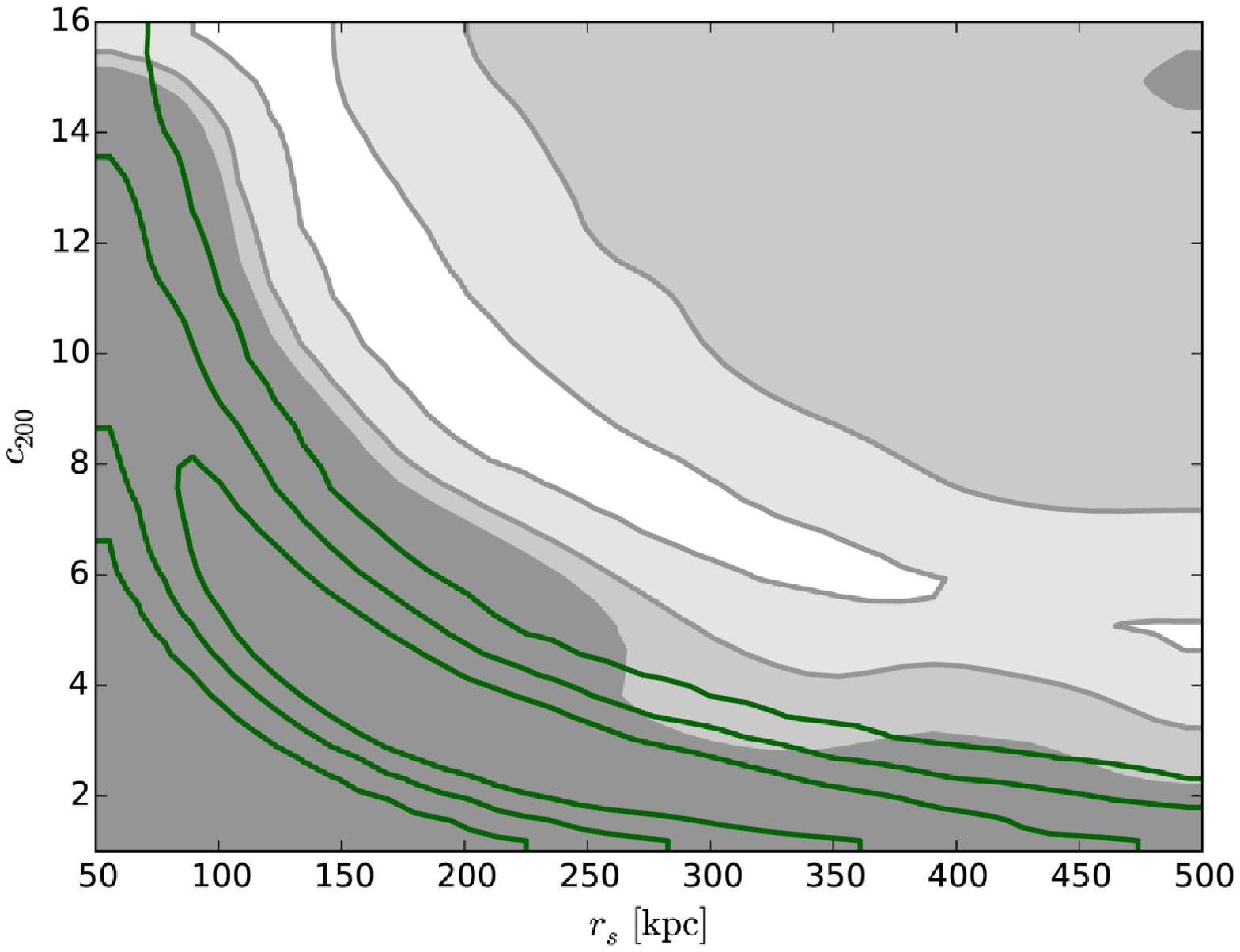}
 \hspace{0.5cm}
\includegraphics[scale=0.5]{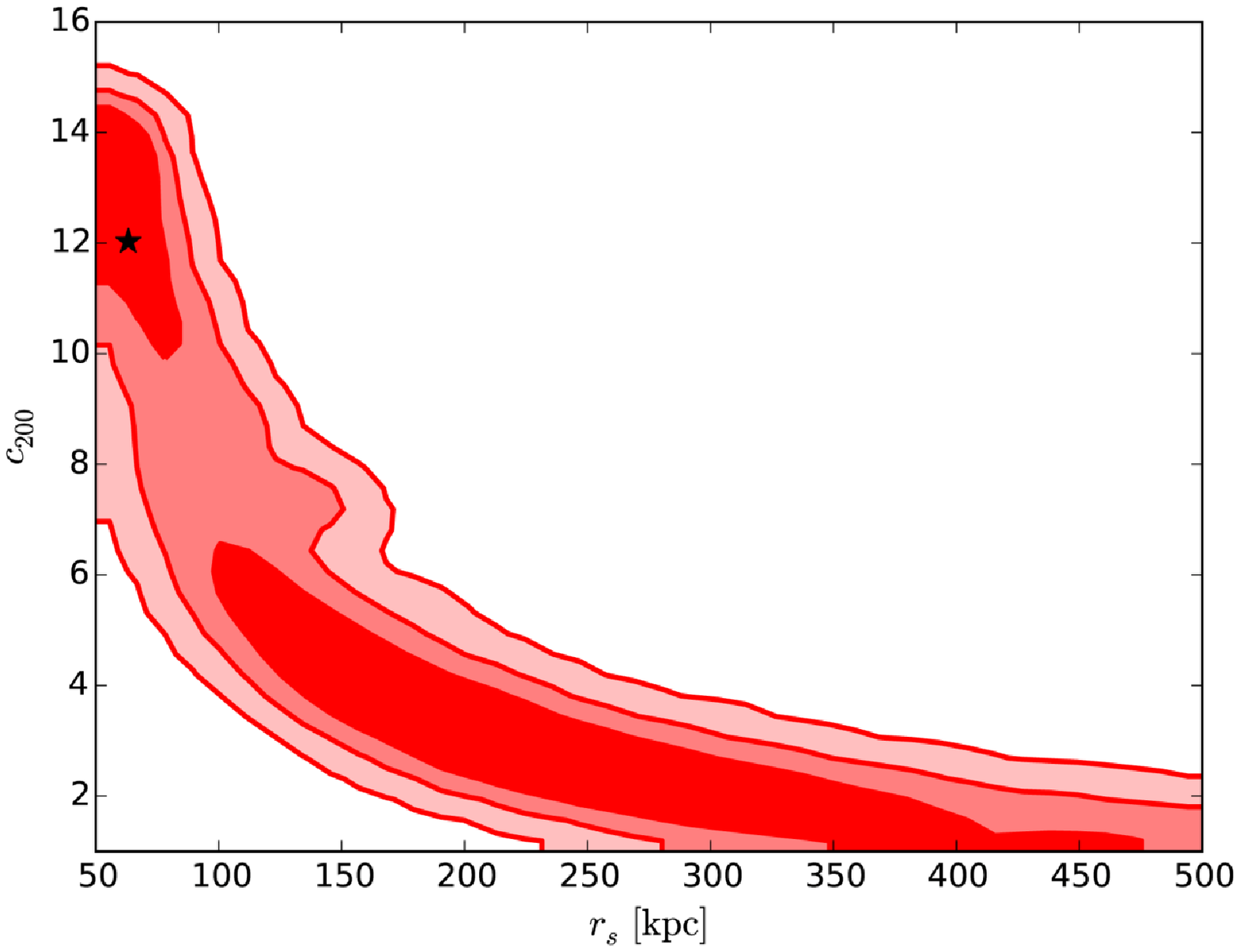}
\caption{Joint distributions (scale radius and concentration) for the spherical model. \textit{ Left panel.-} From dark to white grey-filled contours are 1, 2 and 3-$\sigma$ regions from the \textrm{SL\,Model}, and green contours stand for the 68, 95, and 99\% confidence levels for the \textrm{Dyn\,Model}.  \textit{ Right panel.-} Red-filled contours is the result of the \textrm{SL+Dyn\,Model},  with the best solution depicted with a  black star.}
\label{figA2} \end{center} 
\end{figure}

To compare the spherical \textrm{Dyn\,Model} with a spherical lensing model, we construct an additional model using LENSTOOL.  We set the ellipticity and the position angle equal to zero, and we use the same constraints than those used in the elliptical case. Since lensing spherical models tend to be poorly constrained, the parameter $c_{200}$ is free to range between 1 $\leq$ $c_{200}$ $\leq$ 16, avoiding large unphysical values and reducing the possible solutions in the $r_s$-$c_{200}$ parameter space. Finally we also set  $\sigma_{ij}$ = 3 $\arcsec$, in order to obtain a reduced $\chi^{2}$ near unity. We show the result in  Fig.~\ref{figA2}. For clarity, the colors are reversed with respect to the plots in the main text, grey-filled contours depict the result of the spherical \textrm{SL\,Model}, and green contours the result of the \textrm{Dyn\,Model}.
 
\begin{figure}[h!]\begin{center}
\includegraphics[scale=0.6]{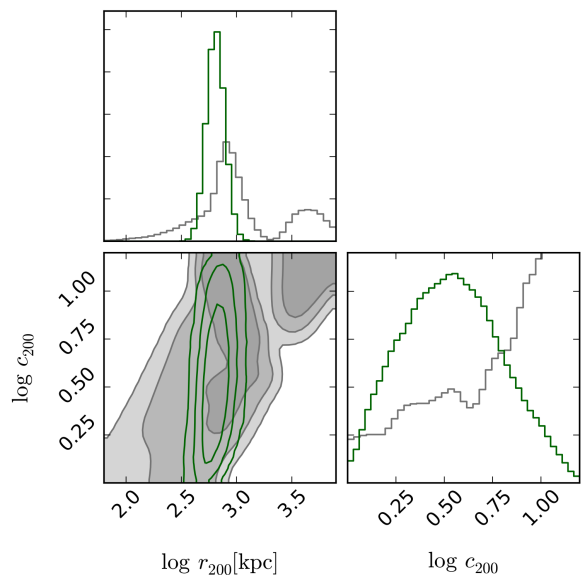}
\caption{PDFs and contours of the parameters log\,$c_{200}$ and log\,$r_{200}$. From dark to white grey-filled contours are 1, 2 and 3-$\sigma$ regions from the \textrm{SL\,Model}, and green contours stand for the 68, 95, and 99\% confidence levels for the \textrm{Dyn\,Model}.}
\label{figA3} \end{center} 
\end{figure}

To highlight further the result of the comparison between models, we present in Fig.~\ref{figA3} the solutions in the log\,$c_{200}$$-$log\,$r_{200}$ space. Note that exists a clear tendency to greater values of concentration in the SL model, with a bimodal distribution in $r_{200}$. The log\,$c_{200}$$-$log\,$r_{200}$ parameter space also makes evident the existence of a possible bimodal solution in the combined model, which is consistent with the result depicted in the right panel of Fig.~\ref{figA2}.

\end{appendix}



 \end{document}